\documentclass{article}

\usepackage{arxiv}

\usepackage[utf8]{inputenc} 
\usepackage[T1]{fontenc}    
\usepackage{hyperref}       
\usepackage{url}            
\usepackage{booktabs}       
\usepackage{amsfonts}       
\usepackage{nicefrac}       
\usepackage{microtype}      
\usepackage{lipsum}
\usepackage{graphicx}
\usepackage{amsmath} 
\usepackage{bm}

\usepackage[font=footnotesize]{caption}
\usepackage[font=footnotesize]{subcaption}
\graphicspath{ {./images/} }

\title{The effect of a toroidal opinion space on opinion bi-polarisation}

\author{
 Frank P.\ Pijpers \\
  Statistics Netherlands\\
  The Hague\\
   \& \\
  Korteweg-de Vries Institute for Mathematics\\
  University of Amsterdam\\
  Amsterdam\\
  \texttt{f.p.pijpers[at]uva.nl} \\
  \And
 Benedikt V.\ Meylahn \\
  Korteweg-de Vries Institute for Mathematics,\\
  Institute of Physics, \\
   Dutch Institute for Emergent Phenomena\\
  University of Amsterdam\\
  Amsterdam\\
  \texttt{b.v.meylahn[at]uva.nl} \\
   \And
 Michel R.H.\ Mandjes \\
  Korteweg-de Vries Institute for Mathematics\\
  University of Amsterdam\\
  Amsterdam\\
  \& \\
  Mathematical Institute\\
  Leiden University\\
  Leiden\\
  \texttt{m.r.h.mandjes[at]uva.nl} \\
}

\begin{document}
\maketitle
\begin{abstract}
Many models of opinion dynamics include measures of distance between opinions. Such models are susceptible to boundary effects where the choice of the topology of the opinion space may influence the dynamics. In this paper we study an opinion dynamics model following the seminal model by Axelrod, with the goal of understanding the effect of a toroidal opinion space. To do this we systematically compare two versions of the model: one with toroidal opinion space and one with cubic opinion space. 

In their most basic form the two versions of our model result in similar dynamics (consensus is attained eventually). However, as we include bounded confidence and eventually per agent weighting of opinion elements the dynamics become quite contrasting. The toroidal opinion space consistently allows for a greater number of groups in steady state than the cubic opinion space model. Furthermore, the outcome of the dynamics in the toroidal opinion space model are more sensitive to the inclusion of extensions than in the cubic opinion space model. 

\end{abstract}




\section{Introduction}
Stylized models of opinion dynamics are vast and varied. One of the central questions in this line of literature is what enables (or perhaps even encourages) opinion polarization. For an overview of opinion dynamics modelling we refer the interested reader to the reviews~\cite{Castellano2009,Flache2017,Noorazar2020,Jusup2022,Liu2026}.

One of the common assumptions implemented (though, seldom explicitly discussed) in these models is that the opinions held by agents exist in a space with boundaries and extremes. This may take the shape of some discrete set such as $\{-1,1\}$ (for example in~\cite{Holley1975,Sznajd2000,Galam2002}) or a finite interval $[-1,1]$ (in for example~\cite{DeGroot1974,Hegselmann2002,Chan2024}). If one considers a range of topics or opinion elements (with extremal values) we get a multidimensional opinion space resulting in a hypercube.

Recently some researchers have started to question the hidden assumption of extremes. For example Chan \textit{et al.}~\cite{Chan2024} study opinion formation under the assumption that it is not the nominal opinions that matters, but rather the opinions \textit{relative} to each other which influence the outcome of an interaction between individuals. The resulting dynamics are rich, illustrating the possibility of consensus formation, as well as periodic patterns, and polarization. Caponigro \textit{et al.}~\cite{Caponigro2015} and Aydo\v{g}du \textit{et al.}~\cite{Aydogdu2017}, study opinion formation as differential equations with opinions placed on general manifolds including $\mathbb{R}^2,\mathbb{S}^2$, and $\mathbb{T}^2$. By enriching the opinion space they find possibilities for `dancing' equilibria where relative to one another the opinions do not change, yet as a whole they move around on the sphere.

We continue in this line of investigation by comparing a version of the Axelrod~\cite{Axelrod1997} model where the space in which the opinion vectors are embedded is toroidal to one in which the space is cubic. The toroidal opinion space removes the notion of an extreme opinion. We systematically study the effect of the assumption on the opinion space by adding elements to both models and comparing the resulting dynamics. 

Our main contribution is a systematic comparison of the dynamics in the toroidal and cubic opinion spaces, aimed at understanding how the underlying opinion space influences collective behavior. We find that, in the absence of model extensions, the two spaces produce very similar dynamics. However, once extensions are introduced, their differences become more pronounced. In particular, the toroidal opinion space generally supports a greater diversity of steady-state opinions, marking a shift away from the consensus outcome observed in both models without extensions. These results highlight that the choice of opinion space — often an implicit modeling assumption in the literature — can substantially affect the resulting dynamics.

Studying the difference between toroidal opinion spaces and cubic ones also relates to assumptions on how flexible we are as humans. Where in the cubic space a jump from one extreme to the other has only one possible path, through the centre. In the toroidal opinion space there are two ways to get to there. This added mobility however doesn't seem to result in more consensus as one might expect. Rather it gives the agents the mobility to hide away from one another and form more isolated groups.

In the next section we present the basic model discussing how it is inspired by, and yet subtly different from, the seminal work of Axelrod~\cite{Axelrod1997}. \S\ref{sec:results} contains the description of the extensions to the model as well as the presentation of the resulting dynamics. We close the paper in \S\ref{sec:discussion} with a discussion of our work and its significance.

\section{The model}\label{sec:basic}
The model we study is heavily influenced by the seminal model of culture-formation by Robert Axelrod~\cite{Axelrod1997}. We study the model by computer simulation. Note that in our simulation experiments we always use a grid as the population structure (or social network). Performing similar experiments on any other population structure should not pose any difficulty. The grid is a natural starting point when it comes to studying structured populations, and allows us to focus on the effects of the opinion space rather than having to disentangle those from the effects of the network structure itself.

We do this heeding the caution of~\cite{Meylahn2024c} who, studying the model of Banisch and Olbrich~\cite{Banisch2019}, illustrate that small changes in the details of the agent-agent interaction can have profound consequences on the global dynamics. In fact these consequences can be considered bigger than the specifics of the network upon which these interactions take place. In that example, making the interaction symmetric with respect to the agents changed the global dynamics from absorbing to ergodic, while the effect of the network is only one of time-scales. As such we aim to understand the nature of the switch from cubic to toroidal opinion spaces. Studying how this change interplays with changes in the network topology is left for future work.

\subsection{The basic model}

We model a population of $N$ agents where each agents $i\in\{1,\ldots, N\}$  is endowed with an opinion profile $s_i$ (initially to be drawn uniformly at random). The opinion profile is a vector of length $M$ and each element of this vector takes integer values in $\{1,2,\ldots,q\}$. Thus $s_i\in\{1,\ldots,q\}^M$. The population structure (in our case a grid) is represented by the adjacency matrix $A$ where $A_{ij}=1$ implies that agents $i$ and $j$ are connected by an edge.

In our base model per discrete time step:
\begin{itemize}
    \item An agent $i$ is selected uniformly at random.;u This agent selects one of their neighbours $j$ ($A_{ij}=1$) uniformly at random.
    \item The distance between their opinions $d_{ij}$ is calculated.
    \item At probability $f(d_{ij})$ they \textit{interact}, meaning agent $i$ adjusts their opinion to be more like that of agent $j$ according to \[s_i^k\leftarrow s_i^k \pm 1,\] for the $k$ at which the distance between their opinions is largest, and such that $d_{ij}$ decreases.
\end{itemize}
The function $f(\cdot)$ is defined as a monotonic decreasing function:
\begin{equation} \label{eq:fa}
f(d) = 1-\left(\frac{d}{qM- M}\right)^\alpha.
\end{equation}

There are a couple of important differences between our model and that of Axelrod. The first is that we define the similarity (and therewith the interaction probability) of opinions to be related to the \textit{distance} between them rather than a proportion of shared views. In the Axelrod model the interaction probability is a function of the number of opinion elements where the agents share the same view. A related difference is the opinion updating mechanism. The agents in our model approach one another by `stepping' in the direction in one opinion element whereas the agents in the Axelrod model did what we would consider a jump to copy the value of one of the elements where they differ in views. In the Axelrod model however, this need not be considered a `jump' because the values opinions can take are not ordinal in that model.

Throughout the paper we will systematically compare results and dynamics obtained when distance between between agents views is calculated as the (Manhattan) distance $d_{ij}$ on a cube

\begin{equation}\label{eq:disC}
    d^C_{ij}:=\sum_{k=1}^M|s_i^k-s_j^k|, 
\end{equation}
and on a torus
\begin{equation}\label{eq:disT}
    d^T_{ij}:=\sum_{k=1}^M\frac{q-1}{\lfloor q/2\rfloor}\min(|s_i^k-s_j^k|,q-| s_i^k-s_j^k|).
\end{equation}
The factor $\frac{q-1}{\lfloor q/2\rfloor}$ in the toroidal distance is a normalization factor making the distances in the two spaces comparable. In particular it is chosen such that the maximum attainable distance in the cubic and the toroidal space is the same:~$qM-M$.

By design, in (\ref{eq:fa}) the probability of interaction when the agents have the same opinion vector is 1 however the result of an interaction does not change the state of the system (agent's views cannot become more similar if they are already identical). Whereas, if opinions are maximally apart ($d=qM-M$), there is no chance of interaction. We consider this qualitative similarity to the Axelrod model desirable. Dynamics will only stop completely once all agents have the same opinion or are placed on `opposite' ends of the torus. In the Axelrod model there were more possible steady states being any configuration in which each pair of neighbouring edges shares no (or all) views. This drastic cut in the number of steady states is because the interaction between two agents happens at positive probability as long as their views are not maximally apart.

\subsection{Extensions to the model}
To thoroughly compare the dynamics resulting from a toroidal and cubic opinion space we study two extensions. The first is \textit{bounded confidence} (see for example~\cite{Deffuant2000, Hegselmann2002} and Section IV.B.5 of~\cite{starnini2025} for a recent review) whereby a maximal distance is defined beyond which agents consider each other `too far apart' and thus share no social influence (probability of interaction becomes zero). The second is a further augmentation of the opinion space by \textit{personal opinion element weighting} per agent. This way each agent decides for themself the relative importance of opinion elements when calculating the distance between their views and those of their neighbour. Finally, to check whether the results are robust to different network topologies we add the extension of network rewiring. This changes the 2-dimensional grid into something of a small-world network by specifying a rewiring probability and rewiring each edge at that probability before commencing the model dynamics. These extensions are described in detail as they appear in the paper.

The main paper formulates the dynamics of the system in terms of probabilities for a change of state of any node and carries out the probabilistic simulations. In recent years some of the tools that have been developed in the field of statistical physics also have been brought to bear on the problem of opinion dynamics. By formulating 
transition probabilities on the basis of a Hamiltonian of the interaction, using either Glauber dynamics \cite{Glau1963} or the Metropolis-Hastings algorithm \cite{Metroetal53,Hast70}, many of the general results of Hamiltonian dynamical systems can be carried over immediately into this new setting. In appendix B it is presented briefly what Hamiltonian gives rise to the transition probabilities used, including bounded confidence. While the results of the main paper can be understood without this alternative paradigm, the Hamiltonian approach shown in appendix B gives valuable additional insights. 

\section{Results}\label{sec:results}
We simulate the model described above for 100 agents arranged on a $10\times 10$ grid, having 3-dimensional opinions ($M=3$) taking values in $\{1,\ldots,8\}$ ($q=8$). In order to investigate the timescales of the resulting dynamics we study the absorption time $\tau$:
\begin{equation}
    \tau:=\min \Bigg\{t: \forall (i,j) \text{ with } A_{ij}=1,
    \begin{cases}
        d_{ij}(t)=0, \text{or }\\
        d_{ij}(t): f(d_{ij})=0
    \end{cases} \Bigg\}.
\end{equation}
After $\tau$ no more dynamics can take place because for all pairs of connected agents either the probability of them interacting is zero or the outcome of an interaction would not change the state of the system. To study the emergence of opinion bi-polarisation we study the number of `species' over time $t$:
\begin{equation}
    S(t):= |\{s_i(t) \mid i\in N\}|.
\end{equation} 
That is the number of unique opinion profiles in the population at time $t$. When $S(\tau)=2$ we know that opinion bi-polarisation is taking place. It should be noted that this is a useful measure despite one limitation: it does not indicate that the number of groups $S(t)$ are contiguous. It is possible (but unlikely) that there are 2 species at absorption time but these are dispersed across the population structure in more than two groups. 

All parameter settings for the basic and the extended models were repeated for 1000 iterations up until a maximum time step of 250\,000.

\subsection{Basic model}
\begin{figure}[htb!]
    \centering
    \begin{subfigure}{0.5\textwidth}
    \centering
        \includegraphics[width=0.95\textwidth]{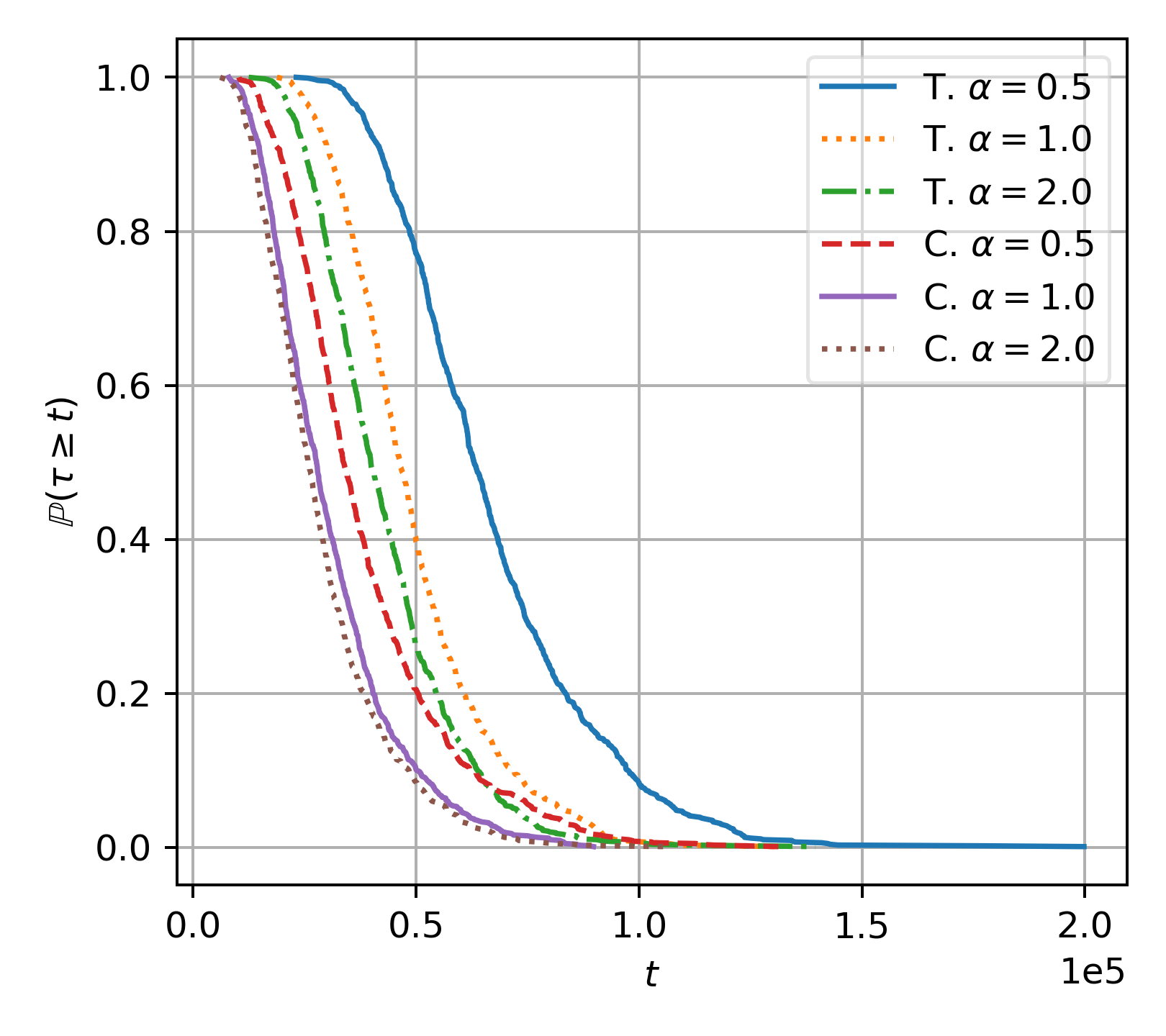}
        \caption{Survival probability}\label{fig:Survival}
    \end{subfigure}%
    \begin{subfigure}{0.5\textwidth}
    \centering
        \includegraphics[width=0.95\textwidth]{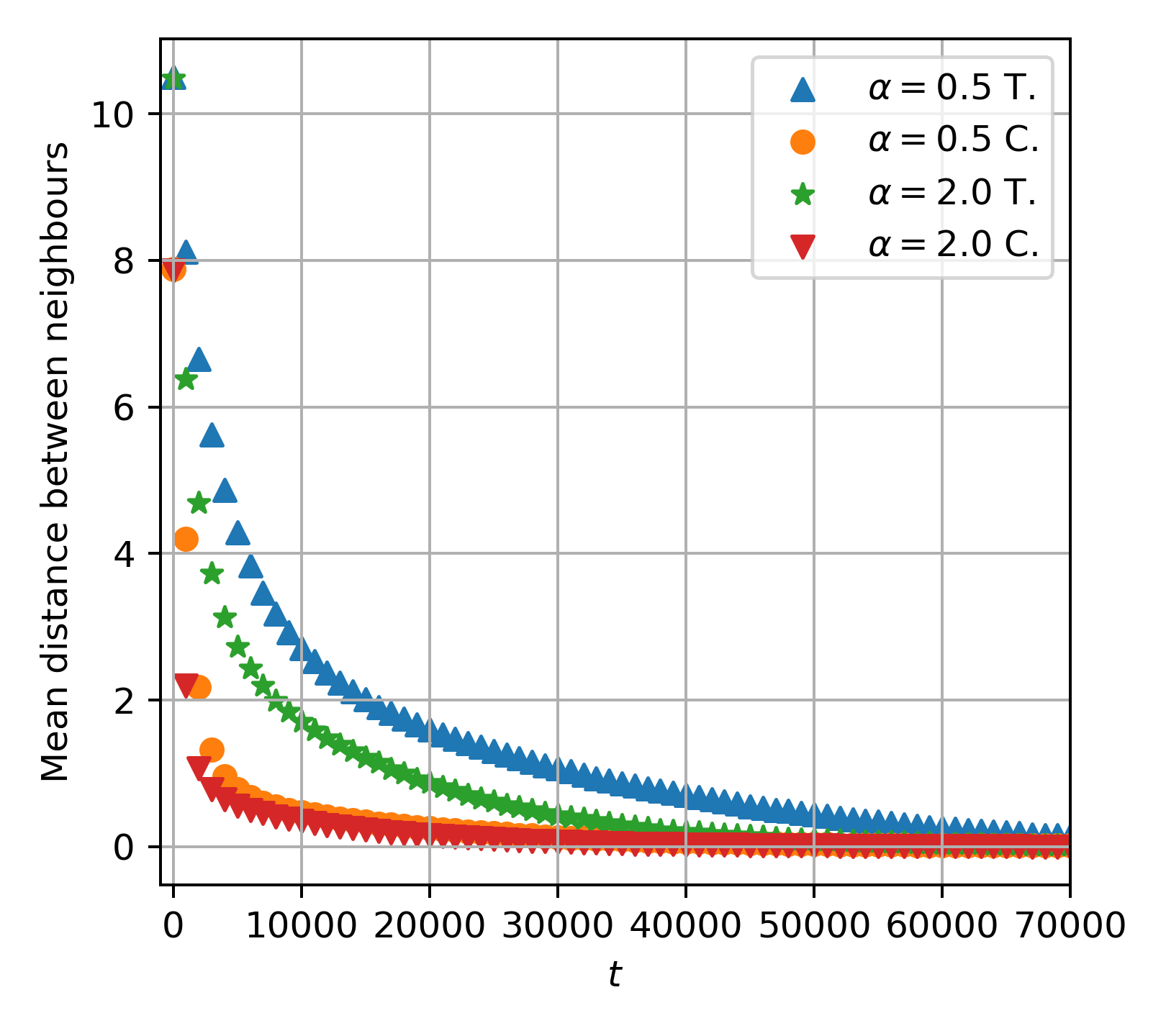}
        \caption{Mean distance between neighbours}\label{fig:mean_dist}
    \end{subfigure}%
    \caption{(a) Survival probability over time for both the toroidal (T.) and the cubic (C.) opinion space for different values of $\alpha$ from the interaction function (\ref{eq:fa}). (b) The mean distance between neighbours against time measured every 1000 time steps for $\alpha=0.5,2$. 1000 iterations were run for each parameter setting up to a maximum of 250\,000 simulated time steps.}
\end{figure}

\begin{figure}[htb]
    \centering
    \includegraphics[width=0.45\linewidth]{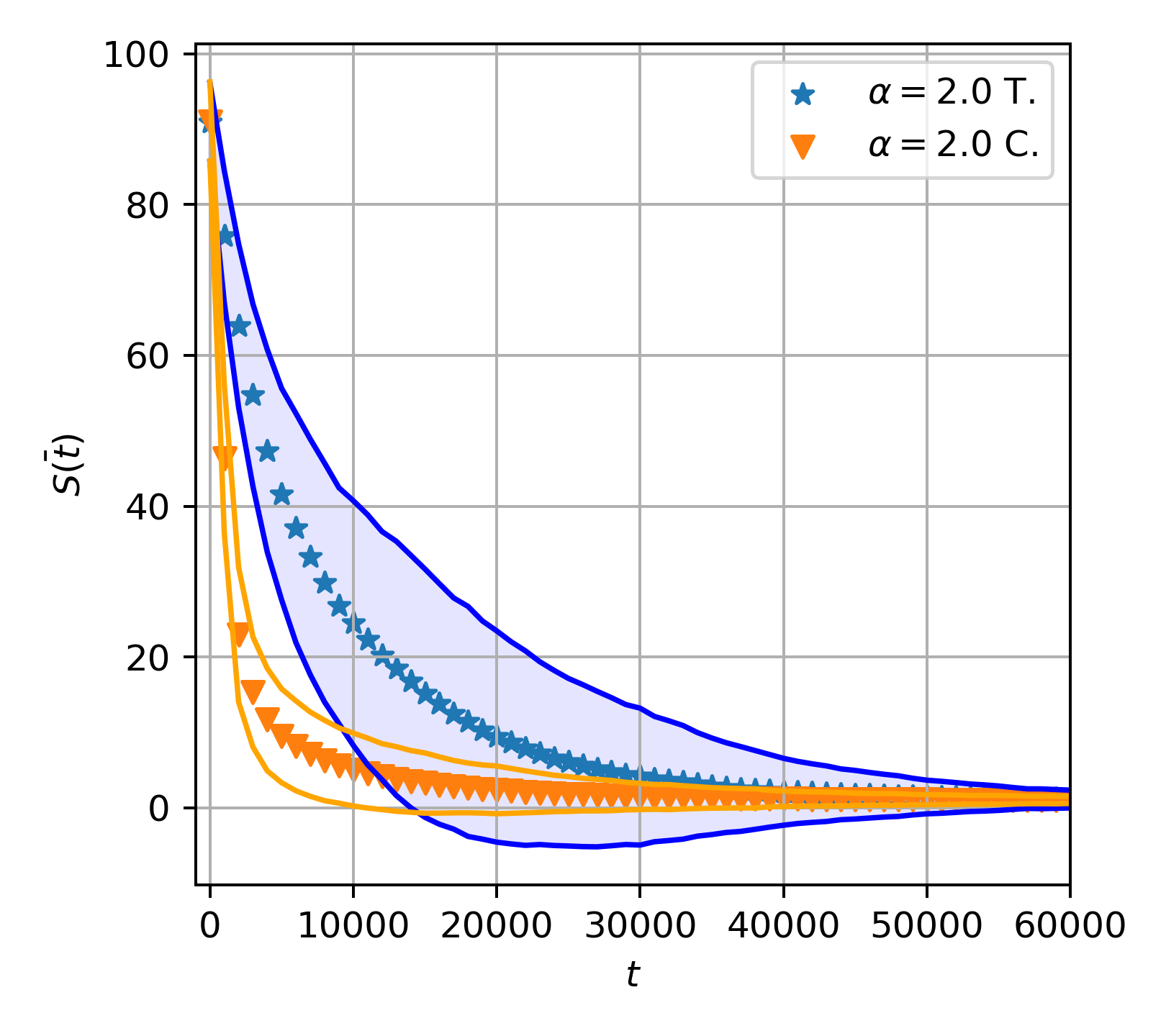}
    \caption{The number of groups over time for the toroidal and cubic opinion space with $\alpha=0.2$. The bars shown indicate the sample mean plus/minus 2 standard deviations. We see that the differences in the dynamics disappear after roughly 15\,000 time steps. For legibility we only plot until time step 30\,000.}
    \label{fig:groups}
\end{figure}

It is clear from the survival probability plotted in Figure~\ref{fig:Survival} that in the basic model the time to steady state is consistently less in the cubic than in the toroidal opinion space. However, in Figure~\ref{fig:mean_dist} showing the mean (over simulation runs) of the average (within a simulation run) distance between neighbours in the network over time, we see that both models converge to consensus as the distance shrinks to zero (indeed all 1000 simulation runs for both models result in $S(\tau)=1$). From both of these plots it is clear that the dynamics reach steady state faster in (a) the cubic opinion space than the toroidal opinion space, and (b) for greater $\alpha$ than for lower $\alpha$. The convergence to consensus is re-illustrated in the plot of the number of groups over time $S(t)$ in Figure~\ref{fig:groups} which tends to 1 (plotted only for $\alpha =2$ for clarity).

The convergence to consensus can be explained heuristically by the interaction function which is only zero at the maximum distance. The initial opinions are spread uniformly and as soon as agents start to interact they will approach each other. The chance that all agents converge to one of the two extremes is incredibly low. In the toroidal space there are more configurations which may allow stable bi-polarization, but these too will hardly ever be reached as they are delicate and depend on the combination of the starting distribution of beliefs as well as the order of interactions. If the agents at opposite ends of the torus interact with (and step towards) moderate agents before the converse happens, then the polarized outcome fails.

\begin{figure}[htb!]
    \centering
    \begin{subfigure}{0.32\textwidth}
    \centering
        \includegraphics[height=0.8\textwidth]{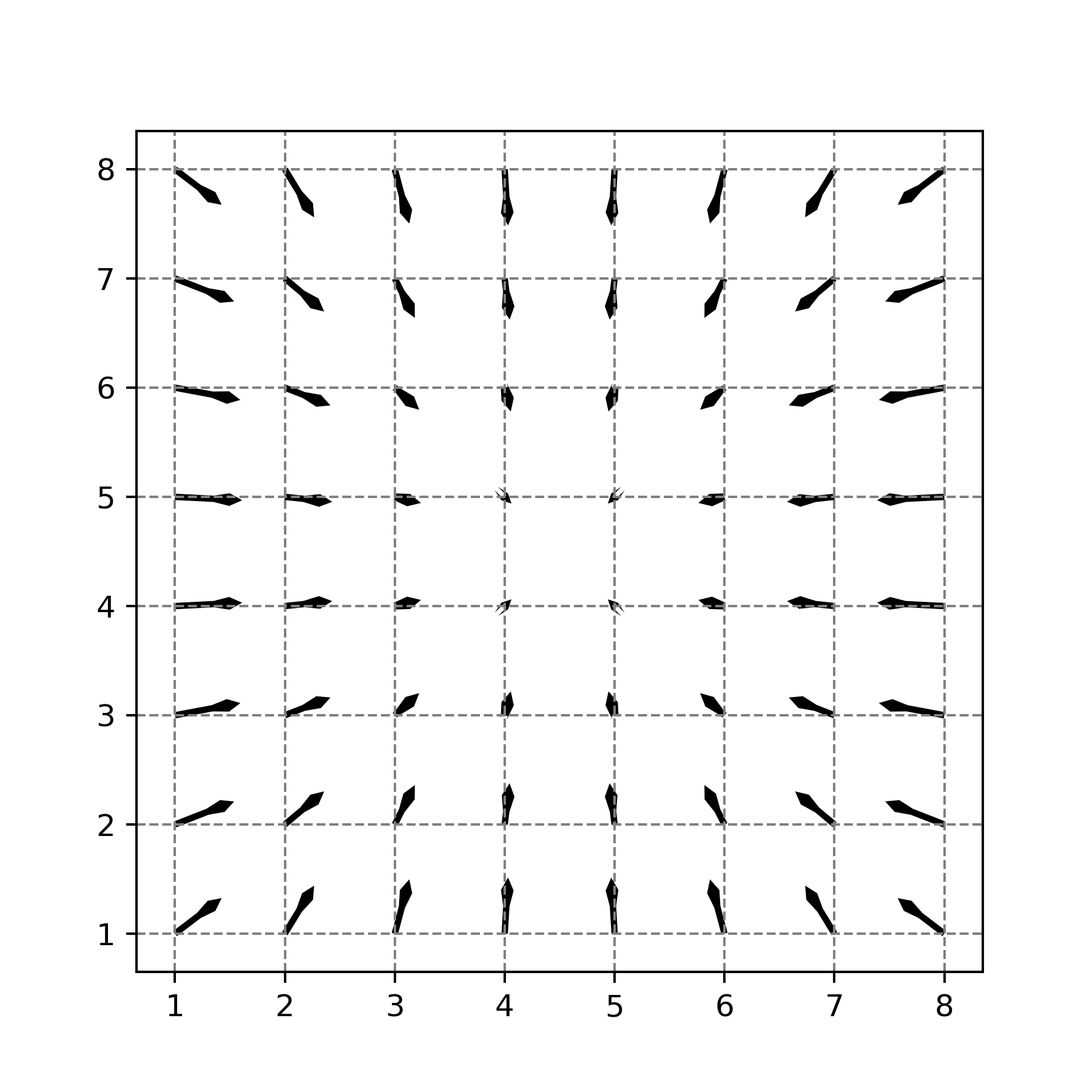}
        \caption{Cubic opinion space}\label{fig:Field_C}
    \end{subfigure}%
    \begin{subfigure}{0.32\textwidth}
    \centering
        \includegraphics[height=0.8\textwidth]{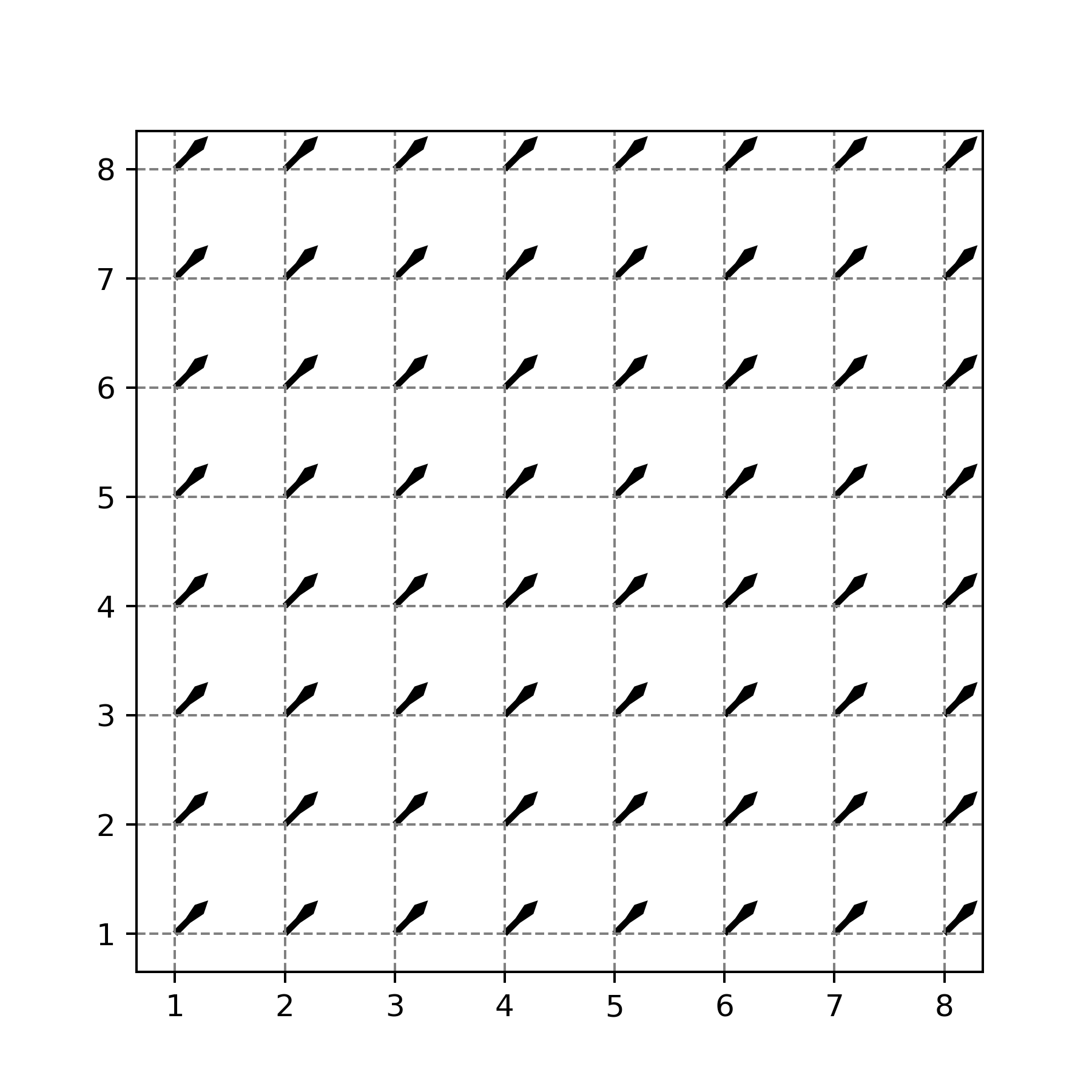}
        \caption{Toroidal opinion space}\label{fig:Field_T}
    \end{subfigure}%
        \begin{subfigure}{0.33\textwidth}
    \centering
        \includegraphics[height=0.65\textwidth]{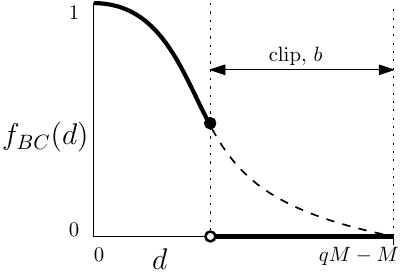}
        \caption{Clip illustration}\label{fig:clip}
    \end{subfigure}%
    \caption{The `pull' felt by individuals in a fully connected population with one individual on each possible belief in (a) Cubic opinion space and (b) in Toroidal opinion space. Note here we set $M=2$ for clarity. (c) Illustration of how the bounded confidence is applied to the interaction function $f(d)$, and the interpretation of the `clip'.}\label{fig:Field}
\end{figure}

In Figure~\ref{fig:Field} we plot the expected direction of motion for an individual in a fully connected population in which each combination of views is held by one agent. In these figures, we have set $M=2$ instead of 3 for ease of visualization. This could be considered the mean-field expected movement with a dense population dispersed over the entire opinion space. Note that the expected direction in the cubic opinion space (Figure~\ref{fig:Field_C}) is toward the centre of the opinion space, contracting the total number of views. This explains why consensus is reached quicker in the cubic opinion space version of the model. In the toroidal opinion space (Figure~\ref{fig:Field_T}) the expected direction is the same for each agent. The fact that there is an expected motion is related to how ties are broken in the model: When comparing the distances on each view, the lower indices win ties. Furthermore, when comparing whether to increment the view in the positive or negative direction ties are broken in the positive direction. If these ties were broken by a fair coin flip there would be no expected motion (all arrows having length zero). Both of these scenarios however, would explain that the dynamics on the toroidal opinion space take longer to settle down as the contraction is not expected right from the start.


We conclude that the cubic and the toroidal versions of this model form an adequate base for comparison. The dynamics resulting from the cubic and the toroidal opinion space have mainly quantitative differences. The toroidal model takes longer to reach a steady state, however both models \textit{consistently} result in consensus when initial opinions are spread uniformly. The results of the basic model illustrate that qualitatively the behaviour is not changed by the specific value of $\alpha$. In order to avoid overly long simulations and computational cost we continue with the case where $\alpha=2$.

\subsection{Bounded confidence}
The first addition to the model is that of bounded confidence. That is the idea that agents whose views are `too far apart' (or perceived as such by the agents) and no longer \textit{interact} bringing their views closer together. Including an element of bounded confidence in an Axelrod-type model is not uncommon. For example, MacCarron \textit{et al.}~\cite{MacCarron2020} study an Axelrod model in which the traits are ordinal and include a bounded confidence in the interaction likelihood \textit{and} the effect of the interaction. Instead of interacting at a probability proportional to the number of shared traits, the probability of interaction in their model is proportional to the number of traits for which the views of agents are within the \textit{agreement threshold}. This tends to increase the interaction likelihood for which they compensate by allowing changes to opinions only for those opinions in which the agents are also within the agreement threshold. 

We implement the bounded confidence in this model by means of a `clip' $b$ which is subtracted from the maximum distance $d_{\text{max}}$ that the opinions can be apart (in both the toroidal and the cubic opinion space). We modify $f(d)$ to be the interaction probability under bounded confidence $f_{\text{BC}}(d)$ thus (illustrated in Figure~\ref{fig:clip}):
\begin{equation} \label{eq:fa_bc}
f_{\text{BC}}(d) = \begin{cases}
    1-\left(\frac{d}{qM- M}\right)^\alpha\quad &\text{if }d\leq qM-M-b,\\ 
    0 &\text{else.}\\
\end{cases}
\end{equation}

Bounded confidence in this model allows for persistent disagreement between neighbours whose views are at a distance $d> d_{\text{max}}-b$ from each other by setting the probability of their interaction to zero. Initially this effect is small but as $b$ grows a greater number of groups (showcasing in-group agreement and out-group disagreement) become possible in steady state. We remind the reader that the initial opinions are drawn uniformly at random from all possible opinions, which means that for small clip sizes the chance of not resulting in consensus is negligible (estimated at zero by our simulations). However, in theory more steady states emerge as soon as we include some bounded confidence. This would play a bigger role if the process of drawing initial opinions was more biased to the extremes.

\begin{figure}[htb]
    \centering
    \begin{subfigure}{0.33\textwidth}
            \includegraphics[width=0.9\linewidth]{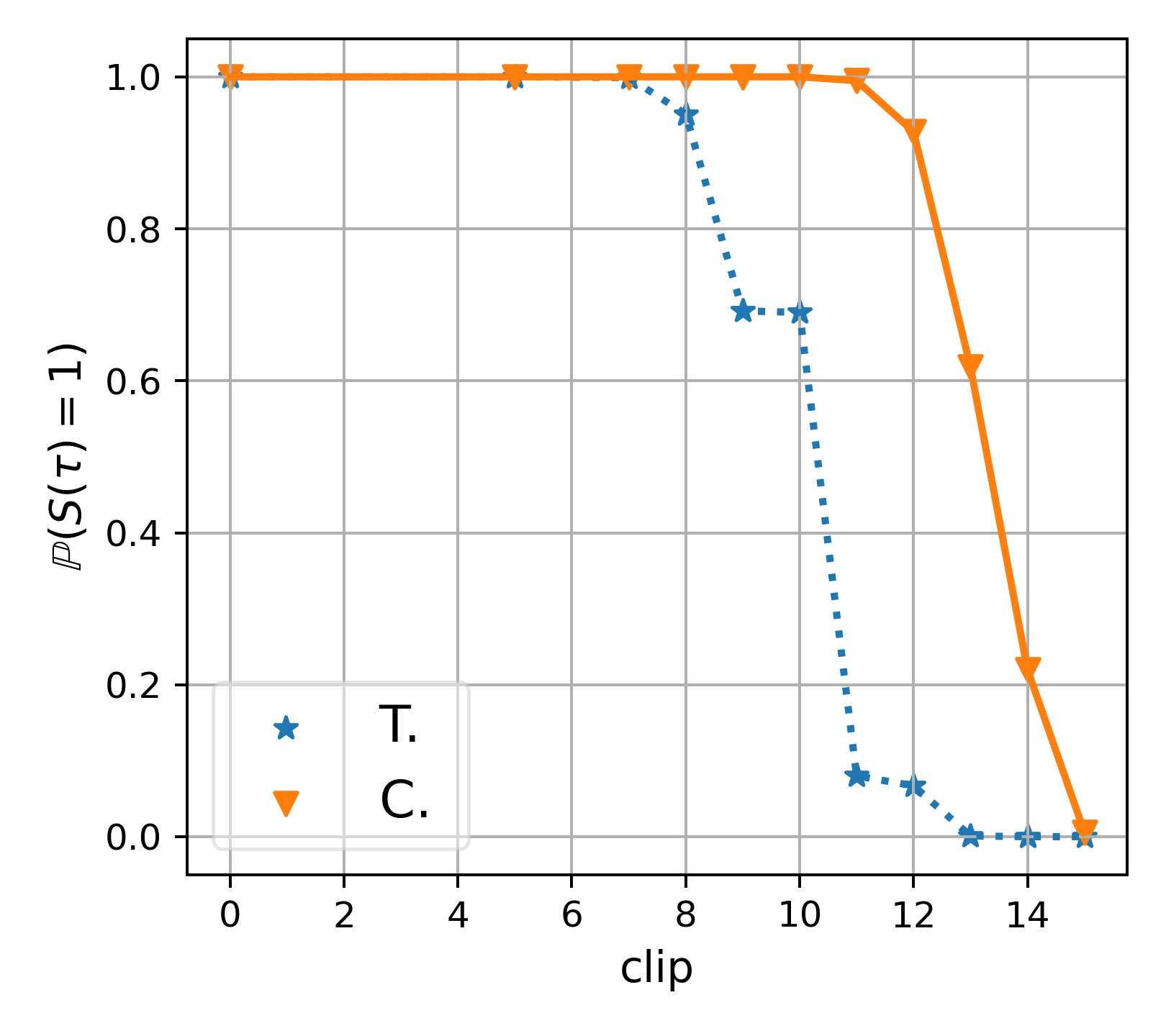}
    \caption{Consensus.}
    \label{fig:con_clip}
    \end{subfigure}%
    \begin{subfigure}{0.33\textwidth}
    \includegraphics[width=0.9\linewidth]{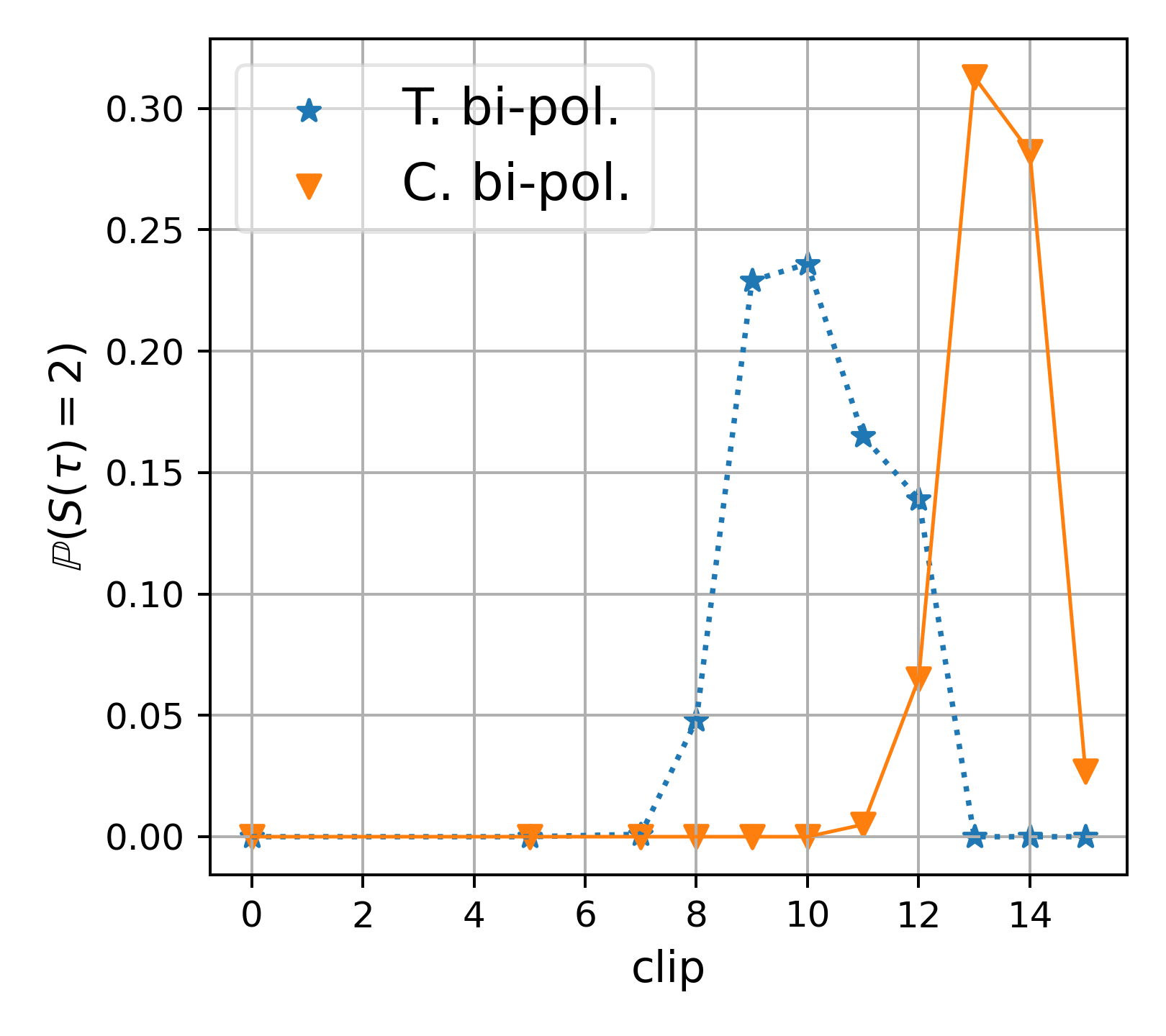}
    \caption{Bi-polarization.}
    \label{fig:bipol}
    \end{subfigure}%
    \begin{subfigure}{0.33\textwidth}
    \includegraphics[width=0.9\linewidth]{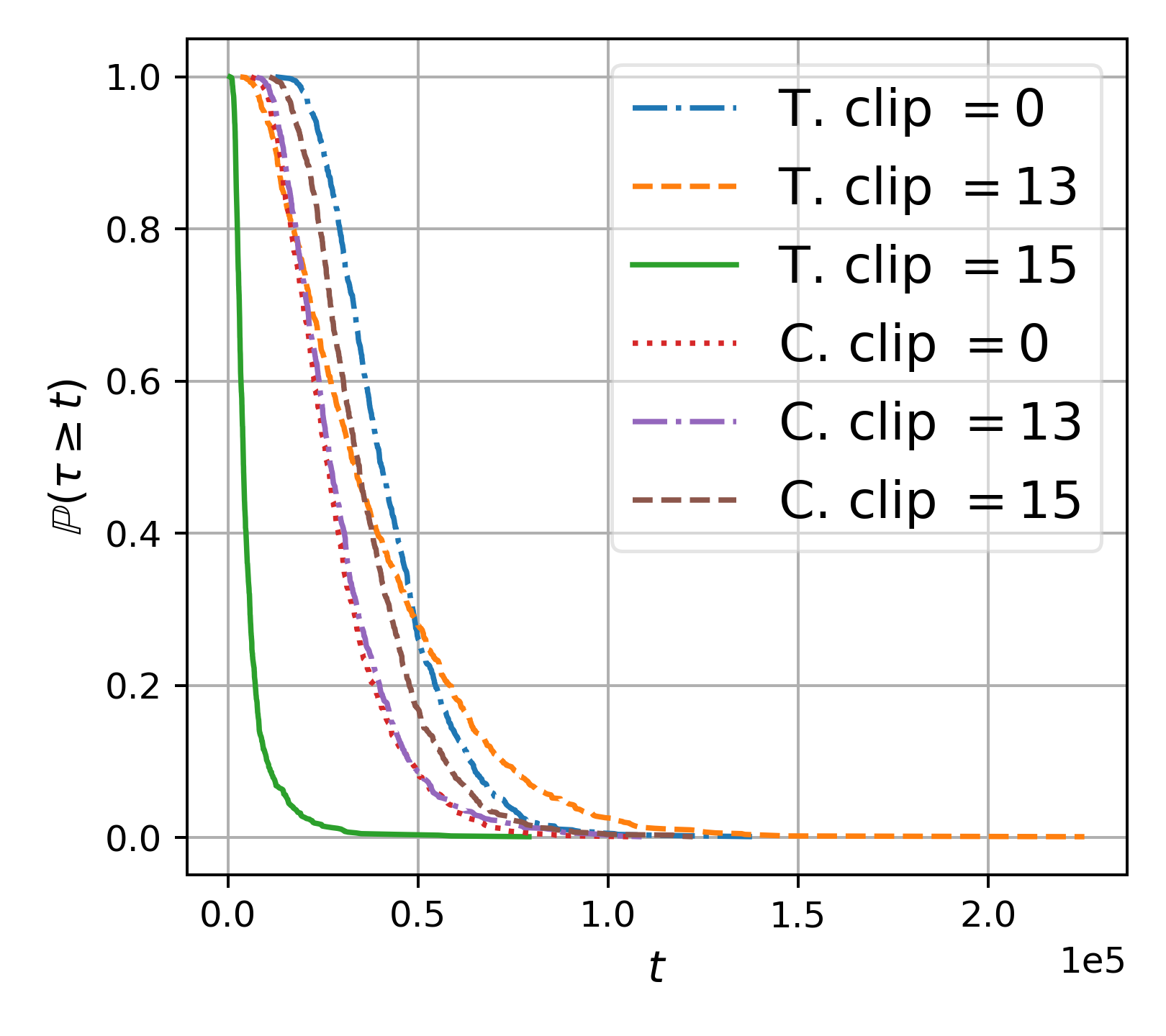}
    \caption{Survival probability.}
    \label{fig:Surv_clip}
    \end{subfigure}%
    \caption{The effect of bounded confidence has on the estimated probability of reaching a consensus and the estimated probability of opinion bi-polarization in both the cubic and toroidal opinion model, as well as the time it takes for dynamics to settle down.}
\end{figure}

\subsubsection{Effect of bounded confidence on the timescales}
In Figure~\ref{fig:Surv_clip} we show the survival probability over time (illustrating the time to absorption) for the cubic and the toroidal opinion space models. In particular we plot the results for bounded confidence ($b$ = 0, 13, 15). We see that the toroidal opinion space model is both the slowest ($b$ = 0) and the fastest ($b$ = 15). Adding a bounded confidence clip of 15 in the toroidal opinion space model, speeds up the dynamics. Conversely, for the cubic opinion space model adding bounded confidence clip of 15 slows down the dynamics. We also plot the values for $b$ = 13 to illustrate some nuances of this effect. In the cubic opinion space, a clip of 13 does change the absorption time significantly. Though, for the toroidal opinion space we see a larger spread of time scales: more simulation runs reaching steady state earlier as well as later than without any bounded confidence.

\subsubsection{Probability of consensus under bounded confidence}
The proportion of simulation runs which end in consensus is plotted in Figure~\ref{fig:con_clip}. Until a clip of $b=6$ both opinion topologies lead exclusively to consensus in the long-term. For the cubic opinion space this continues until $b=11$. The dynamics of the toroidal opinion space on the other hand already start to show results other than consensus at $b=8$. Initially only a small percentage of simulation iterations are not driven to consensus but this number increases quickly at clips $b=9$, and $b=11$. At $b=13$ the dynamics have undertaken a complete makeover showing no iterations ending in consensus in the toroidal opinion space while for the cubic opinion space this percentage is still roughly $60\%$. We see that the changes in the toroidal opinion space model occur in a stepwise fashion. This is explained by the factor of $(q-1)/\lfloor q/2\rfloor=1.75$ used in normalizing the distances which means that not all integer distances are possible.

\subsubsection{Effect of bounded confidence on the number of groups}

In Figure~\ref{fig:bipol} we plot the percentage of simulation iterations for the various clip sizes which result in strict bi-polarization, meaning exactly 2 groups in the steady state. We see that for both the cubic and the toroidal opinion spaces this first increases as the clip size increases, and subsequently decreases. A larger clip means less interaction which allows \textit{more} than two groups to form. Observe that both the probability of consensus and the probability of bi-polarization are decreasing from $b=10$ for the toroidal opinion space and from $b=13$ for the cubic opinion space. This means that the resulting number of groups must be $S(\tau)\geq3$. Initially, between clips $[7,12]$ a toroidal opinion space results in more opinion bi-polarization than the cubic space. This however is turned on its head between clips 13, and 15 where the probability of bi-polarization is much greater in the cubic opinion space than in the toroidal one.

\begin{figure}[ht]
    \centering
    \begin{subfigure}{0.33\textwidth}%
        \includegraphics[width=\textwidth]{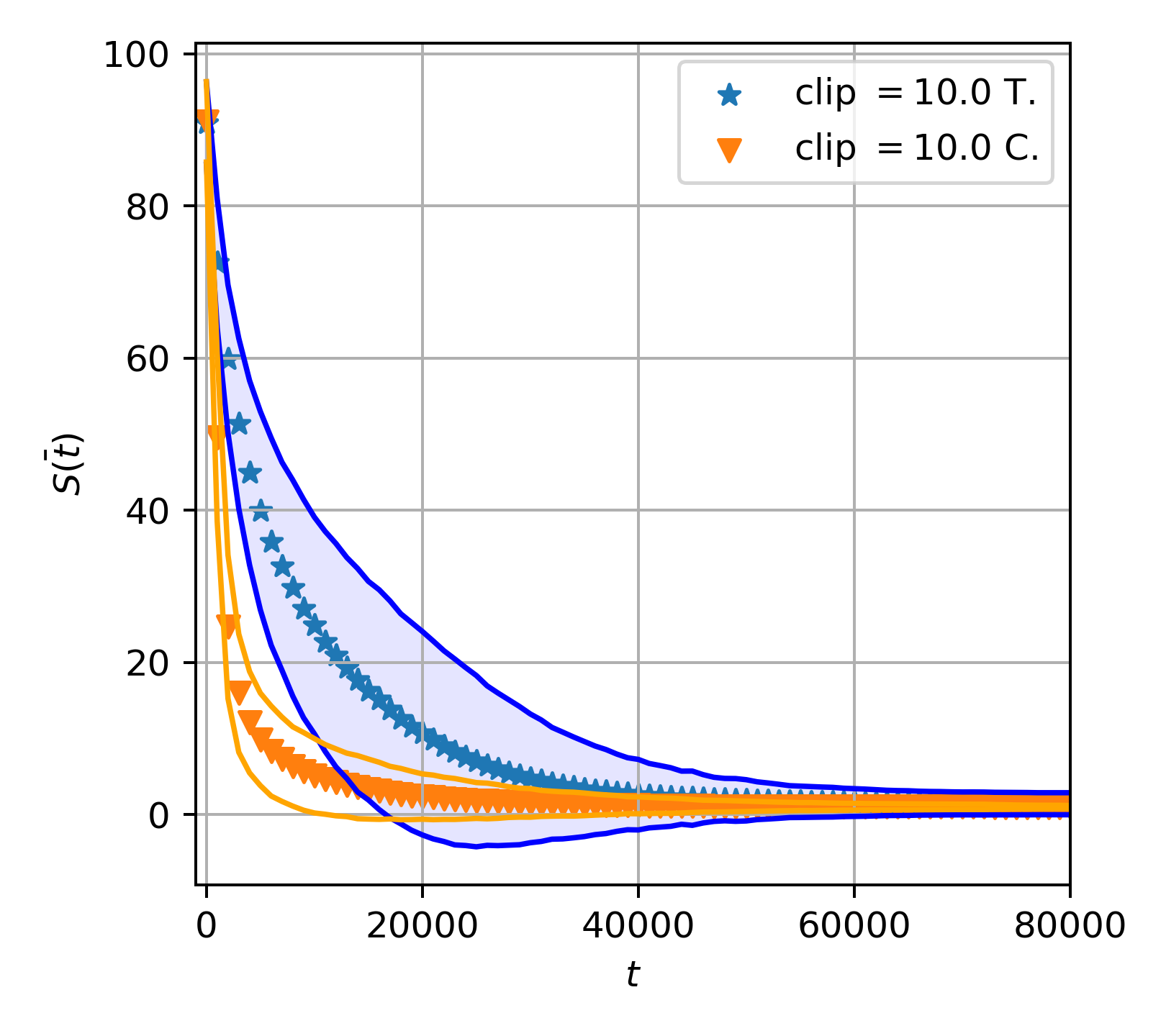}
        \caption{$b=$ 5}
    \end{subfigure}%
    \begin{subfigure}{0.33\textwidth}%
        \includegraphics[width=\textwidth]{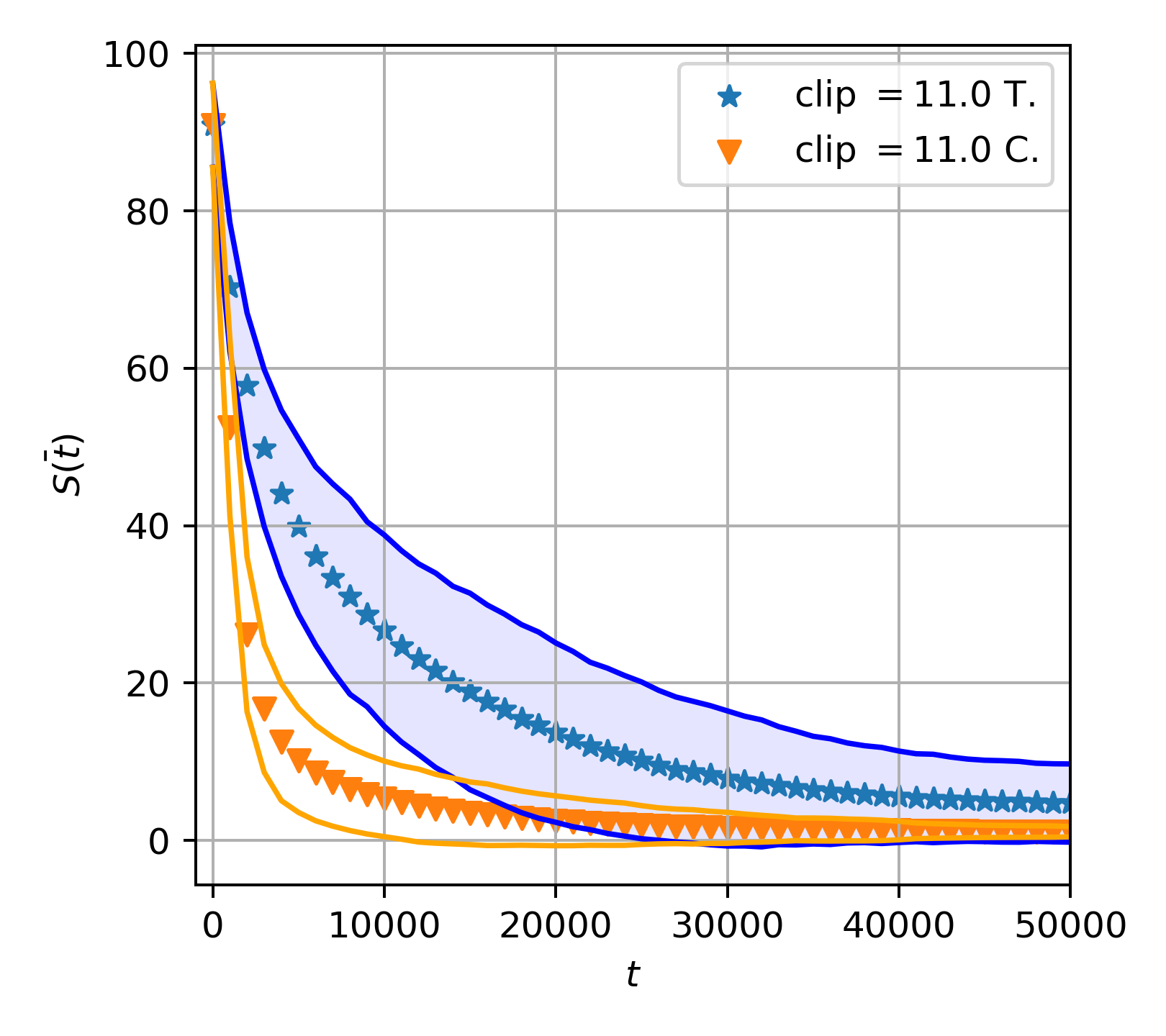}
        \caption{$b=$ 10}
    \end{subfigure}%
    \begin{subfigure}{0.33\textwidth}%
        \includegraphics[width=\textwidth]{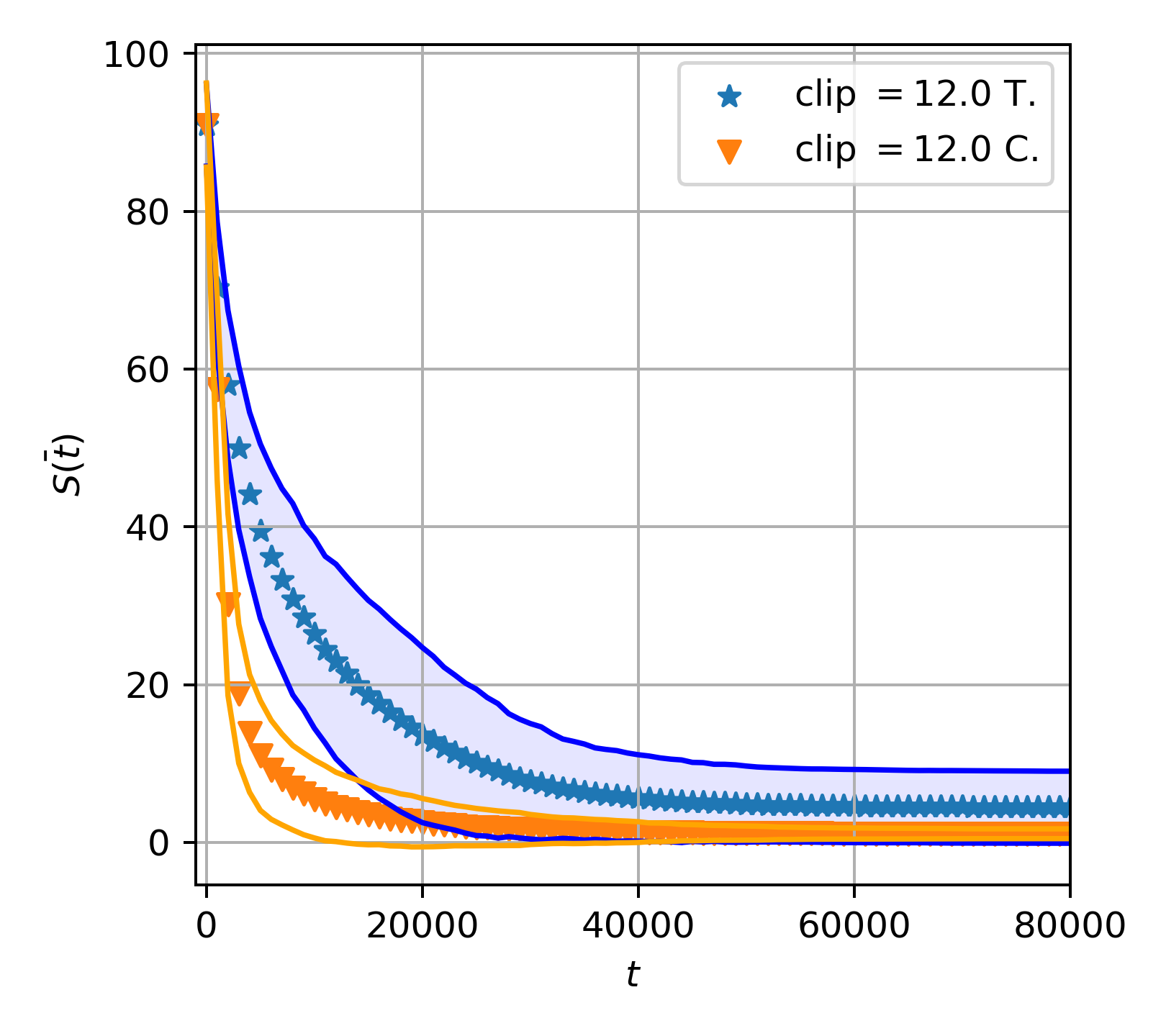}
        \caption{$b=$ 12}
    \end{subfigure}%
    \\
    \begin{subfigure}{0.33\textwidth}%
        \includegraphics[width=\textwidth]{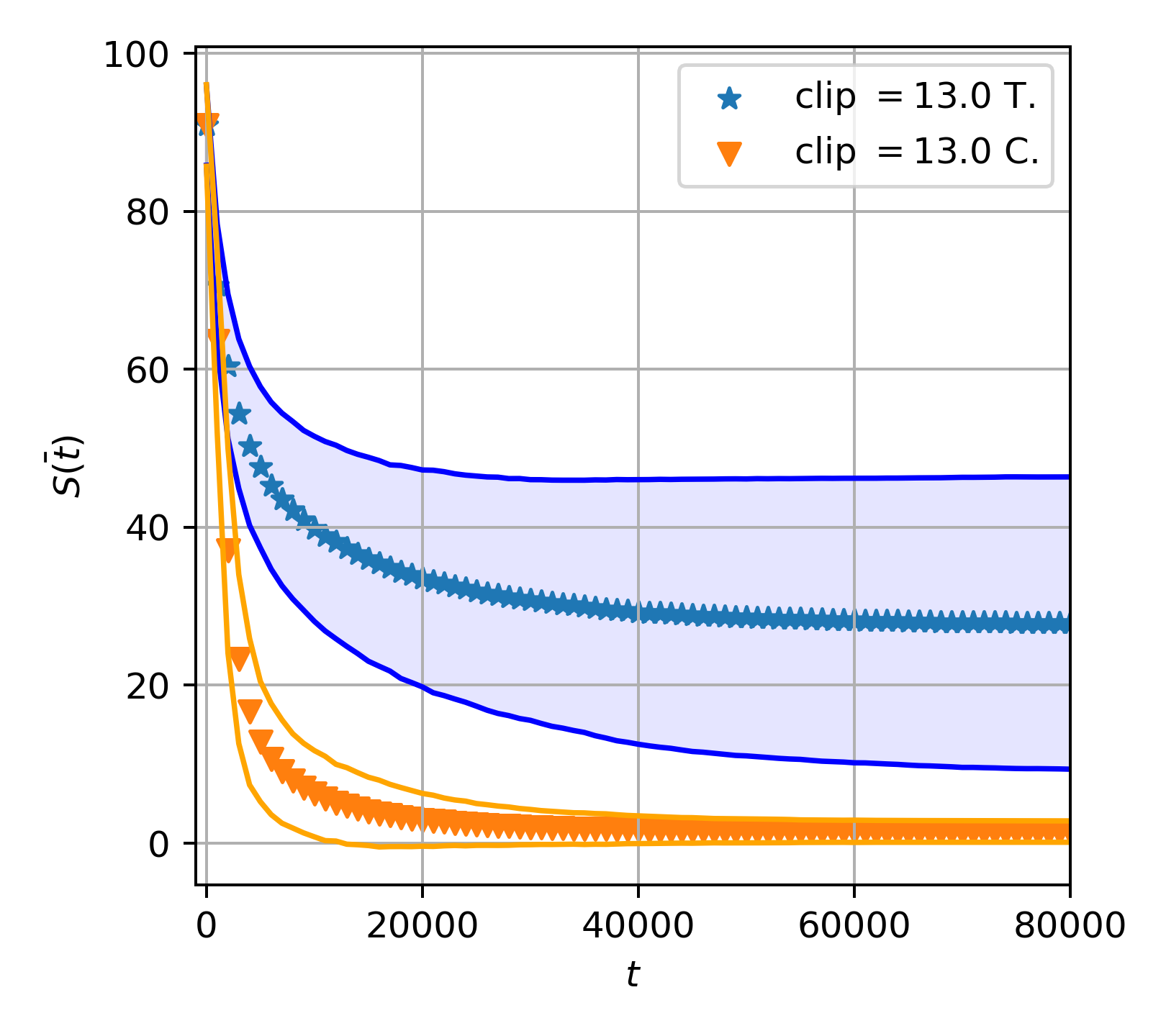}
        \caption{$b=$ 13}
    \end{subfigure}%
    \begin{subfigure}{0.33\textwidth}%
        \includegraphics[width=\textwidth]{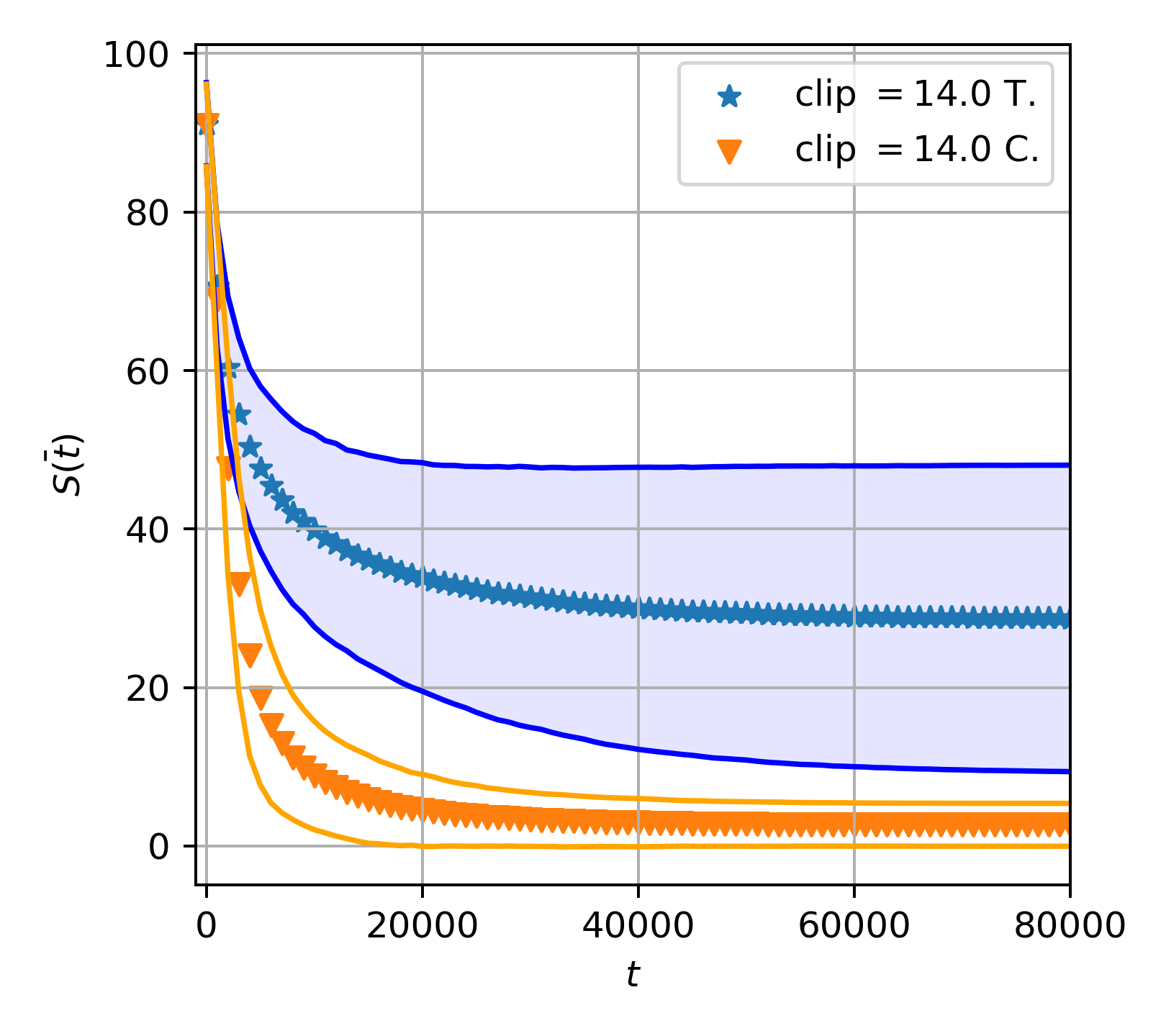}
        \caption{$b=$ 14}
    \end{subfigure}%
    \begin{subfigure}{0.33\textwidth}%
        \includegraphics[width=\textwidth]{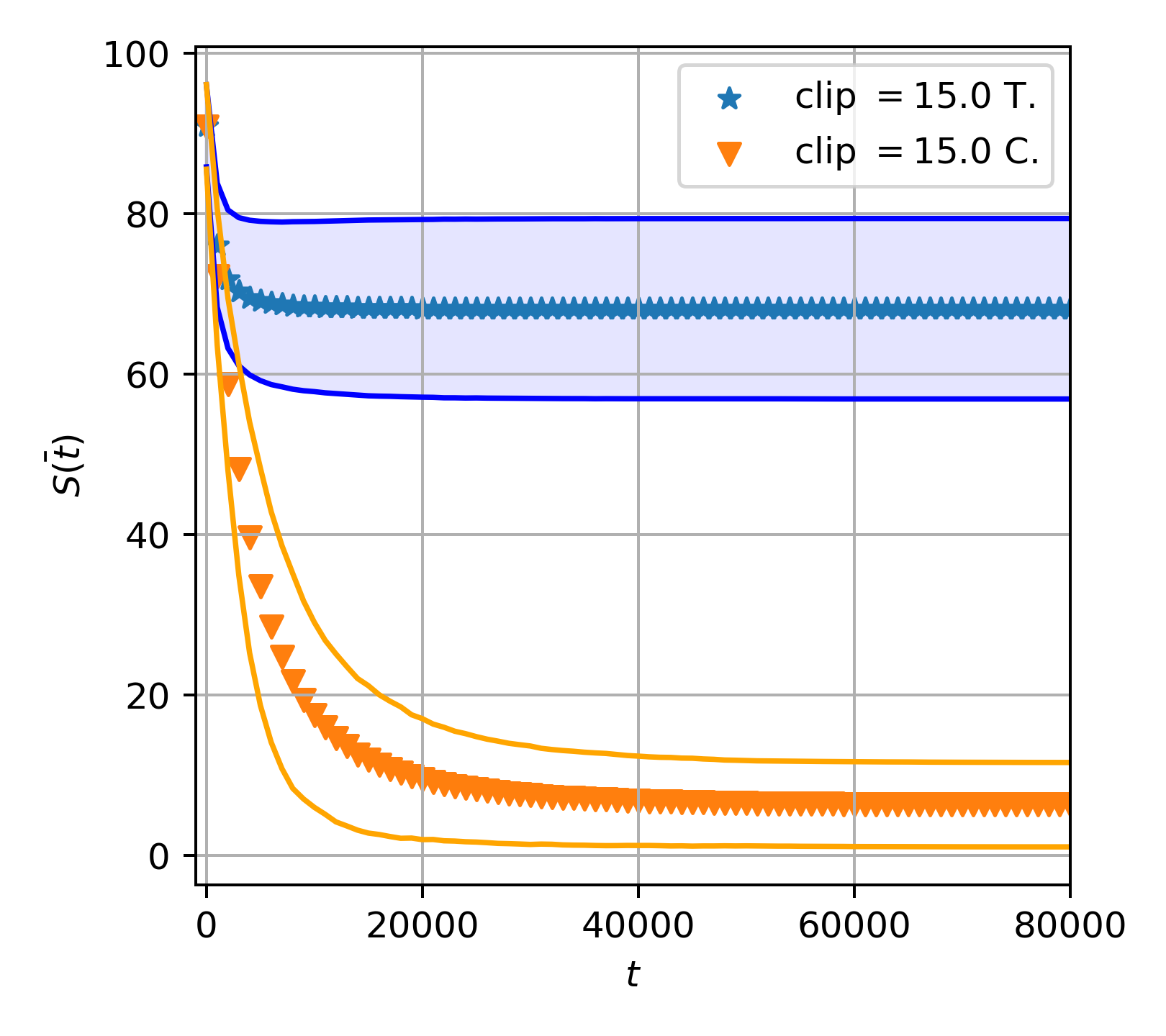}
        \caption{$b=$ 15}
    \end{subfigure}%
    \caption{The mean number of groups over time for the simulations with $b$ in $\{5, 10, 12,13,14, 15\}$ with $\alpha=2$ on cubic and toroidal opinion spaces (marked C.\ and T.\ respectively). We see that the difference between dynamics on a cubic vs a toroidal opinion space grow as the bounded confidence allows for less interaction.}
    \label{fig:NoGrps_clips}
\end{figure}

In Figure~\ref{fig:NoGrps_clips}, plotting the mean number of groups over time for clips 5, 10, 12, 13, 14, and 15 in both the toroidal and cubic opinion model, we see that the difference in the dynamics starts to grow. In particular as the clip size gets bigger (more bounded confidence), the toroidal opinion space allows for a larger number of groups. This is true not only in the transient period, but also for steady state. Furthermore, the variance in the number of groups is also substantially larger for the toroidal opinion space than for the cubic one. The toroidal opinion space lends itself to more steady state configurations under the bounded confidence than the cubic one, which explains the observed effect. 

Two facts stand out: First, that the peak in the probability of bi-polarization is greater in the cubic opinion space than the toroidal (see Figure~\ref{fig:bipol}). Second, that $\mathbb{E}(S(\tau))$ is greater for the toroidal opinion space than for the cubic space, in particular for the largest clip sizes, much greater than 2. Thus we conclude that under bounded confidence the probability of observing opinion bi-polarization is greater in cubic opinion space than in the toroidal opinion space.

\subsection{Topical weights}
The second extension we investigate adds agent specific weights to each topic (element within $s_i$) of their opinion. Huet and Deffuant~\cite{Huet2010}, for example, studied an opinion dynamics model in which one opinion element was considered more important than the other, leading to rich dynamics. Similarly, Baldassari and Bearman~\cite{Baldassarri2007} show how including personal relative interest of agents in certain topics (opinion elements) in the opinion updating mechanism can lead to opinion polarization and organisation into groups based on the polarized topics. Kalinowska and Dybiec~\cite{Kalinowska2023} studied an Axelrod model with topical weights per opinion dimension. Their model used the same similarity measure as in the original Axelrod model, however in the opinion adoption step the mechanisms differ. In their weighted model, the probability of adoption the view of the chosen neighbour was only set to the similarity \textit{if} the similarity between the two agents is greater than the weight of the dimension which is to be updated. They find that the addition of this weighted mechanism increases the number of groups in steady state compared to the original model.

We believe the extension of topical weights to be natural, though it has not seen much attention in the literature until now (barring the two examples mentioned). Furthermore, this extension has been suggested in the discussion of possible future work in~\cite{Pham2022} indicating that we are not alone in our views.

Upon initialisation, each agent randomly draws three weights (summing to 1), one for each opinion element. This is a weighting they use in the calculation of the `distance' to a neighbours opinion when determining the likelihood of an interaction between them. This is how we take into account that for one agent the views related to minority rights may `weigh' more heavily than an issue related to animal product consumption, for example.

In the simulation we do this by the following steps: 
\begin{itemize}
    \item For each agent $i$ ($\forall i\in V$) we draw a pseudo random number vector $\bm{r}^i$ of length $M-1$. Each $r\in \bm{r}^i$ is drawn uniformly $\mathcal{U}_{[0,1)}$ (as performed by the Python function \texttt{numpy.random.rand()}).
    \item We sort this vector such that $r_j^i<r_{j+1}^i$ for all $j\in \{1,\ldots, M-2\}$.
    \item Agent $i$ is assigned the opinion weights vector of length $M$: $\bm{w}_i:=[r^i_1, r^i_2-r^i_1,\ldots, 1-r_{M-1}^i]\top$.
\end{itemize}

The agent $i$ `uses' this weighting to calculate the weighted Manhattan distance between their own opinion $s_i$ and that of the neighbour $j$ they have selected to possibly interact with, on a cube:
\begin{equation}\label{eq:disC_w}
    d^C_{ij}= \sum_{k=1}^M w_i^kM \cdot \mid s_i^k-s_j^k\mid,
\end{equation}
and on a torus:
\begin{equation}\label{eq:disT_w}
    d^T_{ij} = \sum_{k=1}^M w_i^kM\cdot \frac{q-1}{\lfloor q/2\rfloor} \min (\mid s_i^k-s_j^k\mid, q- \mid s_i^k-s_j^k\mid).
\end{equation}
In order to keep distances comparable in the weighted and unweighted versions we need to multiply the weight $w_i^k$ by $M$. The unweighted versions can be considered similar to the weighted version with all weights equal. This may be seen by substituting $w_i^k=1/M$ for all opinions in (\ref{eq:disC_w}) (or (\ref{eq:disT_w})) and noting that (\ref{eq:disC}) (and (\ref{eq:disT}) respectively) results. 

\subsubsection{Probability of consensus and bi-polarization under topical weights}
\begin{figure}[hb!]
    \centering
    \begin{subfigure}{0.33\textwidth}
        \includegraphics[width=0.9\textwidth]{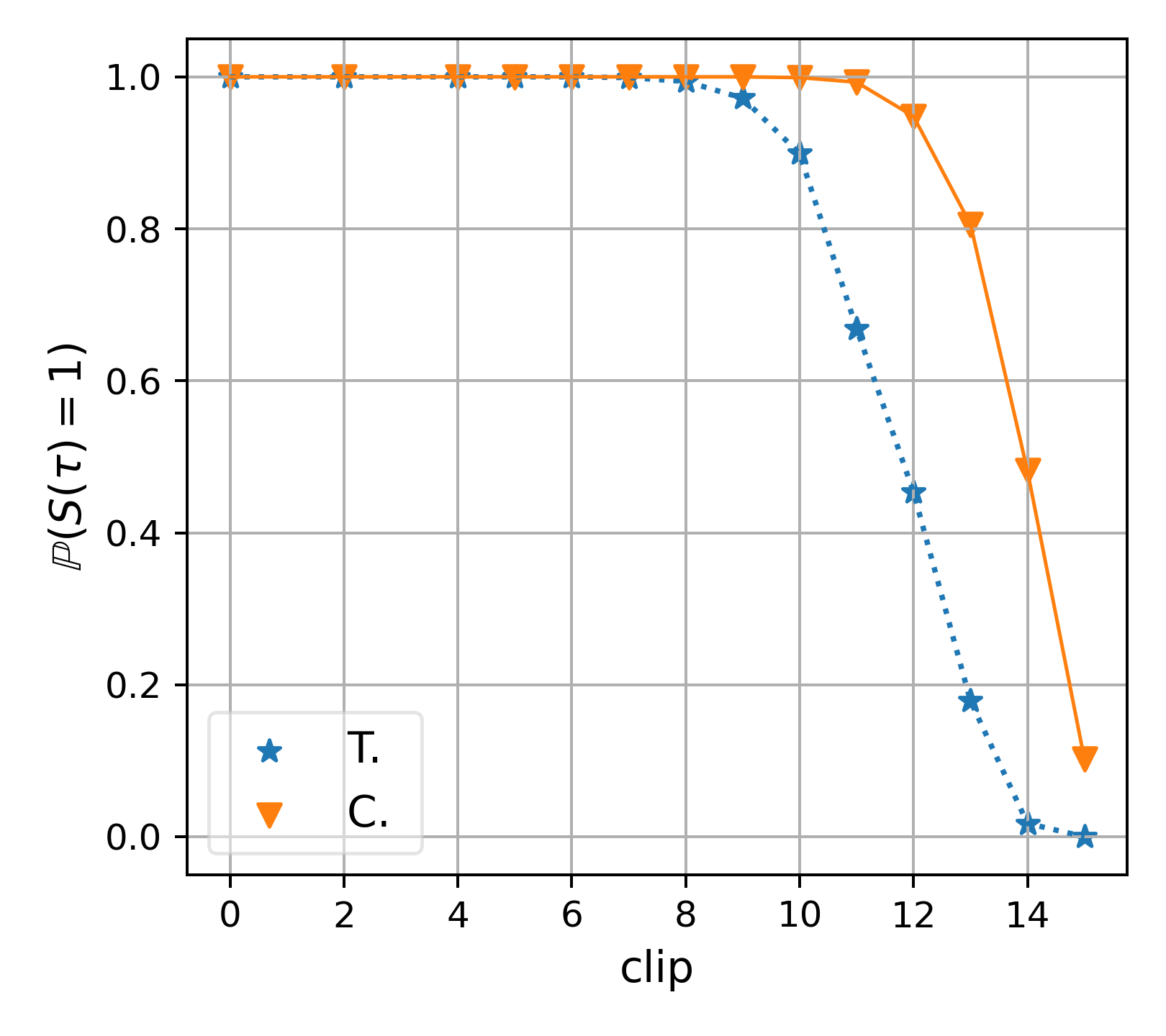}
    \caption{Consensus}
    \label{fig:W_con}
    \end{subfigure}%
    \begin{subfigure}{0.33\textwidth}
        \includegraphics[width=0.9\textwidth]{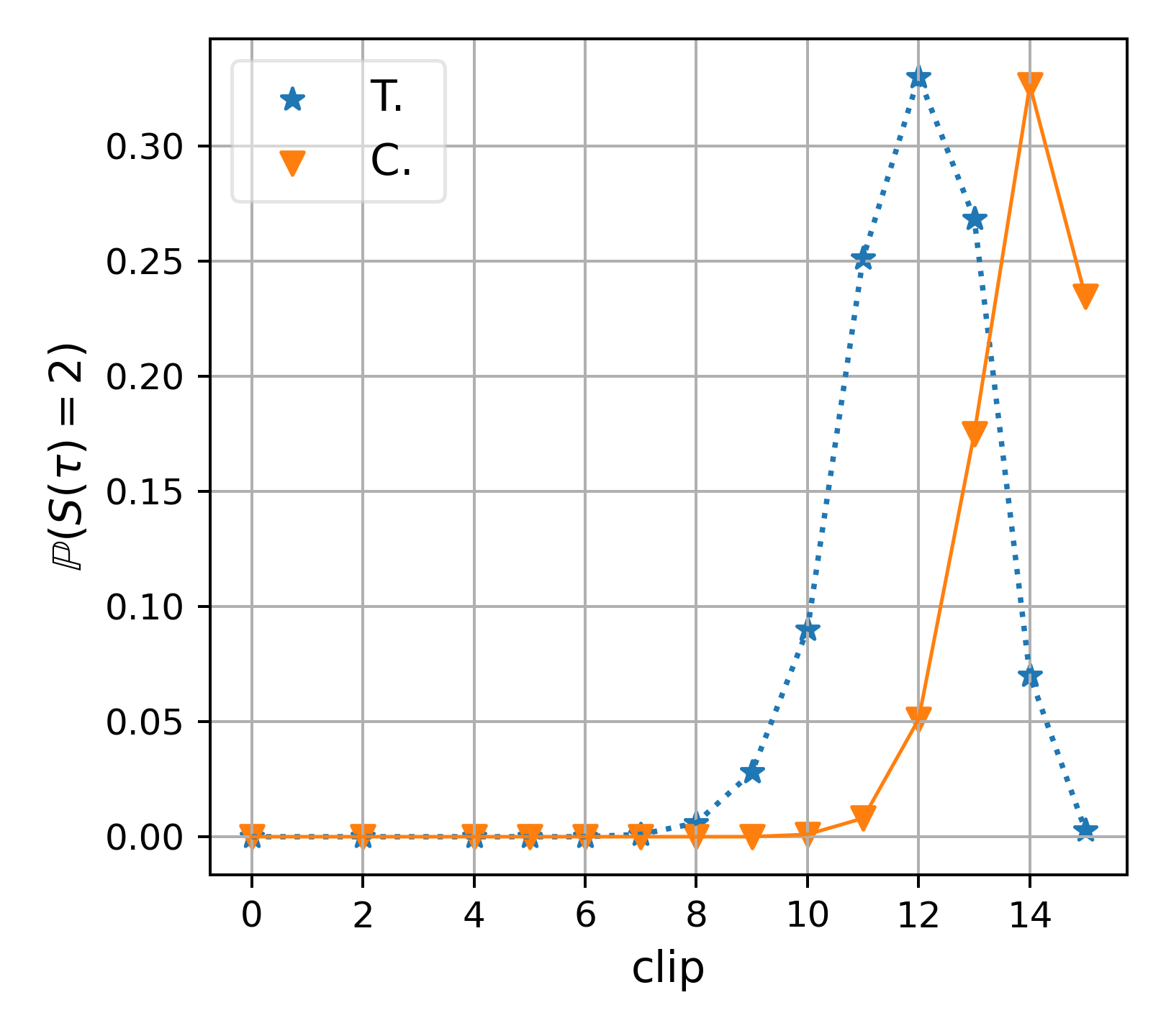}
    \caption{Bi-polarization}
    \label{fig:W_bp}
    \end{subfigure}%
        \begin{subfigure}{0.33\textwidth}
        \includegraphics[width=0.9\textwidth]{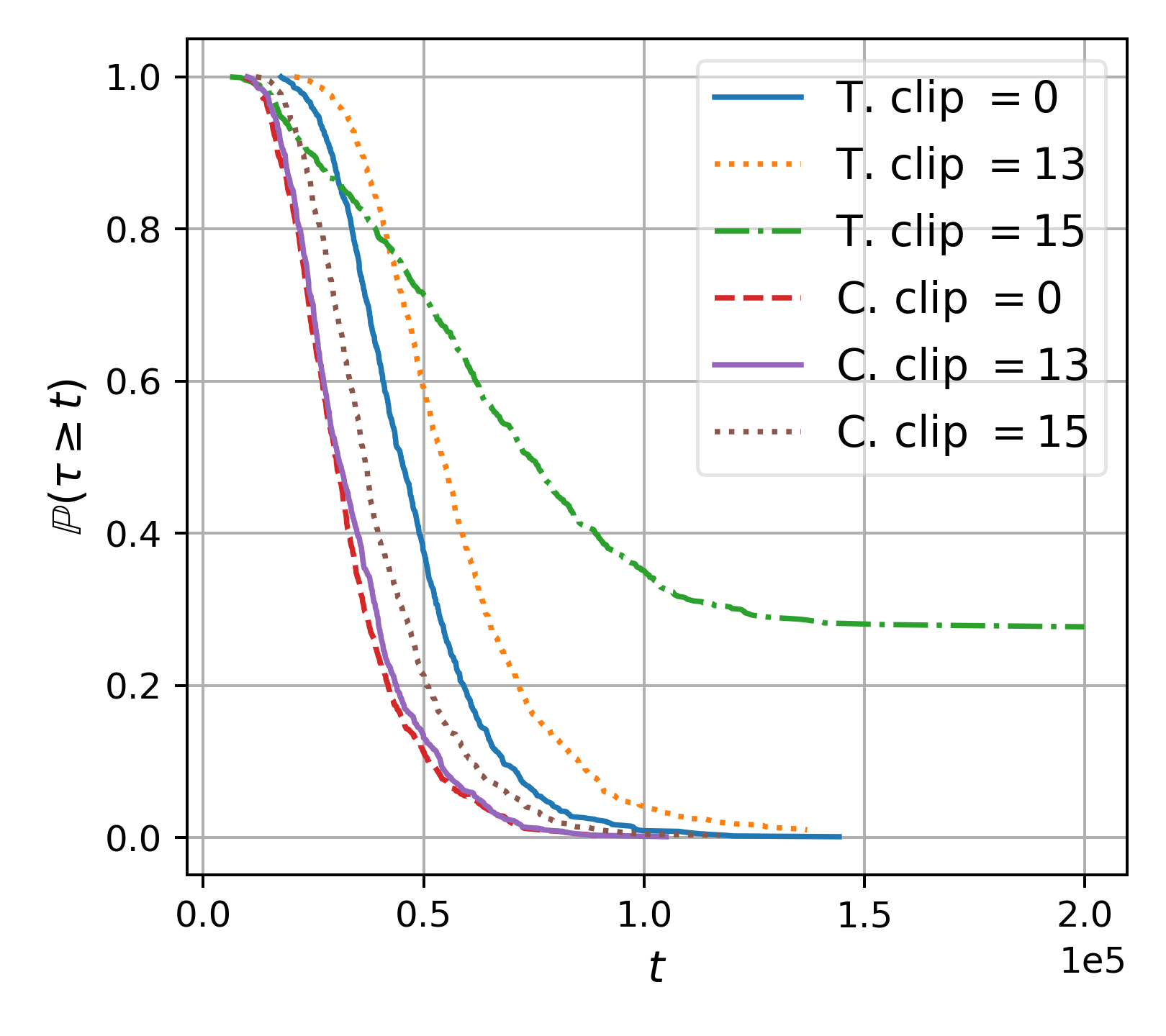}
    \caption{Survival probability}
    \label{fig:W_surv}
    \end{subfigure}%
    \caption{The estimated probability of (a) consensus and (b) opinion bi-polarization under the topical weights extension to the bounded confidence model. The results are plotted in stars and a dotted line for the toroidal opinion space and with down chevrons and a solid line for the cubic opinion space. (c) Distribution of the time to absorption (survival probability). Notice that with this extension, for the first time there are runs which do not absorb in the simulated period.}
\end{figure}

We plot the estimated probability of consensus and the probability of \textit{exactly} two species at steady state in Figures~\ref{fig:W_con} and~\ref{fig:W_bp} with the extension of topical weights in both the cubic and toroidal opinion spaces for various levels of bounded confidence. These two figures may be compared to Figures~\ref{fig:con_clip} and~\ref{fig:bipol} to see how adding topical weights changes the dynamics from the model with only bounded confidence.

Comparing Figures~\ref{fig:W_con} and~\ref{fig:con_clip} we observe that adding personalized weights to topics has only a minor effect on the probability of consensus in the cubic opinion space model. The effect in the toroidal model is larger. In particular, the probability of consensus increases slightly under the addition of weights in both models (though more so in the toroidal opinion space model). This effect can be seen mainly at the larger levels of bounded confidence.


Comparing Figure~\ref{fig:W_bp} to Figure~\ref{fig:bipol} we see that the addition of topical opinion weights `pushes' the peak of probability of bi-polarisation to the right in both the toroidal opinion space, and the cubic opinion space. However, the change is larger in the toroidal opinion space model, being pushed from $b=10$ to $b=12$ and increasing the peaks height from just under 0.25 to over 0.3. In the cubic opinion space model on the other hand the peak only shifts from $b=13$ to $b=14$, and does not significantly change in height.

In \S\ref{sec:distances} we discuss the effect of the topical weighting of opinions on the perceived distance between agent opinions. This provides an explanation for the observed effect in the shifting of peak of the probability of polarization.



\subsubsection{Effect of topical weights on the number of groups over time}
We see the time scales of the dynamics shift under this extension. Plotted in Figure~\ref{fig:W_surv} is the survival probability. We notice that the extension of topical weights increases the time it takes to reach a steady state in the toroidal opinion space model (both at $b=13$, and $b=15$ the absorption time is delayed). In the cubic opinion space model there is almost no effect on the time it takes to reach steady state. Note that not all the simulation runs reached steady state under this extension. The number of runs that did not converge to steady state in simulated time is presented in Table~\ref{tab:no_converge}. From this table too, it is evident that the toroidal opinion space model has a larger change in dynamics than the cubic opinion space model under the addition of topical weights. Note that when calculating the estimated probability of consensus and bi-polarization we use the proportion of converged runs.

\begin{table}[htb!]
    \centering
    \caption{Number of simulation runs (out of 1000) that did not converge to a steady state within the simulated 250\,000 time steps.}
    \label{tab:no_converge}
    \begin{tabular}{l|c|c|c|c|c}
    $b$:  & 11 & 12 & 13 & 14 & 15\\ 
    \hline
        Cubic &  1 & 0 & 0 & 1 & 2\\
        \hline
         Toroidal & 1 & 12 & 9 & 28 & 276
    \end{tabular}
\end{table}

\begin{figure}
    \centering
    \begin{subfigure}{0.33\textwidth}%
        \includegraphics[width=\textwidth]{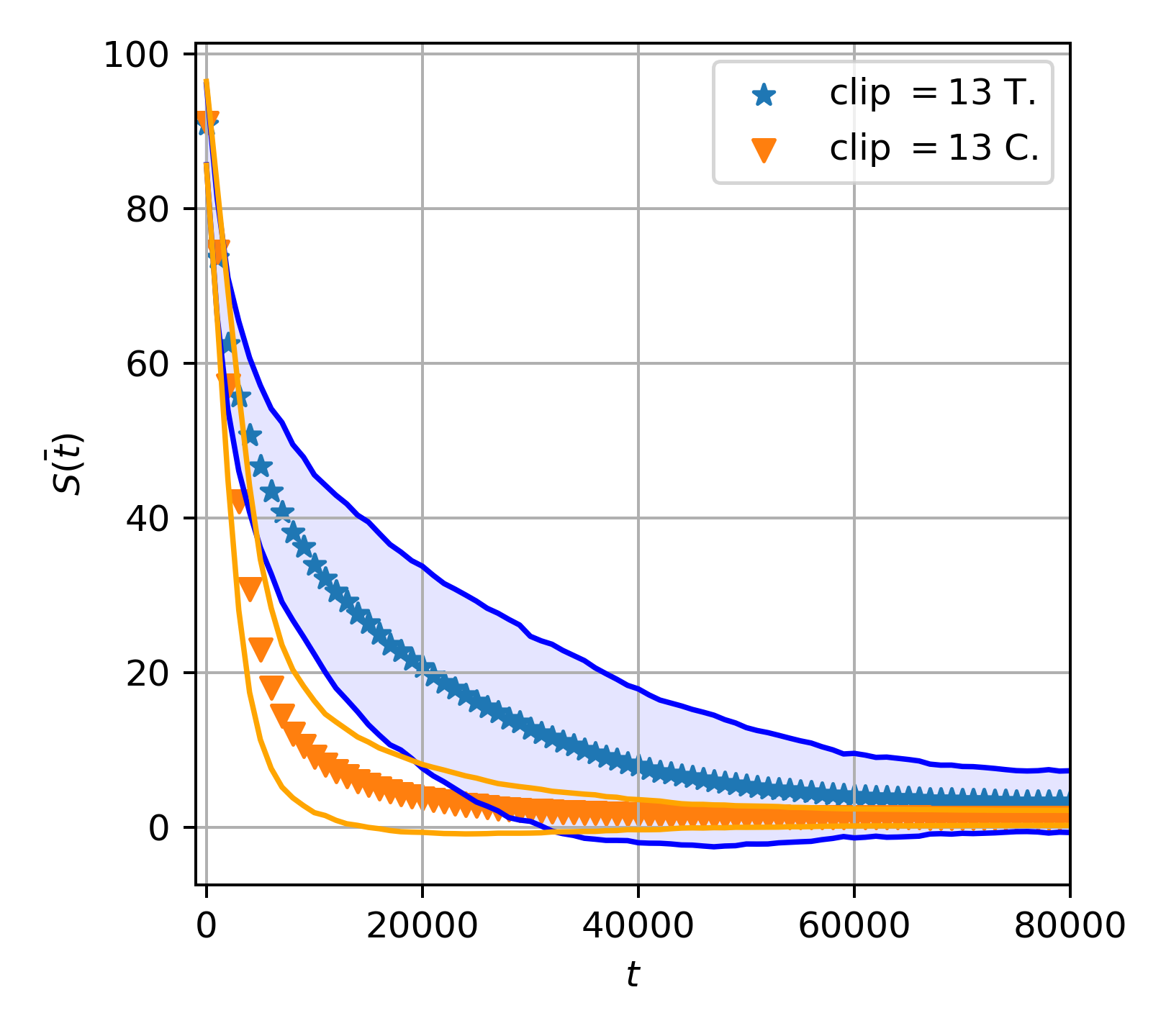}
        \caption{$b=$ 13}
    \end{subfigure}%
    \begin{subfigure}{0.33\textwidth}%
        \includegraphics[width=\textwidth]{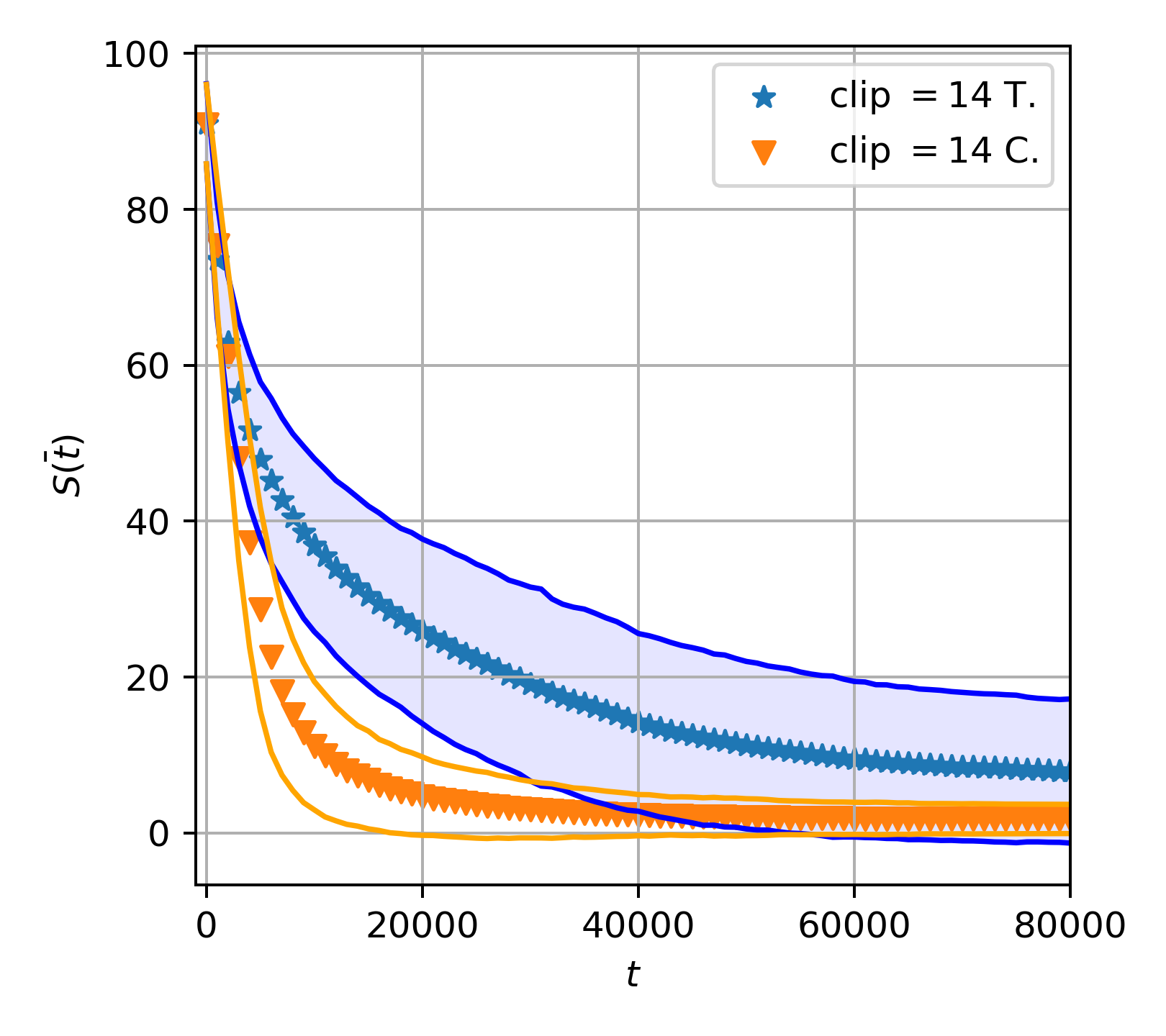}
        \caption{$b=$ 14}
    \end{subfigure}%
    \begin{subfigure}{0.33\textwidth}%
        \includegraphics[width=\textwidth]{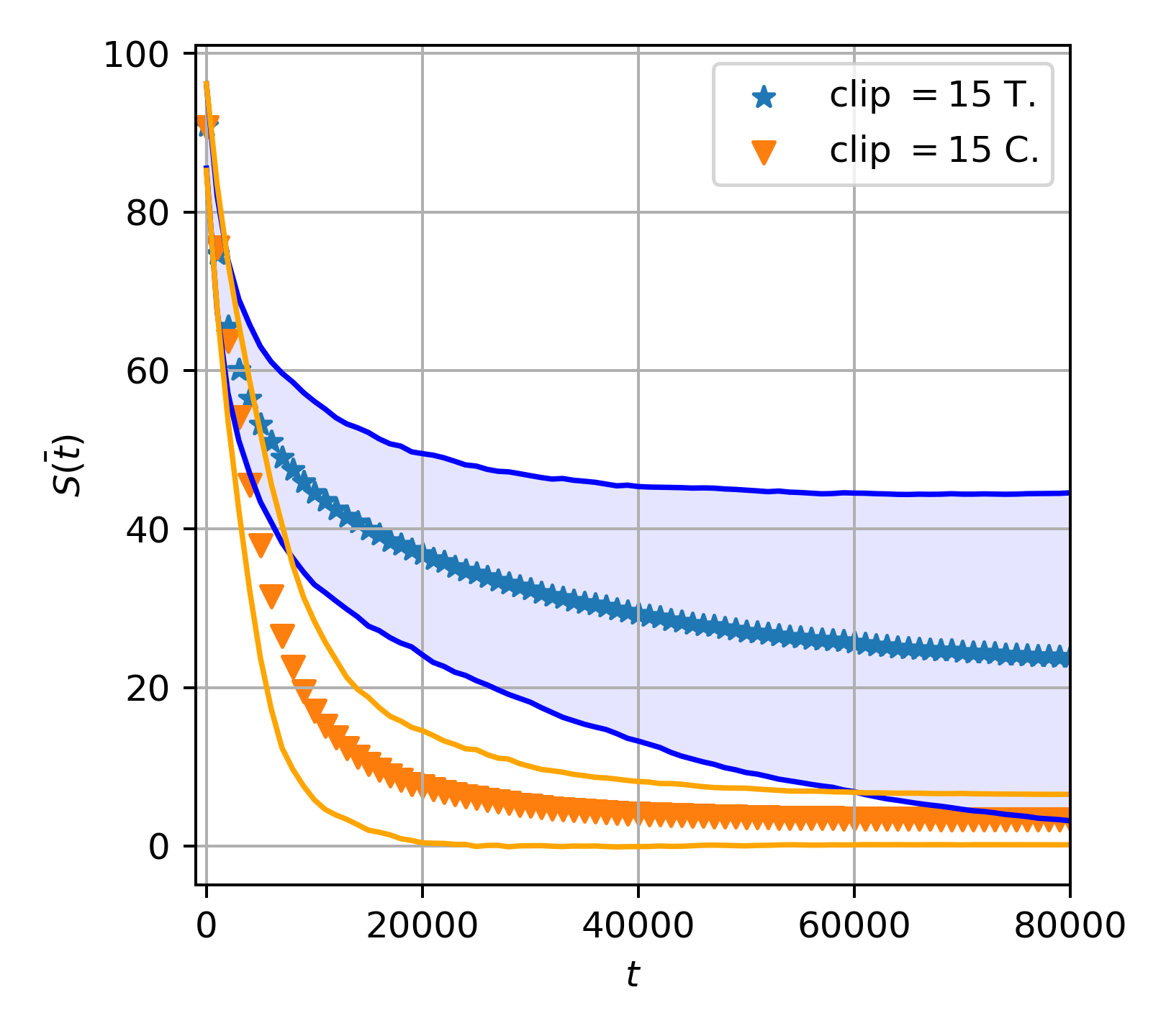}
        \caption{$b=$ 15}
    \end{subfigure}%
    \caption{The mean number of groups over time for the simulations with clip sizes in $\{11,13,15\}$ with $\alpha=2$ on weighted cubic and weighted toroidal opinion spaces (marked C.\ and T.\ respectively). }
    \label{fig:WNoGrps_clips}
\end{figure}

In Figure~\ref{fig:WNoGrps_clips} we plot the number of groups (or opinion `species' present over time in the weighted opinion model with $b=13$, $b=14$, and $b=15$. Similar to in the unweighted model the difference between toroidal and cubic opinion space gets bigger as the confidence gets more bounded. We also see the previously mentioned effect of weighting opinion topics on both models: The number of groups decreases, and this effect is stronger in the toroidal opinion space model than in the cubic opinion space model. Notice that for the $b=15$, the expected number of groups goes way down from more than 60 in without topical weights to just above 20 with topical weights in the toroidal opinion space model. The cubic opinion space model on the other hand has an expected number of groups concentrated much lower than 20.

\subsubsection{Effect of topically weighted opinions on `distance' between agents}\label{sec:distances}
\begin{figure}[htb]
    \centering
    \includegraphics[width=0.4\linewidth]{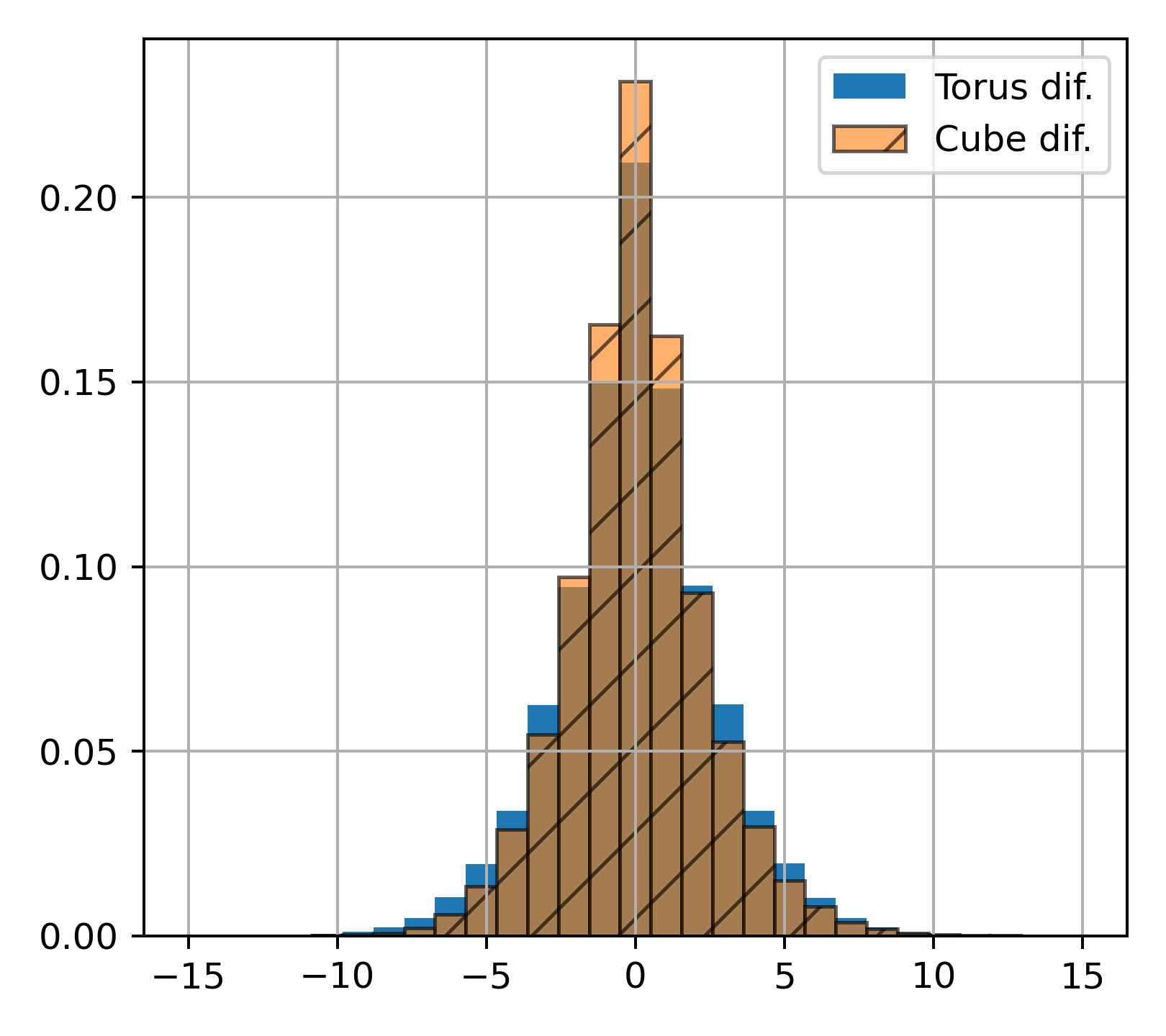}
    \caption{Histogram of the change caused by the uniform weighting of opinions in both cubic and toroidal opinion space. Change is measured: $d_{\text{weighted}}-d_{\text{unweighted}}$, thus a negative change implies that the weighting brought the opinions closer together.}
    \label{fig:change_hist}
\end{figure}

In Figure~\ref{fig:change_hist} we plot the change caused by including the topical weights over various instances. To create this figure we sample 500 agents (with opinions and opinion weightings), then we compare the distance to 499 other agents from the perspective of the first sampled agent under all four distance metrics. This we repeated 1000 times. The resulting histogram shows the change of distance resulting from adding the topical weights. A negative number implies that the distance between the two agents in question shrunk.

From Figure~\ref{fig:change_hist} we see that the change elicited in the toroidal opinion space is more diverse (flatter histogram). More extreme changes in the distance implies that more agents were pulled even closer together, while at the same time there were many agents pushed further apart. This is a possible explanation for why the effect of the topical weights extension has a larger impact on the dynamics of the toroidal opinion space model than the cubic opinion space model.

\begin{figure}[htb]
    \centering
    \begin{subfigure}{0.45\textwidth}
        \includegraphics[width=0.85\textwidth]{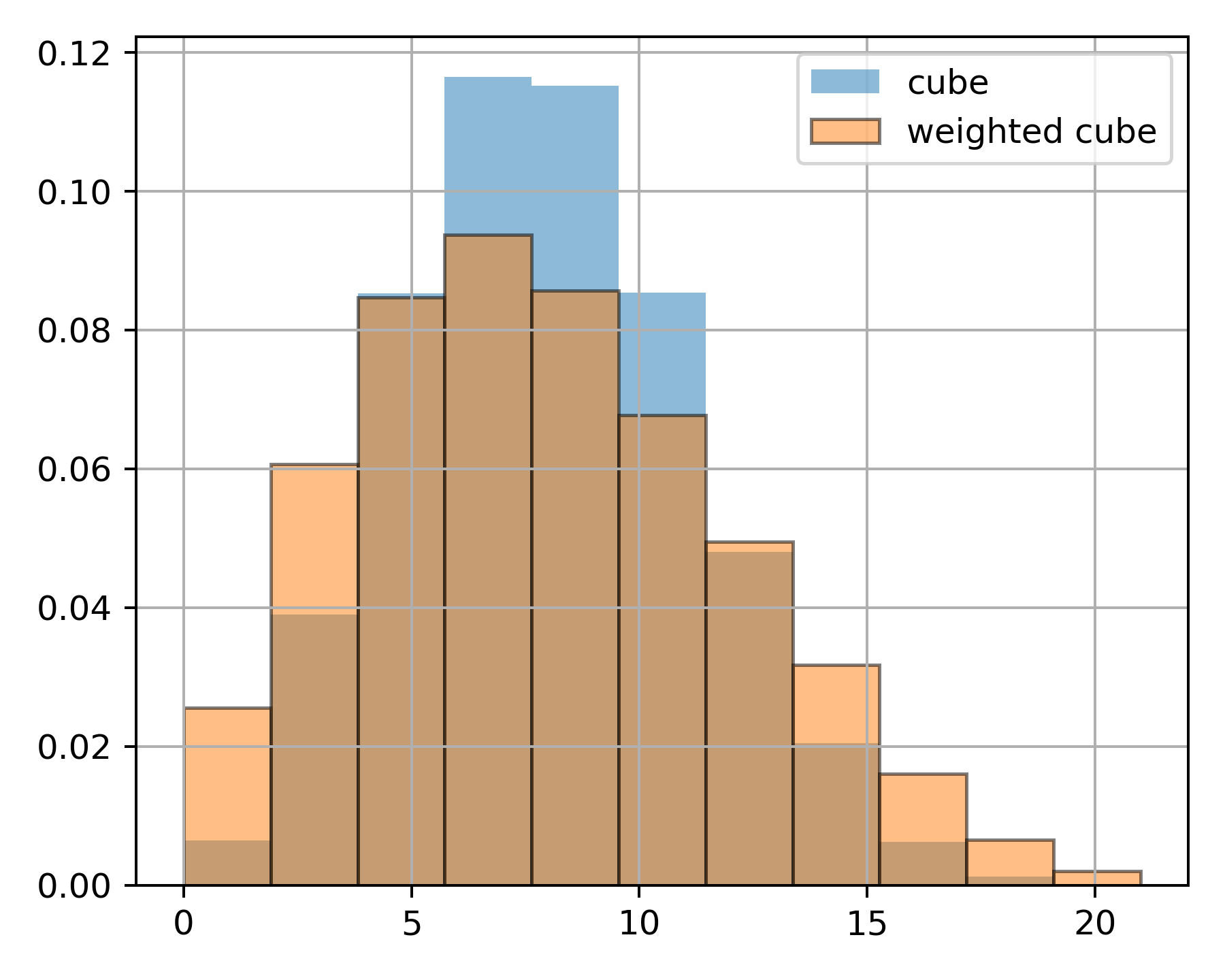}
    \caption{Cubic}
    \label{fig:hist_cube}
    \end{subfigure}%
    \begin{subfigure}{0.45\textwidth}
        \includegraphics[width=0.85\textwidth]{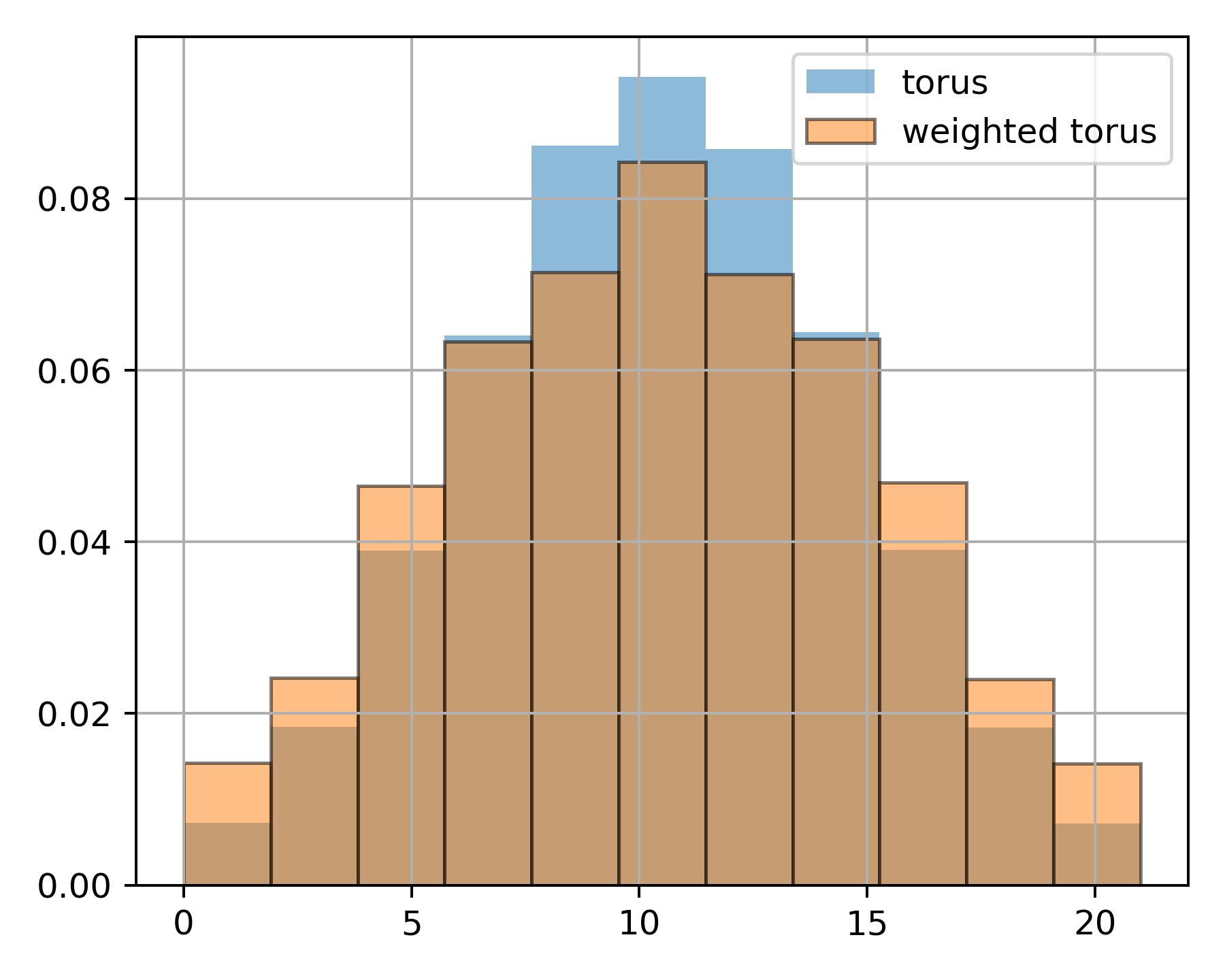}
    \caption{Toroidal}
    \label{fig:hist_tor}
    \end{subfigure}%
    \caption{The distribution of distances under (a) cubic and (b) toroidal opinion space with and without topical weights.}
\end{figure}

Figures~\ref{fig:hist_cube}, and~\ref{fig:hist_tor} show the distributions of opinion distances (weighted and unweighted) in cubic and toroidal opinion space respectively. The effect of the boundaries in the cubic opinion space is exaggerated by the personal opinion weighting modification. The distribution (Figure~\ref{fig:hist_cube}) for the unweighted cubic opinion space is already not symmetric, though this becomes more pronounced under the personal opinion weighting. Conversely, the distribution of distance in the toroidal opinion space (in Figure~\ref{fig:hist_tor}) is symmetric. This symmetry means that \textit{all} agents are equally affected by the changes added, both the bounded confidence and the topical weighting, whereas the agents  on the boundary in a cubic opinion space are more affected than those in the middle, especially by the inclusion of bounded confidence. This explains why the extension to the basic model elicit a greater change in the dynamics of the toroidal opinion space model than the cubic one. 

\subsection{Network rewiring}
In order to determine whether the influence of the toroidal opinion space is consistent over different agent network topologies we run simulations on a rewired grid. We take the regular lattice structure and turn it into a small world network (similar to those studied by Watts and Strogatz~\cite{Watts1998}) by selecting random edges and rewiring them at probability $p\in\{0.1, 0.2, 0.3, 0.4\}$. In doing so we take care to keep the total network connected in order to meaningfully compare probability of consensus upon 1 connected component, though this has the side-effect of slightly changing the probability of observing certain network structures. We implement this by retrying (restarting from the regular grid) if the resulting network turns out to be disconnected at the end of the rewiring procedure.

The effect of such rewiring on the network is well studied. The typical distance between nodes shrinks as the probability of rewiring an edge increases creating a `small-world'. The effect on the dynamics in our model however, is what we elucidate in this section. To do so we run simulations on bounded confidence levels $b\in\{10, 13, 15\}$ in order to enable a range of end states without performing a full search of the parameter space.

\subsubsection{Effect of rewiring on the probability of consensus and the number of groups at steady state}
\label{sec:rewire_con}
\begin{figure}[htb]
    \centering
    \begin{subfigure}{0.33\textwidth}
        \includegraphics[width=0.85\textwidth]{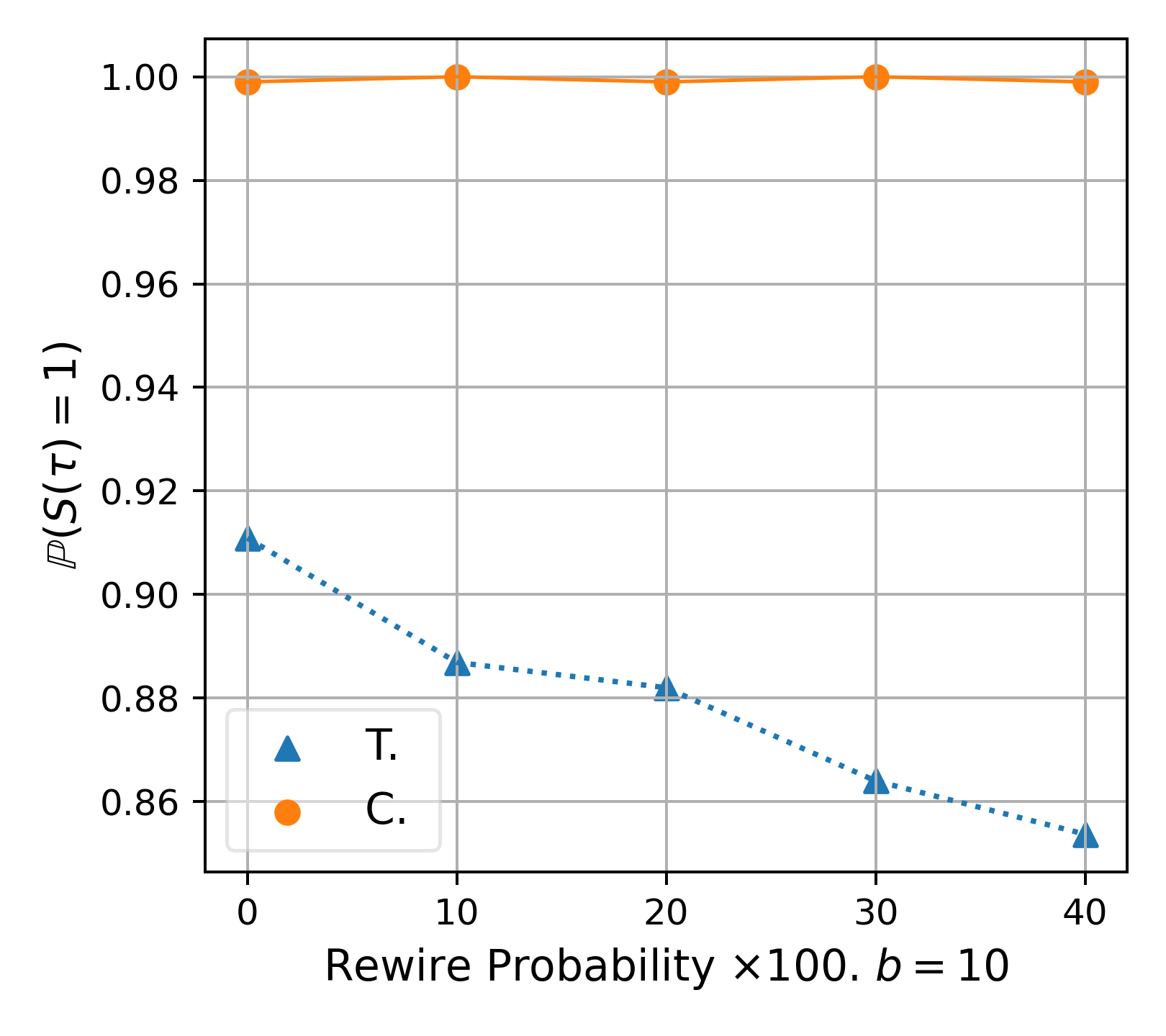}
    \caption{$b=10$}
    \label{fig:W_con_rw_b10}
    \end{subfigure}%
    \begin{subfigure}{0.33\textwidth}
        \includegraphics[width=0.85\textwidth]{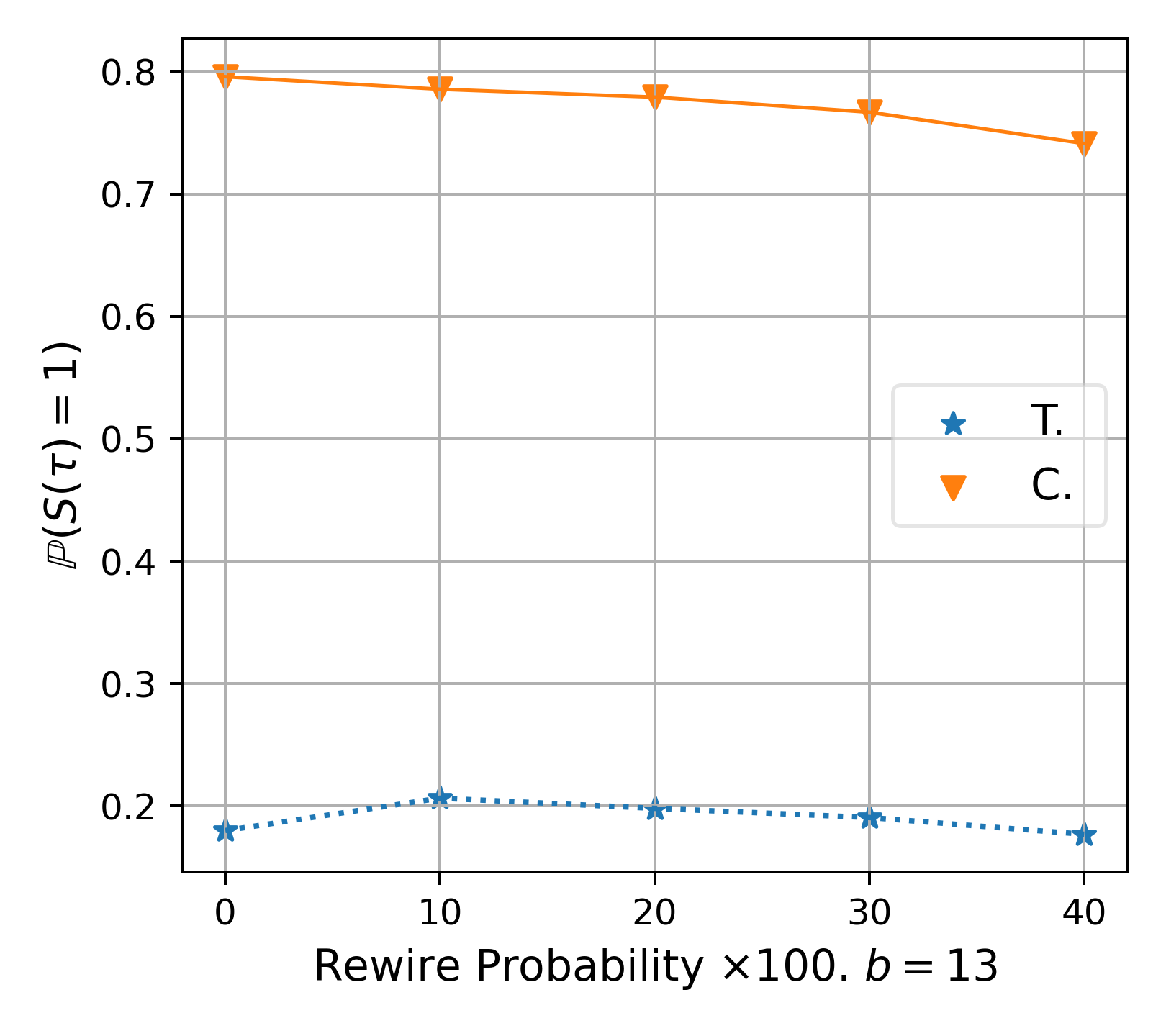}
    \caption{$b=13$}
    \label{fig:W_con_rw_b13}
    \end{subfigure}%
        \begin{subfigure}{0.33\textwidth}
        \includegraphics[width=0.85\textwidth]{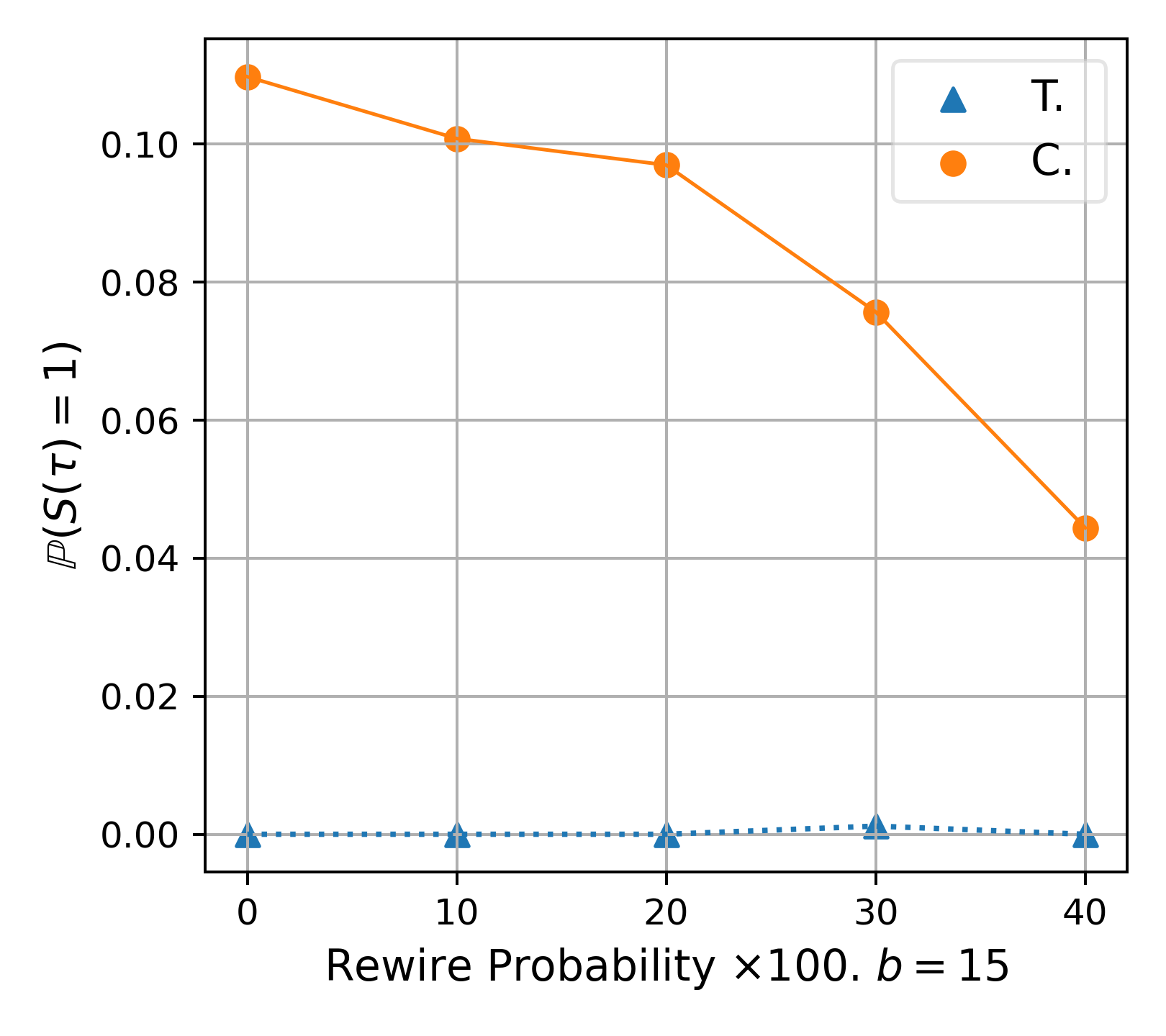}
    \caption{$b=15$}
    \label{fig:W_con_rw_b15}
    \end{subfigure}%
    \caption{The estimated probability of consensus under network rewiring at differing levels of bounded confidence.}
\end{figure}

In Figures~\ref{fig:W_con_rw_b10}--\ref{fig:W_con_rw_b15} we plot the probability of consensus in the toroidal (T.) and cubic (C.) opinion spaces on grids with rewiring as indicated by the $x$-axis. We see that, similarly to the dynamics on networks without rewiring, the effect of this change are felt `earlier' in the toroidal opinion space. That is, the same effect; in this case a reduction of the probability of consensus, is felt at lower levels of bounded confidence in the toroidal opinion space versus the cubic opinion space. This is inline with the finding that the toroidal opinion space is more sensitive to extensions. In particular we see this reduction taking place already at $b=10$ for the toroidal opinion space, and only at $b=15$ in the cubic opinion space.

Note that the effect we observe is unexpected. Rewiring the network edges brings the agents closer together on average. Thus, before conducting these experiments one may suspect that edge rewiring should increase the likelihood of consensus. 

\begin{figure}[htb]
    \centering
    \begin{subfigure}{0.85\textwidth}
        \includegraphics[width=0.9\textwidth]{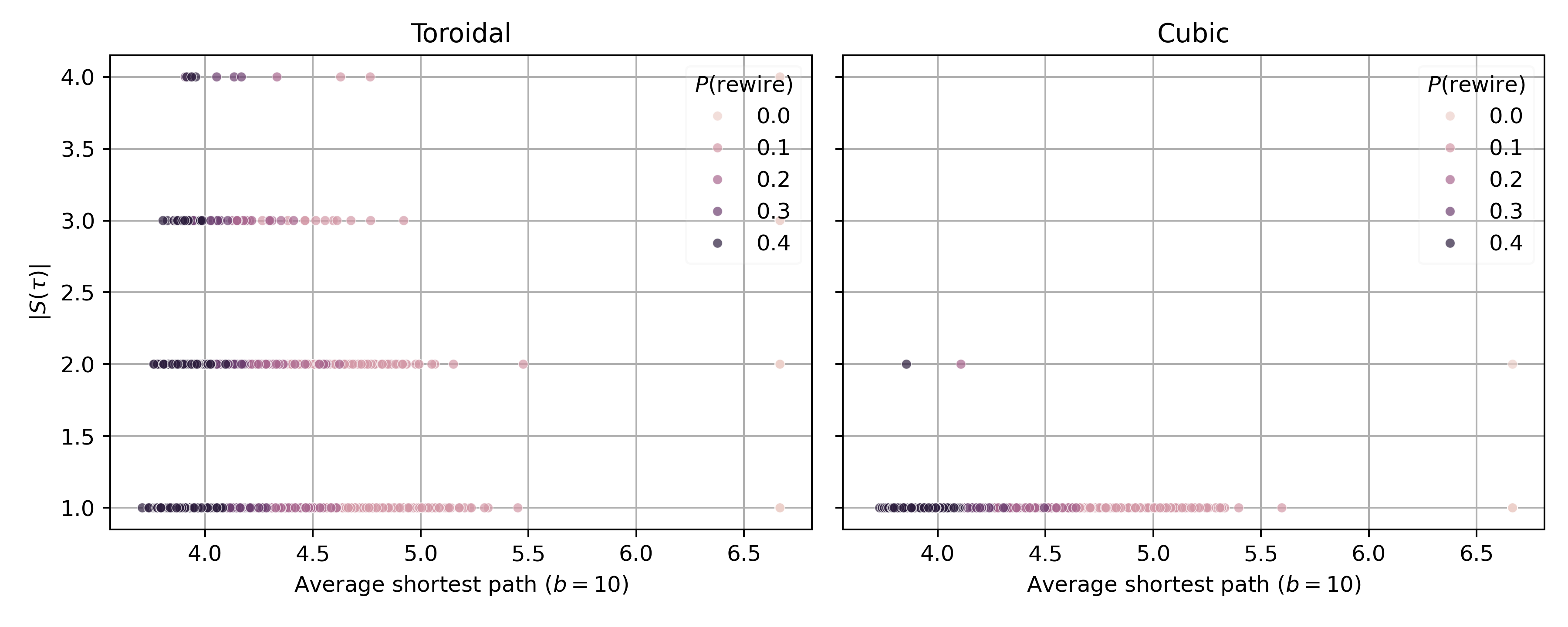}
    \caption{$b=10$}
    \label{fig:S_RW_b10}
    \end{subfigure}\\
        \begin{subfigure}{0.85\textwidth}
        \includegraphics[width=0.9\textwidth]{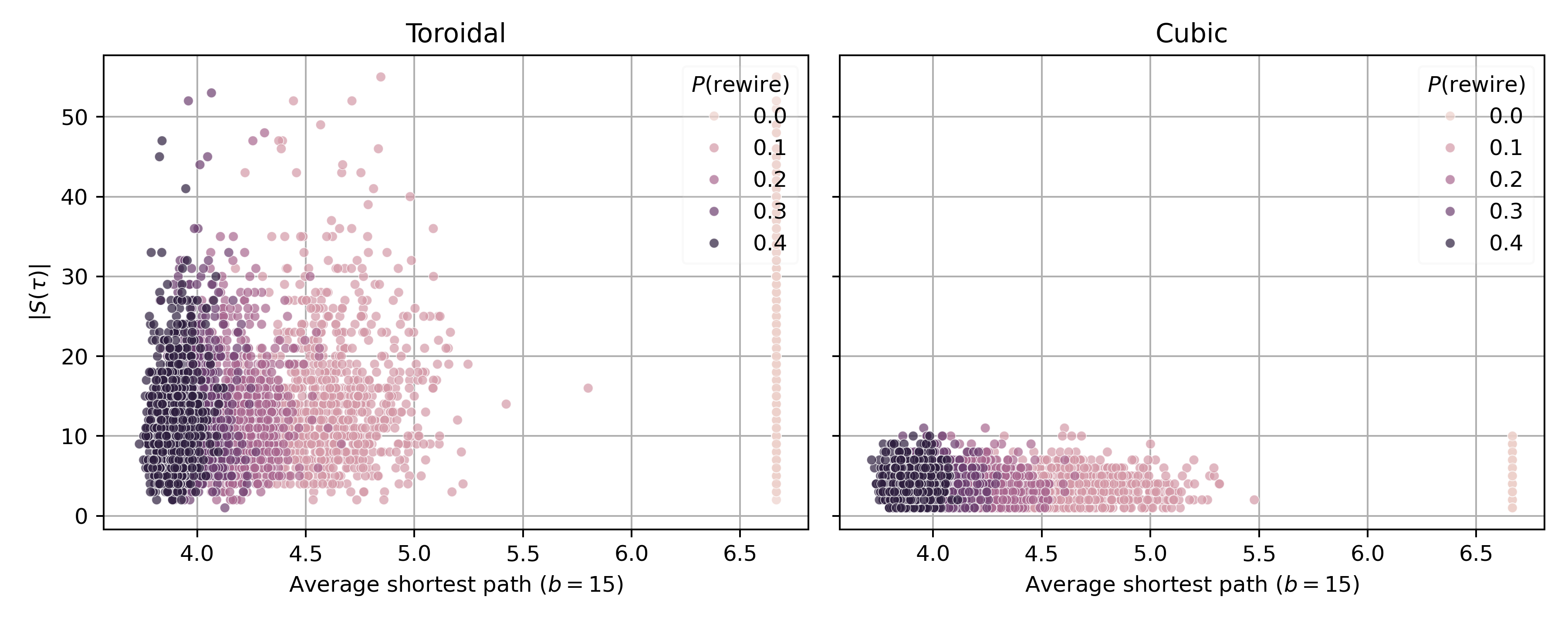}
    \caption{$b=15$}
    \label{fig:S_RW_b15}
    \end{subfigure}%
    \caption{The effect of the average length of the shortest path between vertices in the network on the number of groups at termination of the simulation. That is the per simulation network configuration the average is taken over the shortest paths between vertices.}
    \label{fig:S_sp_rw}
\end{figure}

Further illustrating the effect of the rewiring on the steady state of the dynamics the number of groups in steady steady state ($|S(\tau)|$) is plotted against the average length of the shortest paths between vertices in the network in Figure~\ref{fig:S_sp_rw}. In this figure the hue indicates the probability of rewiring, allowing us to verify that more rewiring decreases the distances between vertices in the network. 

The effect of a decreasing shortest path is weak, but similar to that of more rewiring: shorter distances between vertices, in this model lead to more groups in the steady state and less consensus. We believe this is highly contingent on the starting configuration of the network. Indeed in the opinion dynamics model of~\cite{Meylahn_Searle_2024}, more rewiring leads to consensus more often, which is explained by the fact that disparate clusters are less likely to exist under more rewiring of the circulant network they study. Circulants however have a large number of triangles before rewiring which means that with rewiring there is more connection between already close-knit groups. Rewiring from a 2-dimensional grid means that initially there are no triangles, and so rewiring may decrease the path length, but need not necessarily imply more connection between close-knit divisions of the network. In fact, the rewiring creates opportunities for triangles to form, thus inviting the possibility of closed off groups and more disagreement in the network as a whole. In Appendix~\ref{app:circ} we illustrate the dynamics on rewired circulants for which the starting configuration has many triangles.

\subsubsection{Effect of network rewiring on the time-scales of the dynamics}

\begin{figure}[htb]
    \centering
    \begin{subfigure}{0.85\textwidth}
        \includegraphics[width=0.9\textwidth]{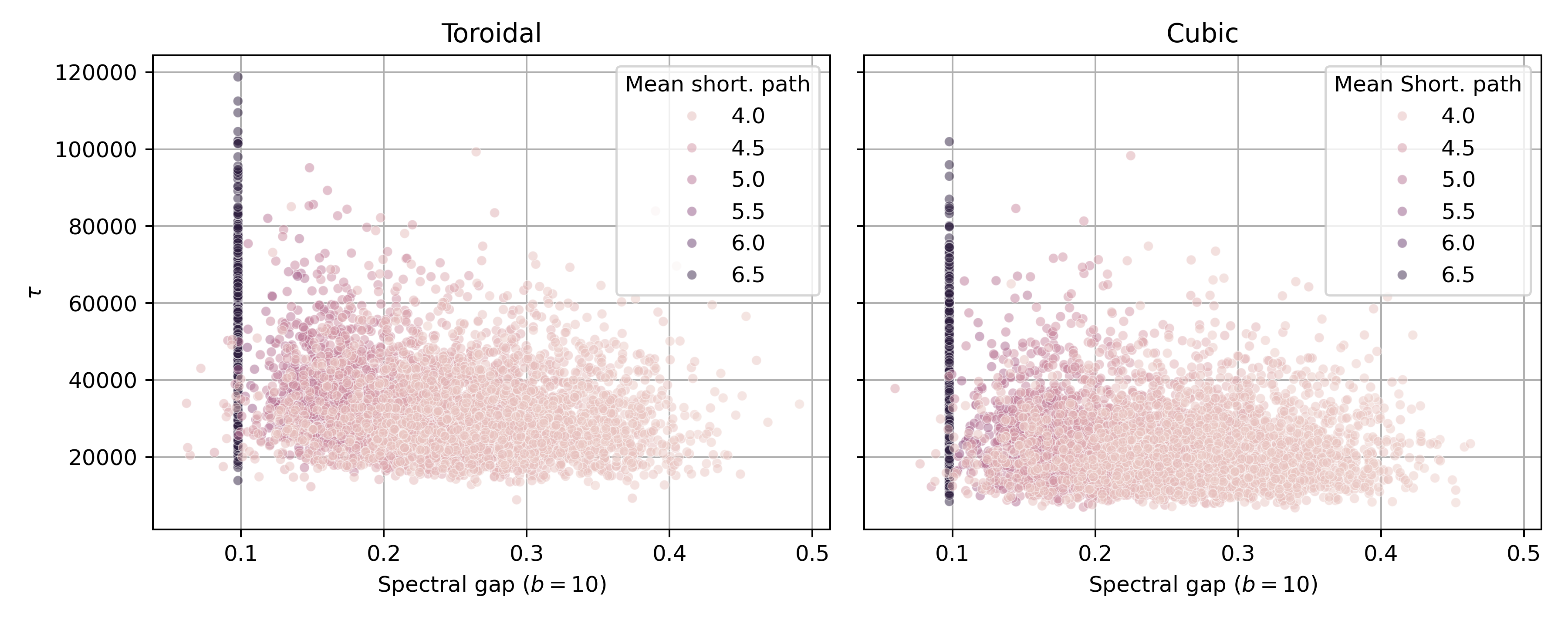}
    \caption{$b=10$}
    \label{fig:Tau_RW_b10}
    \end{subfigure}\\
        \begin{subfigure}{0.85\textwidth}
        \includegraphics[width=0.9\textwidth]{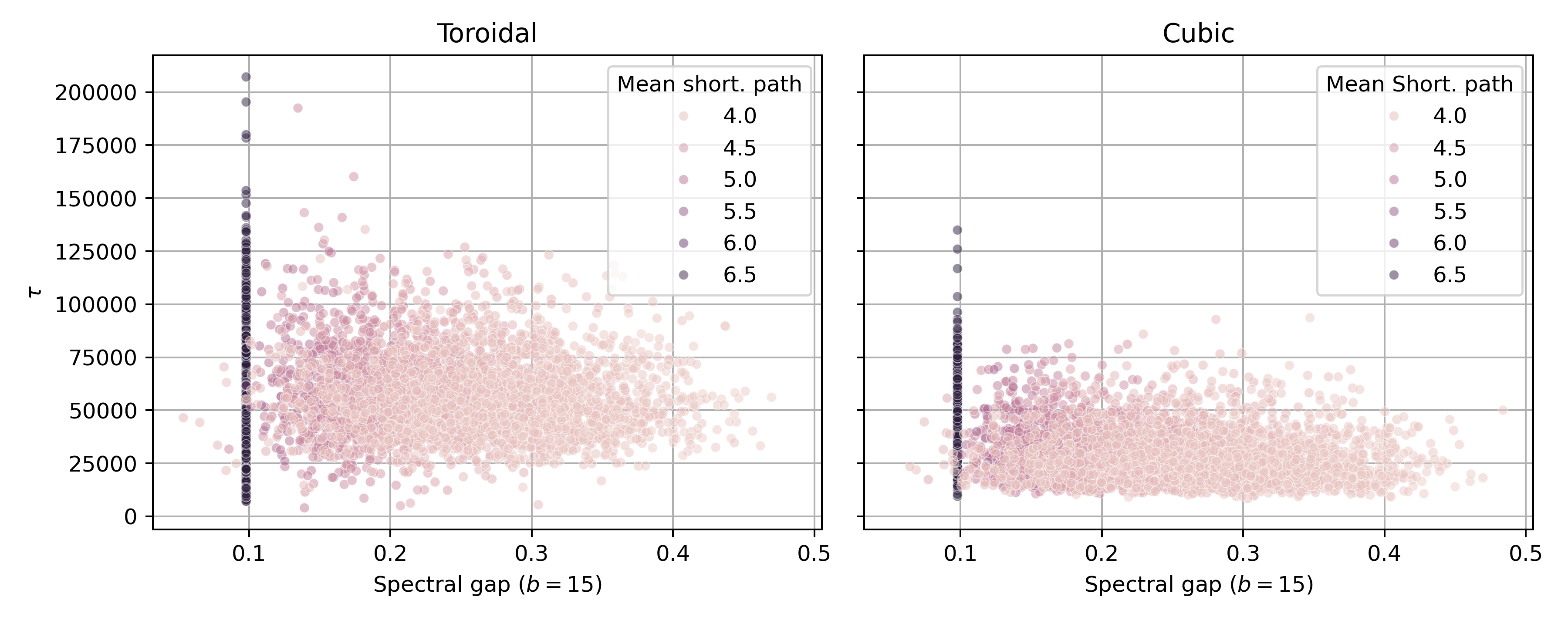}
    \caption{$b=15$}
    \label{fig:Tau_RW_b15}
    \end{subfigure}%
    \caption{The effect of the spectral gap of the network termination time $\tau$ of the simulation.}
    \label{fig:TW_sp_rw}
\end{figure}

The spectral gap (second eigenvalue of Laplacian of the network) has been studied in terms of the effect that it has on the mixing time and the consensus time. In this way the spectral gap acts as a measure of how fast information spreads through a network. In Figures~\ref{fig:Tau_RW_b10}--\ref{fig:Tau_RW_b15} we plot the time to termination of the simulations against the spectral gap on the $x$-axis for different values of the bounded confidence. The expected relationship is present but weak. A greater spectral gap typically implies faster convergence, which is also present in our dynamics. Of course it should be noted that existing results for other models are not expected to apply directly in this model. 

The difference between toroidal and cubic opinion spaces is more subtle in this measure, though as before we consistently see the cloud of points in the cubic opinion space is concentrated at a lower $\tau$ (quicker convergence).

\begin{figure}
    \centering
    \begin{subfigure}{0.33\textwidth}%
        \includegraphics[width=\textwidth]{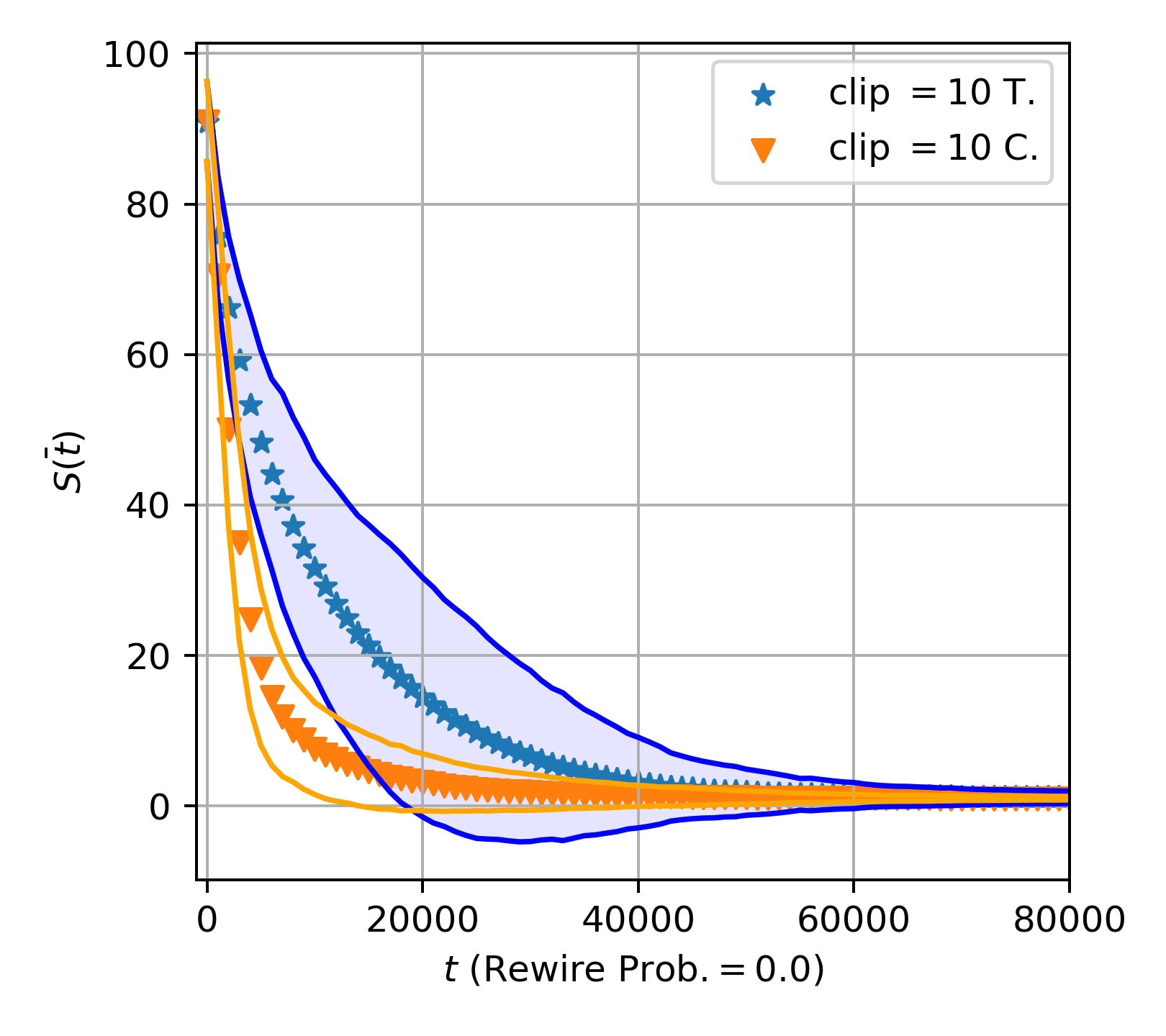}
        \caption{$b=10$, $p=0.0$}
        \label{fig:grps_RW_b10_p0}
    \end{subfigure}%
    \begin{subfigure}{0.33\textwidth}%
        \includegraphics[width=\textwidth]{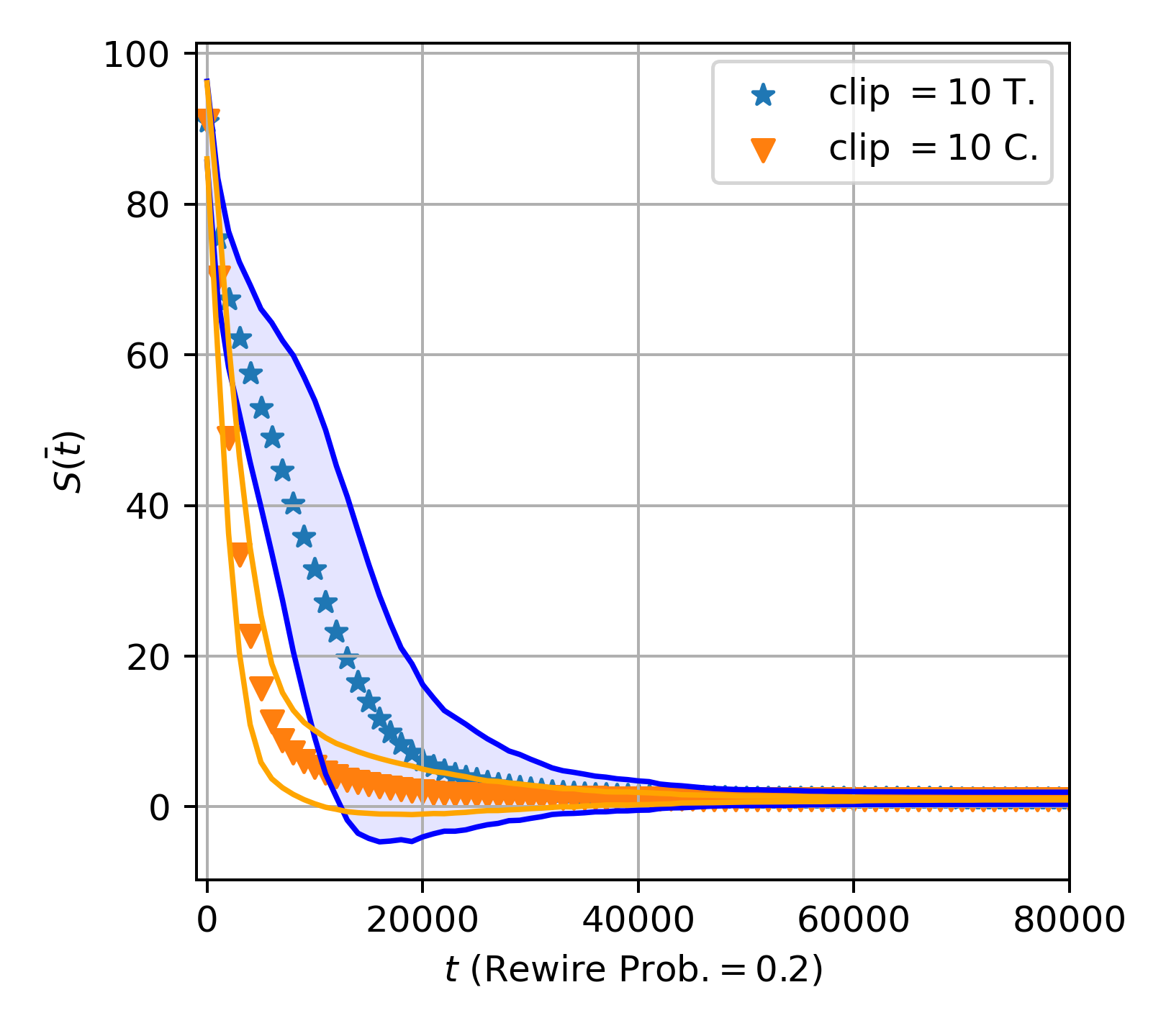}
        \caption{$b=10$, $p=0.2$}
        \label{fig:grps_RW_b10_p2}
    \end{subfigure}%
    \begin{subfigure}{0.33\textwidth}%
        \includegraphics[width=\textwidth]{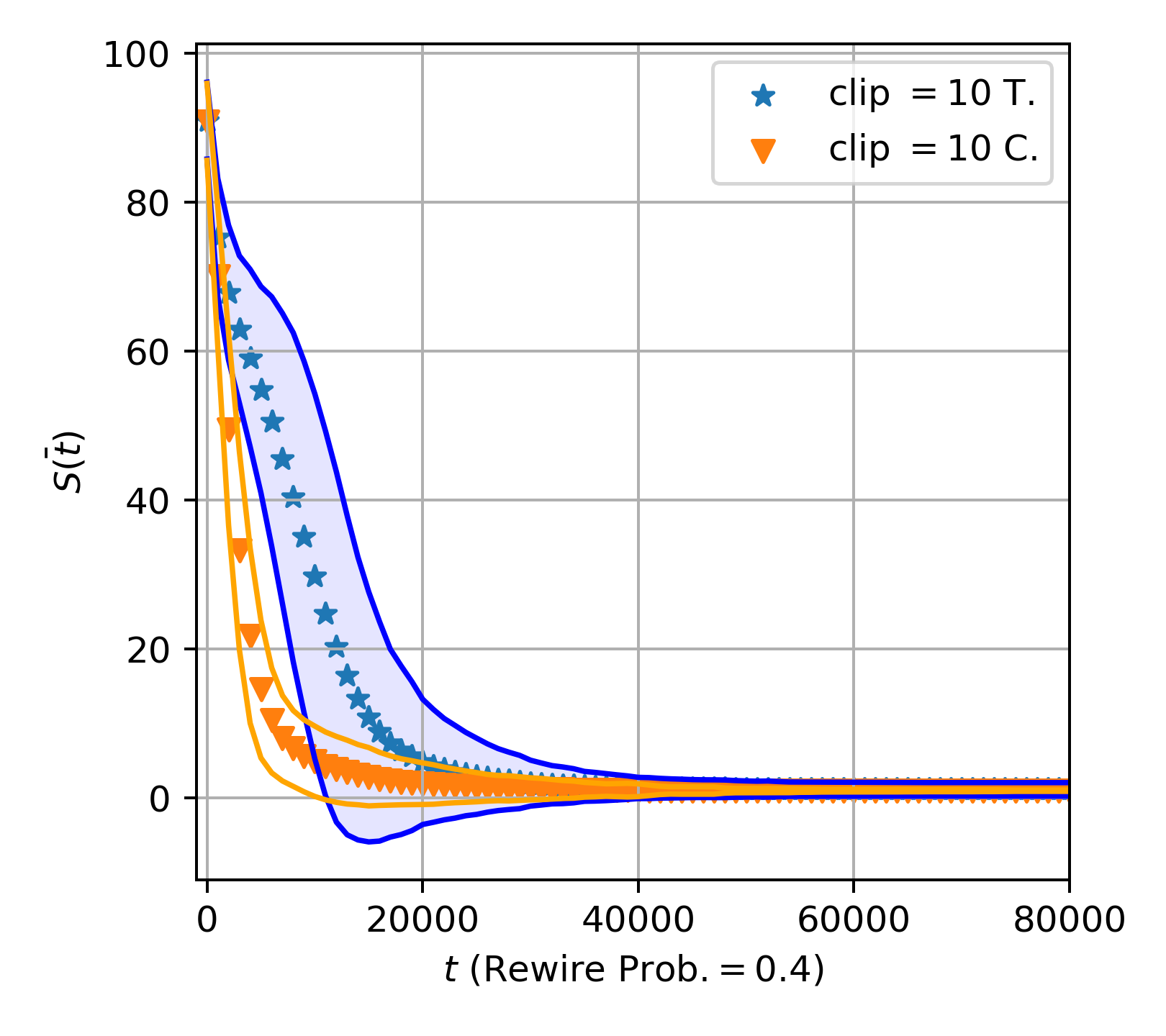}
        \caption{$b=10$, $p=0.4$}
        \label{fig:grps_RW_b10_p4}
    \end{subfigure}\\
    \begin{subfigure}{0.33\textwidth}%
        \includegraphics[width=\textwidth]{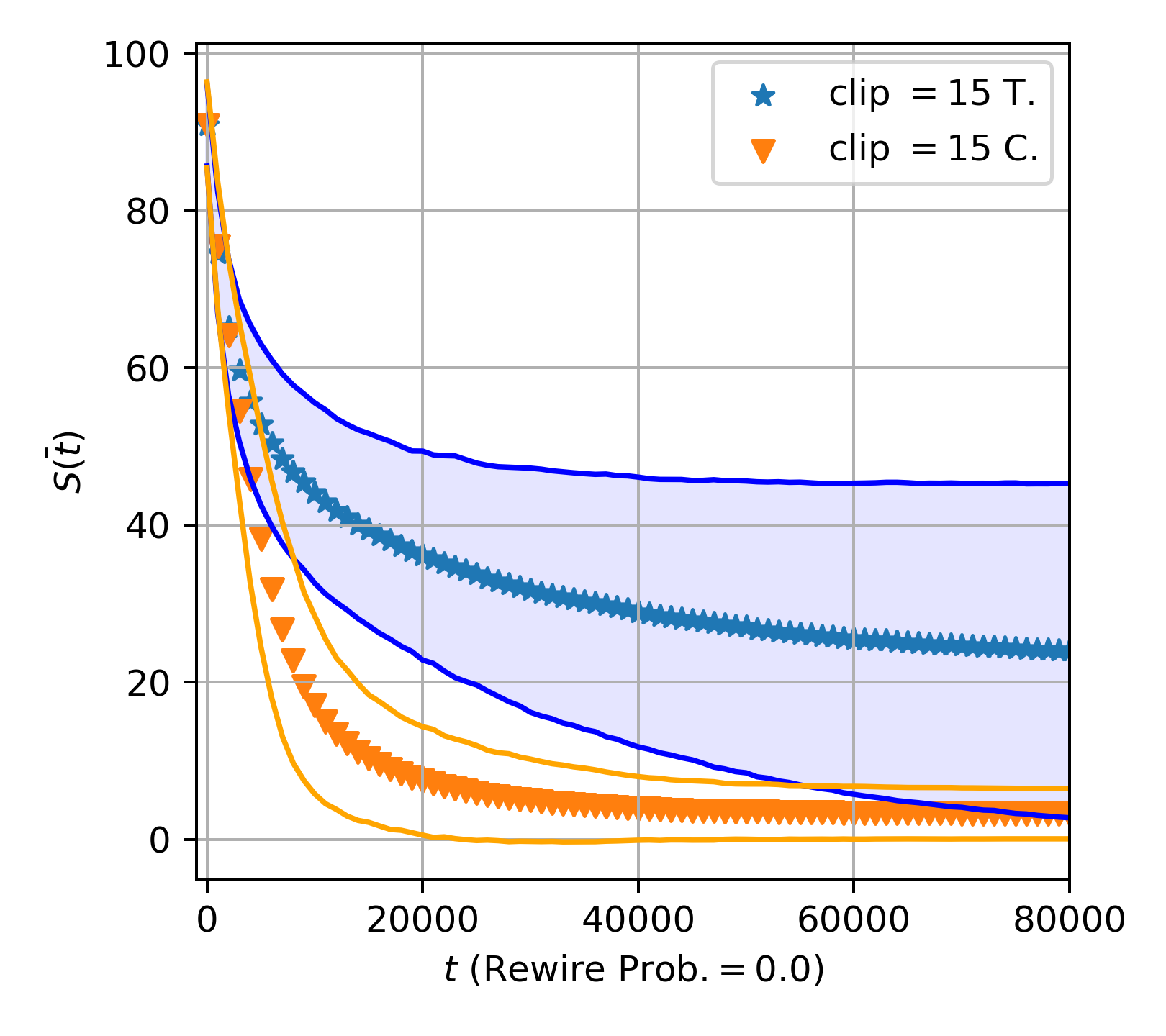}
        \caption{$b=15$, $p=0.0$}
        \label{fig:grps_RW_b15_p0}
    \end{subfigure}%
    \begin{subfigure}{0.33\textwidth}%
        \includegraphics[width=\textwidth]{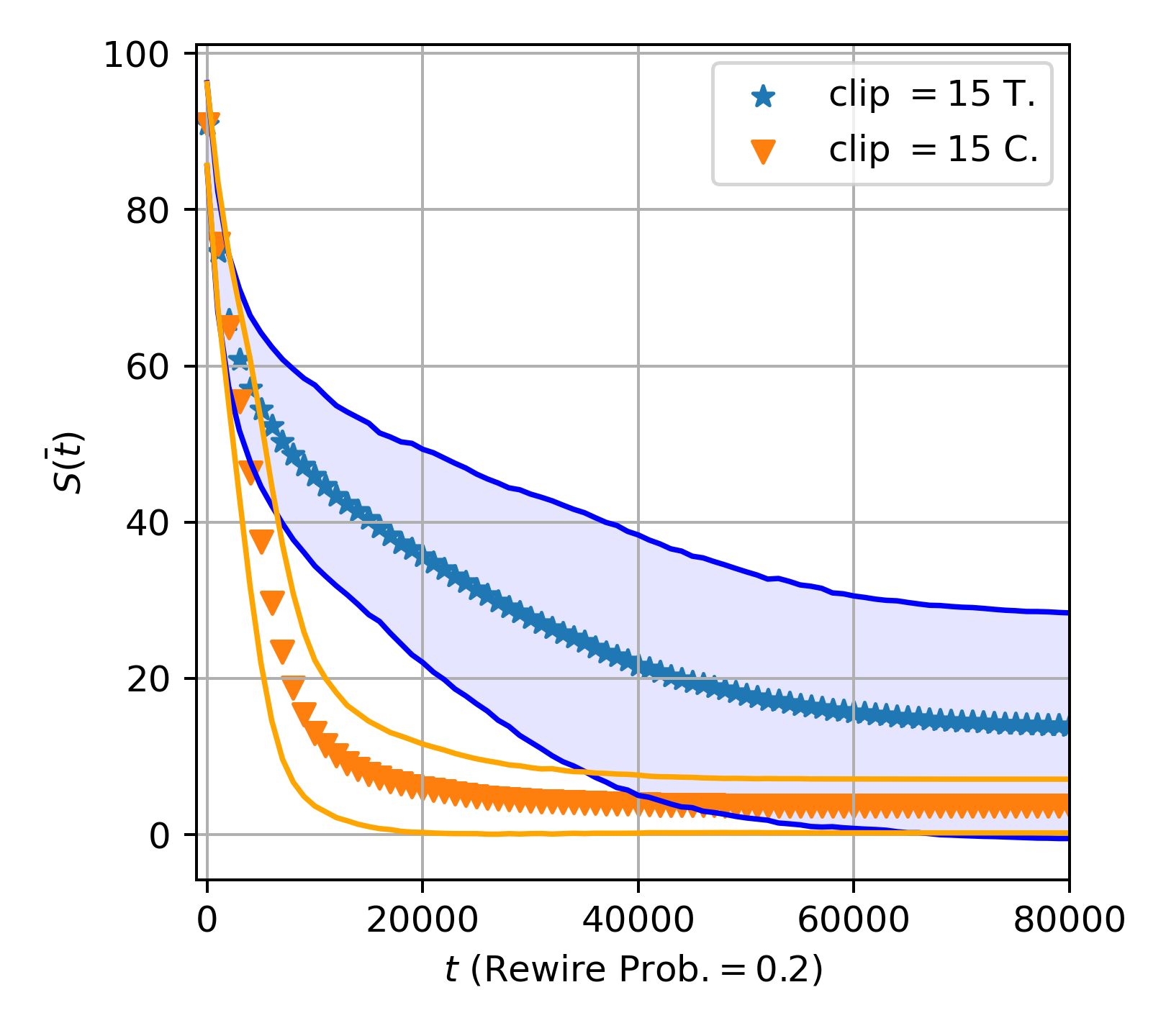}
        \caption{$b=15$, $p=0.2$}
        \label{fig:grps_RW_b15_p2}
    \end{subfigure}%
    \begin{subfigure}{0.33\textwidth}%
        \includegraphics[width=\textwidth]{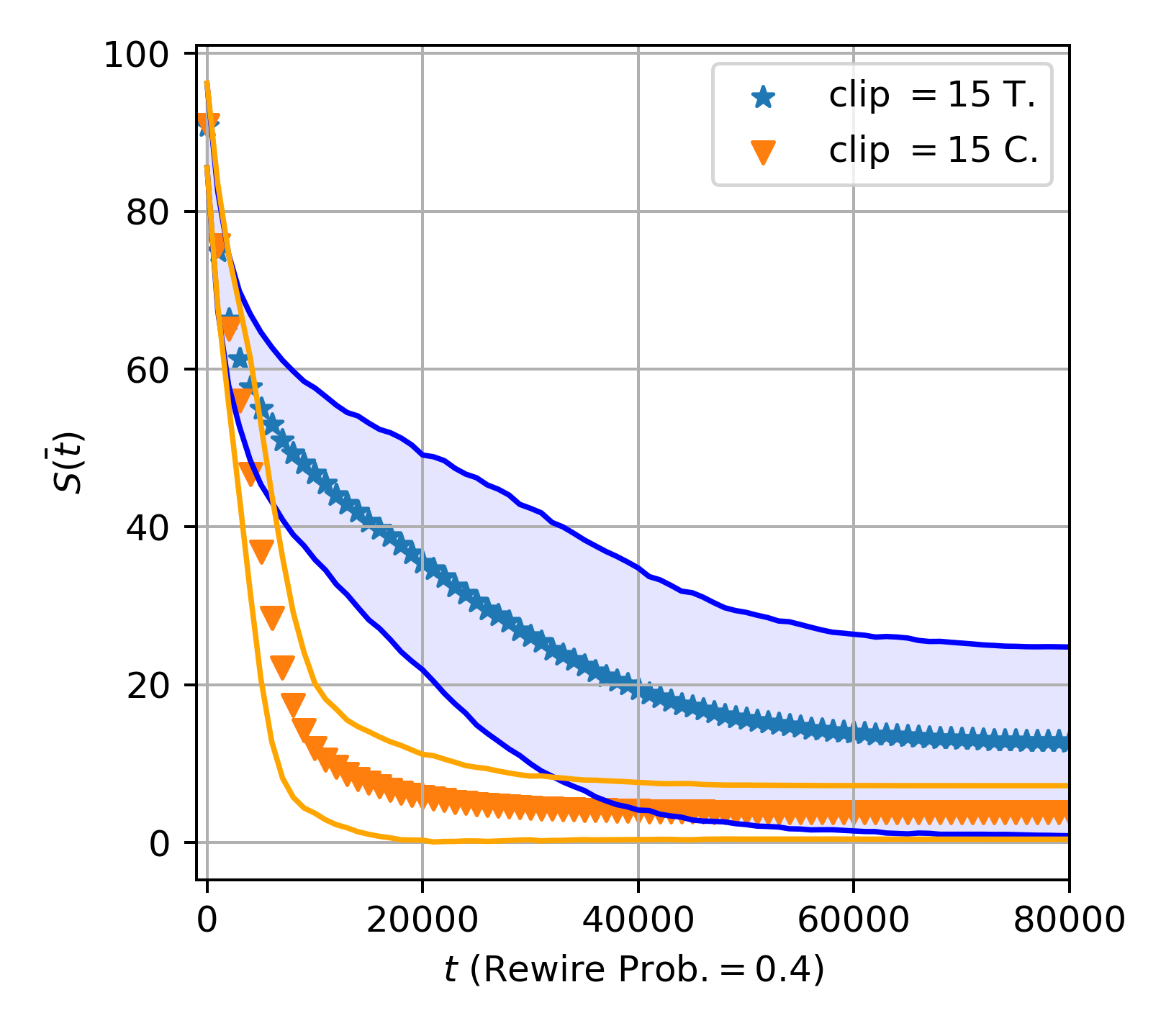}
        \caption{$b=15$, $p=0.4$}
        \label{fig:grps_RW_b15_p4}
    \end{subfigure}%
    \caption{The mean number of groups over time for the simulations with clip sizes in $\{10,13, 15\}$ and rewiring probability $p\in\{0.0, 0.2, 0.4\}$ with $\alpha=2$ on weighted cubic and weighted toroidal opinion spaces (marked C.\ and T.\ respectively).}
    \label{fig:grps_RW_time}
\end{figure}

The dynamics of the number of groups over time are plotted in Figure~\ref{fig:grps_RW_time}. Again we see that the toroidal opinion space is more sensitive to the inclusion of the rewiring extension. The effect being a kind of twisting. We see that the number of groups stays higher for a little longer, and then decreases quickly thereafter. In the last row of this figure we can see that rewiring, while decreasing the likelihood of consensus, at least in the parameter setting with $b=15$, decreases the average number of groups in steady state. This is more akin to what one may expect the effect of rewiring to be. The rewiring may thus shift the network from a loosely connected grid to a fewer number (greater than 1) of better connected clusters.

The corresponding figures showing the effect of rewiring on the time to steady-state for $b=13$ are deferred to Appendix~\ref{app:omitFig} for brevity of the main text. In particular these figures are deferred because they are sufficiently similar to the included figures that their inclusion here is not required.

\section{Discussion}\label{sec:discussion}
The fact that without bounded confidence and personal opinion weighting, the two opinion spaces both always result in consensus in relatively similar time scales illustrates that the two baseline models are suitable for comparison.

Under the addition of bounded confidence a difference between the models emerges: The toroidal opinion space allows for more disagreement between neighbouring agents. The probability of consensus drops to zero faster (as the confidence shrinks), and the number of `opinions species' increases faster than the cubic opinion space. 

Adding topical weights to the bounded confidence extension encourages agreement between neighbouring agents. This pushes the probability of consensus up and the shifts the peak in the probability of polarization to the right (larger clip sizes). We see this effect in both the toroidal and the cubic opinion space models. However, the effect is greater in the toroidal opinion space model. Our conclusions about the effect of adding the bounded confidence and the topical weights extension to the toroidal and cubic opinion space model is robust to changes in the parameter $\alpha$. We demonstrate this in Appendix~\ref{sec:sensitivity}.

There are two main differences between the dynamics in the cubic and toroidal opinion spaces. The first is that the toroidal opinion space consistently allows for a more diverse range of opinions to be sustained in the steady state. At levels of bounded confidence (clip sizes) where consensus is guaranteed in the cubic space, the toroidal opinion spaces shows the emergence of bi-polarization and even $S(\tau)>2.$ When bi-polarization becomes possible and likely in the cubic opinion space, the toroidal opinion space shows opinion fragmentation (many different views held in steady state). The second main difference is that the toroidal opinion space model is more sensitive to changes in the model under the settings we investigated. As already mentioned, the number of groups in steady state (of the toroidal opinion space model) is a lot greater and a lot more diverse than in the cubic opinion space model (with or without the extension of topical weights). Furthermore, as we go from only bounded confidence to bounded confidence and topical weights, the cubic model dynamics hardly change, while the dynamics in the toroidal opinion space undergo a much larger version of the same change. The short explanation for these differences is that all agents in the toroidal opinion space model are affected equally by the addition of bounded confidence and the topical weights. In contrast to this, in the cubic opinion space model, the bounding of confidence affects mainly the agents on the extremes. Agents with moderate views are not influenced until the clip is sufficiently large. An agent with views exactly in the centre of the space for example will not `experience' the bounding of confidence until the the $b>d_{\text{max}}/2$.


As noted, the grid we use as populations structure is a natural starting point. We extend our investigation to include the effect of network rewiring and find that the key results hold true in the face of this extension. The toroidal opinion space allows more groups in steady state that the cubic space even under network rewiring. The toroidal opinion space is also more sensitive to the inclusion of this extension than the cubic opinion space which is inline with the other results in this paper. Furthermore, it has been shown that the specific network topology (whether grid or small-world for example) plays a significantly smaller role than the interaction dynamics for the Axelrod models with ordinal traits~\cite{Dinkelberg2021}. Our results corroborates this finding for the opinion dynamics model we study. 

The effect of the network rewiring itself seems counterintuitive at first though becomes clear when looking at the starting point. More network rewiring in the grid graph creates the opportunity for better defined groups and thus decreases the probability of consensus. This is the opposite finding to that in~\cite{Meylahn_Searle_2024} where in a different opinion dynamics model, rewiring increases the likelihood of consensus. In their paper the rewiring takes place on circulant graphs whereas we focus on rewiring from grid graphs. In Appendix~\ref{app:circ} we briefly demonstrate that also for our model rewiring increases the likelihood of consensus (and generally decreases the number of groups in steady state) when starting from circulant graphs.

We leave investigating whether, as shown by Dinkelberg \textit{et al.}\,\cite{Dinkelberg2021}, and in the current paper, the effect of network topology is not as important as the interaction dynamics for even \textit{more} realistic networks for future work.

We proceed by a short discussion of the difference between the weighted Axelrod model studied by Kalinowska and Dybiec~\cite{Kalinowska2023} and ours. The implementation of weights per opinion functions as a bounded confidence mechanism. If the similarity is not high enough (depending on only the weight of the opinion that may change) then interaction happens at probability zero. Important to highlight here that it depends only on the weight of the \textit{one} opinion dimension which is under consideration for change. This means that the weights of the other opinion dimension do not come into play at all. In our model the similarity measure is \textit{weighted} by the weights on the dimensions of the opinion. This way the whole range of opinions plays a role, weighted by how important an agent considers them to be.

Most models in the opinion dynamics literature operate on a space with extremes. Our investigation shows that the dynamics resulting from a model may be quite sensitive to the choice of opinion space. In particular where some extension appear almost non-influential to the cubic opinion space, they can have a large impact in the toroidal opinion space. The choice of opinion space is important and should be discussed explicitly.

\subsection*{Data availability}
The code for the simulation model in this paper is available at \url{https://github.com/Benephfer/Toroidal-and-cubic-opinion-space-dynamics}.
\bibliographystyle{ieeetr}  
\bibliography{references}  

\begin{thebibliography}{10}

\bibitem{Castellano2009}
C.~Castellano, S.~Fortunato, and V.~Loreto, ``Statistical physics of social dynamics,'' {\em Review of Modern Physics}, vol.~81, no.~2, pp.~591--646, 2009.

\bibitem{Flache2017}
A.~Flache, M.~M\"{a}s, T.~Feliciani, E.~Chattoe-Brown, G.~Deffuant, S.~Huet, and J.~Lorenz, ``Models of social influence: {T}owards the next frontiers,'' {\em Journal of Artificial Societies and Social Simulation}, vol.~20, no.~4, 2017.
\newblock article no. 2.

\bibitem{Noorazar2020}
H.~Noorazar, K.~Vixie, A.~Talebanpour, and Y.~Hu, ``From classical to modern opinion dynamics,'' {\em International Journal of Modern Physics C}, vol.~31, no.~7, 2020.
\newblock article no. 2050101.

\bibitem{Jusup2022}
M.~Jusup, P.~Holme, K.~Kanazawa, M.~Takayasu, I.~Romić, Z.~Wang, S.~Geček, T.~Lipić, B.~Podobnik, L.~Wang, W.~Luo, T.~Klanjšček, J.~Fan, S.~Boccaletti, and M.~Perc, ``Social physics,'' {\em Physics Reports}, vol.~948, pp.~1--148, 2022.

\bibitem{Liu2026}
S.~Liu, Z.~Wu, and L.~Mart\'inez, ``An overview of opinion polarization: models, drivers, and strategic solutions,'' {\em Information Processing \& Management}, vol.~63, no.~2, Part A, p.~104433, 2026.

\bibitem{Holley1975}
R.~A. Holley and T.~M. Ligget, ``Ergodic theorems for weakly interacting infinite systems and the voter model,'' {\em The Annals of Probability}, vol.~3, no.~4, pp.~643--663, 1975.

\bibitem{Sznajd2000}
K.~Sznajd-Weron and J.~Sznajd, ``Opinion evolution in closed community,'' {\em International Journal of Modern Physics C}, vol.~11, no.~06, pp.~1157--1165, 2000.

\bibitem{Galam2002}
S.~Galam, ``Minority opinion spreading in random geometry,'' {\em European Physics Journal B}, vol.~25, pp.~403--406, 2002.

\bibitem{DeGroot1974}
M.~H. DeGroot, ``Reaching a consensus,'' {\em Journal of the American Statistical Association}, vol.~69, no.~345, pp.~118--121, 1974.

\bibitem{Hegselmann2002}
R.~Hegselmann and U.~Krause, ``Opinion dynamics and bounded confidence models, analysis and simulation,'' {\em Journal of Artificial Societies and Social Simulation}, vol.~5, no.~3, 2000.

\bibitem{Chan2024}
K.~Chan, R.~Duivenvoorden, A.~Flache, and M.~Mandjes, ``A relative approach to opinion formation,'' {\em The Journal of Mathematical Sociology}, vol.~48, no.~1, pp.~1--41, 2024.

\bibitem{Caponigro2015}
M.~Caponigro, A.~C. Lai, and B.~Piccoli, ``A nonlinear model of opinion formation on the sphere,'' {\em Discrete and Continuous Dynamical Systems}, vol.~35, no.~9, pp.~4241--4268, 2015.

\bibitem{Aydogdu2017}
A.~Aydoğdu, S.~T. McQuade, and N.~P. Duteil, ``Opinion dynamics on a general compact riemannian manifold,'' {\em Networks and Heterogeneous Media}, vol.~12, no.~3, pp.~489--523, 2017.

\bibitem{Axelrod1997}
R.~Axelrod, ``The dissemination of culture: A model with local convergence and global polarization,'' {\em Journal of Conflict Resolution}, vol.~41, no.~2, pp.~203--226, 1997.

\bibitem{Meylahn2024c}
B.~V. Meylahn and J.~M. Meylahn, ``How social reinforcement learning can lead to metastable polarisation and the voter model,'' {\em PLoS ONE}, vol.~19, no.~12, p.~e0313951, 2024.

\bibitem{Banisch2019}
S.~Banisch and E.~Olbrich, ``Opinion polarization by learning from social feedback,'' {\em The Journal of Mathematical Sociology}, vol.~43, no.~2, pp.~76--103, 2019.

\bibitem{Deffuant2000}
G.~Deffuant, D.~Neau, F.~Amblard, and G.~Weisbuch, ``Mixing beliefs among interacting agents,'' {\em Advances in Complex Systems}, vol.~3, pp.~87--98, 2000.

\bibitem{starnini2025}
M.~Starnini, F.~Baumann, T.~Galla, D.~Garcia, G.~Iñiguez, M.~Karsai, J.~Lorenz, and K.~Sznajd-Weron, ``Opinion dynamics: Statistical physics and beyond,'' 2025.
\newblock arXiv:2507.11521.

\bibitem{Glau1963}
R.~J. Glauber, ``Time-dependent statistics of the ising model,'' {\em Journal of Mathematical physics}, vol.~4, pp.~294--307, Feb. 1963.

\bibitem{Metroetal53}
N.~Metropolis, A.~W. Rosenbluth, M.~N. Rosenbluth, A.~H. Teller, and E.~Teller, ``Equations of state calculations by fast computing machines,'' {\em Journal of Chemical Physics}, vol.~21, p.~1087 – 1092, 1953.

\bibitem{Hast70}
W.~K. Hastings, ``Monte {C}arlo sampling methods using {M}arkov chains and their applications,'' {\em Biometrika}, vol.~57, p.~97 – 109, 1970.

\bibitem{MacCarron2020}
P.~MacCarron, P.~J. Maher, S.~Fennell, K.~Burke, J.~P. Gleeson, K.~Durrheim, and M.~Quayle, ``Agreement threshold on {A}xelrod’s model of cultural dissemination,'' {\em PLOS ONE}, vol.~15, pp.~1--13, 06 2020.

\bibitem{Huet2010}
S.~Huet and G.~Deffuant, ``Openness leads to opinion stability and narrowness to volatility,'' {\em Advances in Complex Systems}, vol.~13, no.~03, pp.~405--423, 2010.

\bibitem{Baldassarri2007}
D.~Baldassarri and P.~Bearman, ``Dynamics of political polarization,'' {\em American Sociological Review}, vol.~72, no.~5, pp.~784--811, 2007.

\bibitem{Kalinowska2023}
Z.~Kalinowska and B.~Dybiec, ``Weighted axelrod model: Different but similar,'' {\em Physica A: Statistical Mechanics and its Applications}, vol.~630, p.~129281, 2023.

\bibitem{Pham2022}
T.~M. Pham, J.~Korbel, R.~Hanel, and S.~Thurner, ``Empirical social triad statistics can be explained with dyadic homophylic interactions,'' {\em Proceedings of the National Academy of Sciences}, vol.~119, no.~6, p.~e2121103119, 2022.

\bibitem{Watts1998}
D.~J. Watts and S.~H. Strogatz, ``Collective dynamics of `small-world' networks,'' {\em Nature}, vol.~393, no.~6684, pp.~440--442, 1998.

\bibitem{Meylahn_Searle_2024}
B.~V. Meylahn and C.~Searle, ``Opinion dynamics beyond social influence,'' {\em Network Science}, vol.~12, no.~4, p.~339–365, 2024.

\bibitem{Dinkelberg2021}
A.~Dinkelberg, P.~MacCarron, P.~J. Maher, and M.~Quayle, ``Homophily dynamics outweigh network topology in an extended {A}xelrod’s cultural dissemination model,'' {\em Physica A: Statistical Mechanics and its Applications}, vol.~578, p.~126086, 2021.

\bibitem{maasetal2020}
H.~L.~J. van~der Maas, J.~Dalege, and L.~Waldorp, ``The polarization within and across individuals: the hierarchical ising opinion model,'' {\em Journal of Complex Networks}, vol.~8, May 2020.

\bibitem{Mullin1997}
W.~J. Mullin, ``Bose-einstein condensation in a harmonic potential,'' {\em J. of Low Temperature Physics}, vol.~106, p.~615–641, Mar. 1997.

\end{thebibliography}
\appendix
\section{Sensitivity spot checks}\label{sec:sensitivity}
In order to test the robustness of the findings we run simulations at other values of $\alpha$ for a limited set of values of bounded confidence clip in both the bounded confidence extension and the topical weights extension. 

In particular we perform spot checks for the clip values $b=12,15$. In Tables~\ref{tab:SA_con} and~\ref{tab:SA_bp} observe that the effect on the two opinion spaces is consistent:
\begin{itemize}
    \item Adding bounded confidence allows for outcomes other than consensus (seen in Table~\ref{tab:SA_con} where the probability of consensus in all models is $<1$).
    \item Adding topical weights to the bounded confidence models increases the likelihood of consensus in both the toroidal and the cubic opinion space models (see Table~\ref{tab:SA_bp}). Furthermore this effect is stronger in the toroidal model than in the cubic model. 
    \item Adding topical weights to the toroidal opinion space model increases the probability of bi-polarization.
    \item Adding topical weights to the cubic opinion space model shifts the peak of the probability of bi-polarization to the right. This is noticeable by a small decrease in the probability of bi-polarization at $b=12$ and an increase at $b=15$ across all values of $\alpha$.
\end{itemize}

\begin{table}[h!]
    \centering
    \begin{tabular}{c|c|c|c|c|c|c|c|c|c}
        & \multicolumn{4}{c|}{Cube} &\multicolumn{4}{c}{Torus} \\
        $\alpha$ & \multicolumn{2}{c|}{$b=12$} & \multicolumn{2}{c|}{$b=15$} & \multicolumn{2}{c|}{$b=12$} & \multicolumn{2}{c}{$b=15$}\\
        & BC & TW & BC & TW & BC & TW & BC & TW \\
        \hline
       0.5 & 0.904 & 0.939 & 0.005 & 0.0901 & 0.04 & 0.3418 & 0.0 & 0.0 \\
       1.0 & 0.92 &  0.9489 & 0.008 &  0.1298 & 0.038  & 0.4039 & 0.00 & 0.00\\
       2.0 & 0.928 & 0.948 & 0.006 & 0.1022 & 0.067 & 0.4534 & 0.0 & 0.00
    \end{tabular}
    \caption{Probability of consensus under the bounded confidence (BC) and weighted topics (TW) for $\alpha\in\{0.5,1,2\}$.}
    \label{tab:SA_con}
\end{table}
\begin{table}[h!]
    \centering
    \begin{tabular}{c|c|c|c|c|c|c|c|c|c}
        & \multicolumn{4}{c|}{Cube} &\multicolumn{4}{c}{Torus} \\
        $\alpha$ & \multicolumn{2}{c|}{$b=12$} & \multicolumn{2}{c|}{$b=15$} & \multicolumn{2}{c|}{$b=12$} & \multicolumn{2}{c}{$b=15$}\\
        & BC & TW & BC & TW & BC & TW & BC & TW \\
        \hline
       0.5  & 0.09 & 0.0590 & 0.031 & 0.1972 & 0.098 & 0.3531 & 0.00 & 0.00\\
       1.0 & 0.077 & 0.0480 & 0.03 & 0.2354 & 0.144 & 0.3215 & 0.00 & 0.00\\
       2.0 & 0.065 & 0.051 & 0.0270 & 0.2345 & 0.139 & 0.3300 & 0.00 & 0.0028
    \end{tabular}
    \caption{Probability of bi-polarization  under the bounded confidence (BC) and weighted topics (TW) for  for $\alpha\in\{0.5,1,2\}$.}
    \label{tab:SA_bp}
\end{table}

\begin{figure}[h!]
    \centering
    \begin{subfigure}{0.33\textwidth}%
        \includegraphics[width=\textwidth]{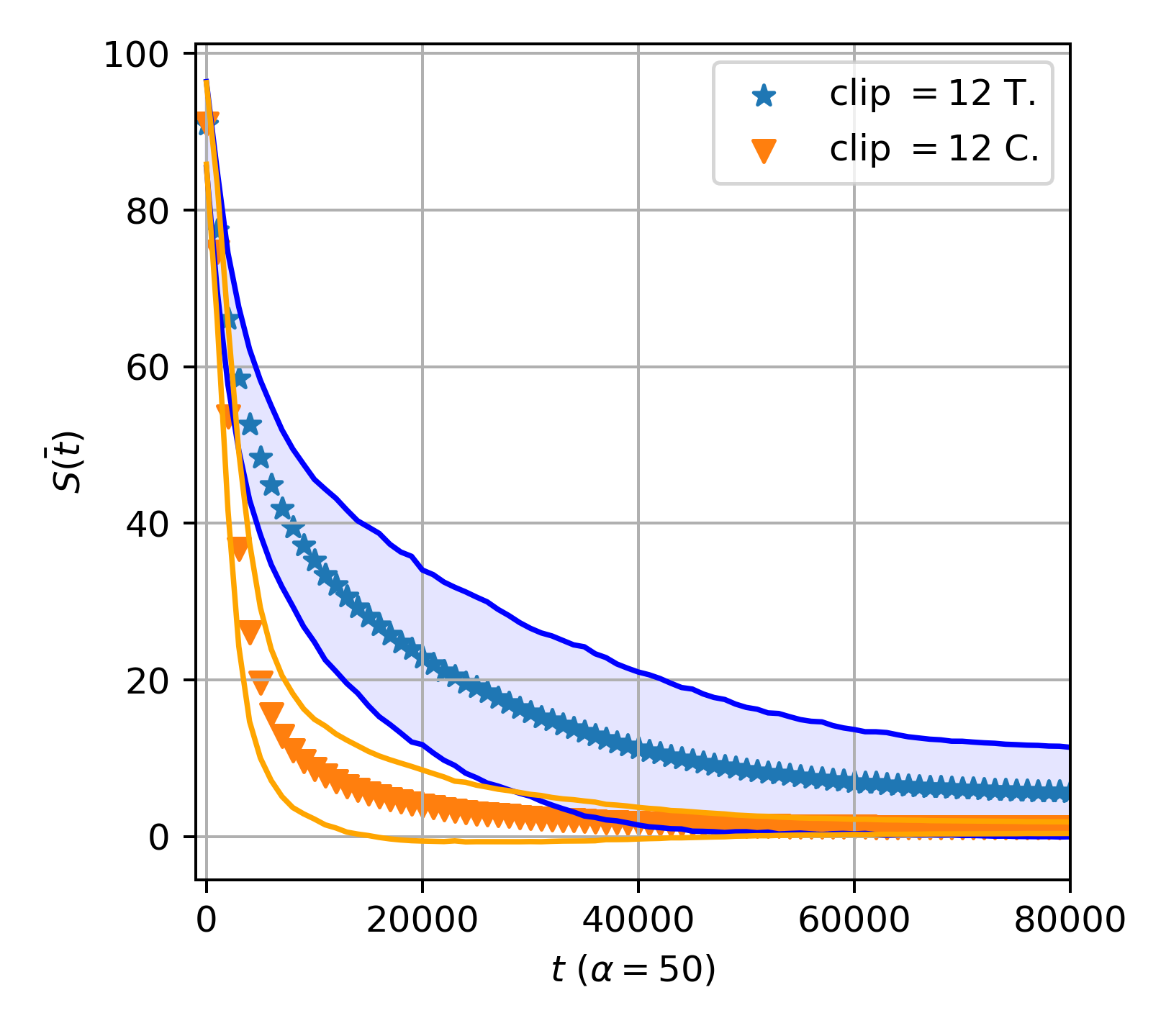}
        \caption{$\alpha=0.5$ (BC)}
    \end{subfigure}%
    \begin{subfigure}{0.33\textwidth}%
        \includegraphics[width=\textwidth]{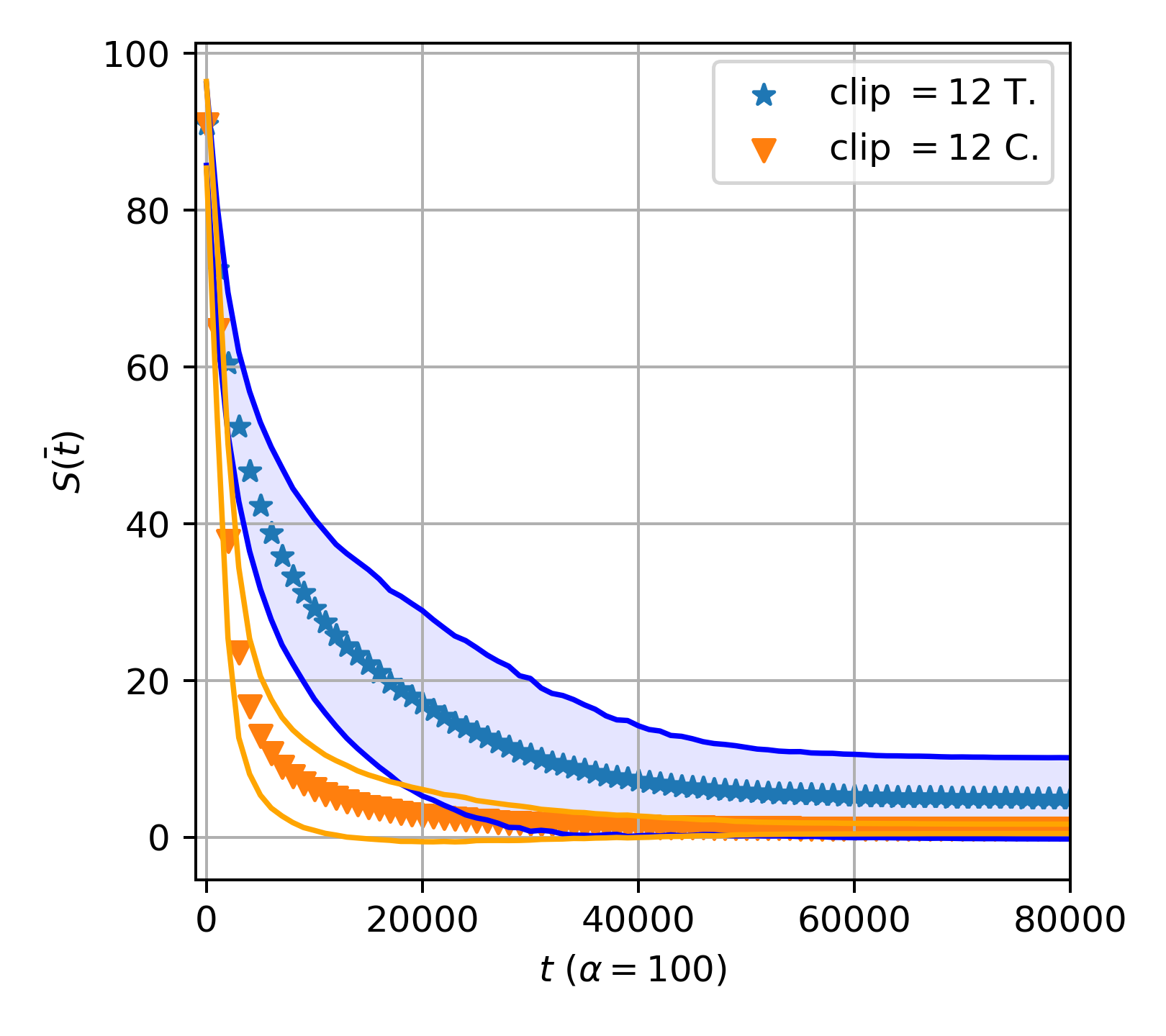}
        \caption{$\alpha=1.0$ (BC)}
    \end{subfigure}%
    \begin{subfigure}{0.33\textwidth}%
        \includegraphics[width=\textwidth]{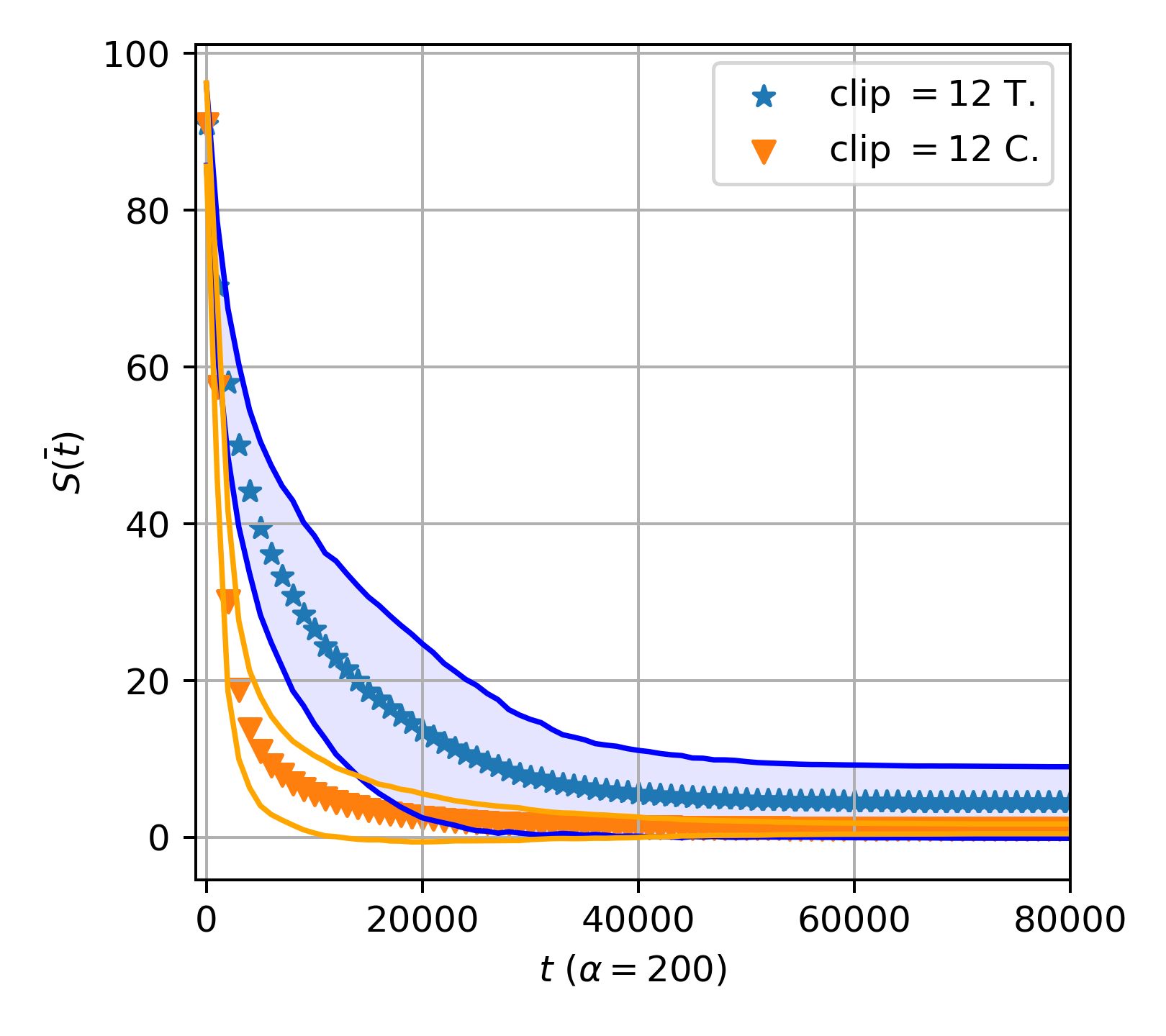}
        \caption{$\alpha=2.0$ (BC)}
    \end{subfigure}\\
    \begin{subfigure}{0.33\textwidth}%
        \includegraphics[width=\textwidth]{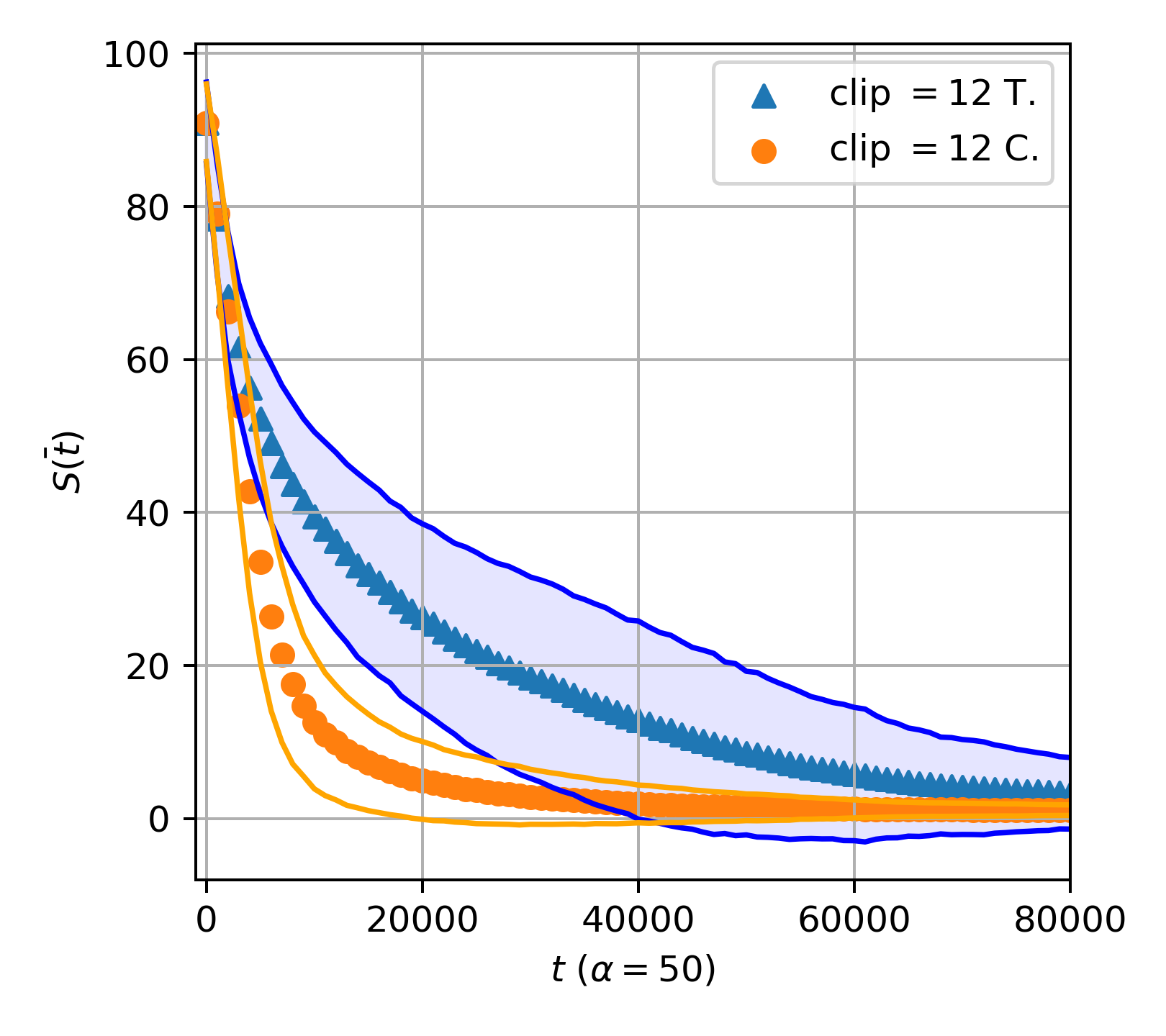}
        \caption{$\alpha=0.5$ (TW)}
    \end{subfigure}%
    \begin{subfigure}{0.33\textwidth}%
        \includegraphics[width=\textwidth]{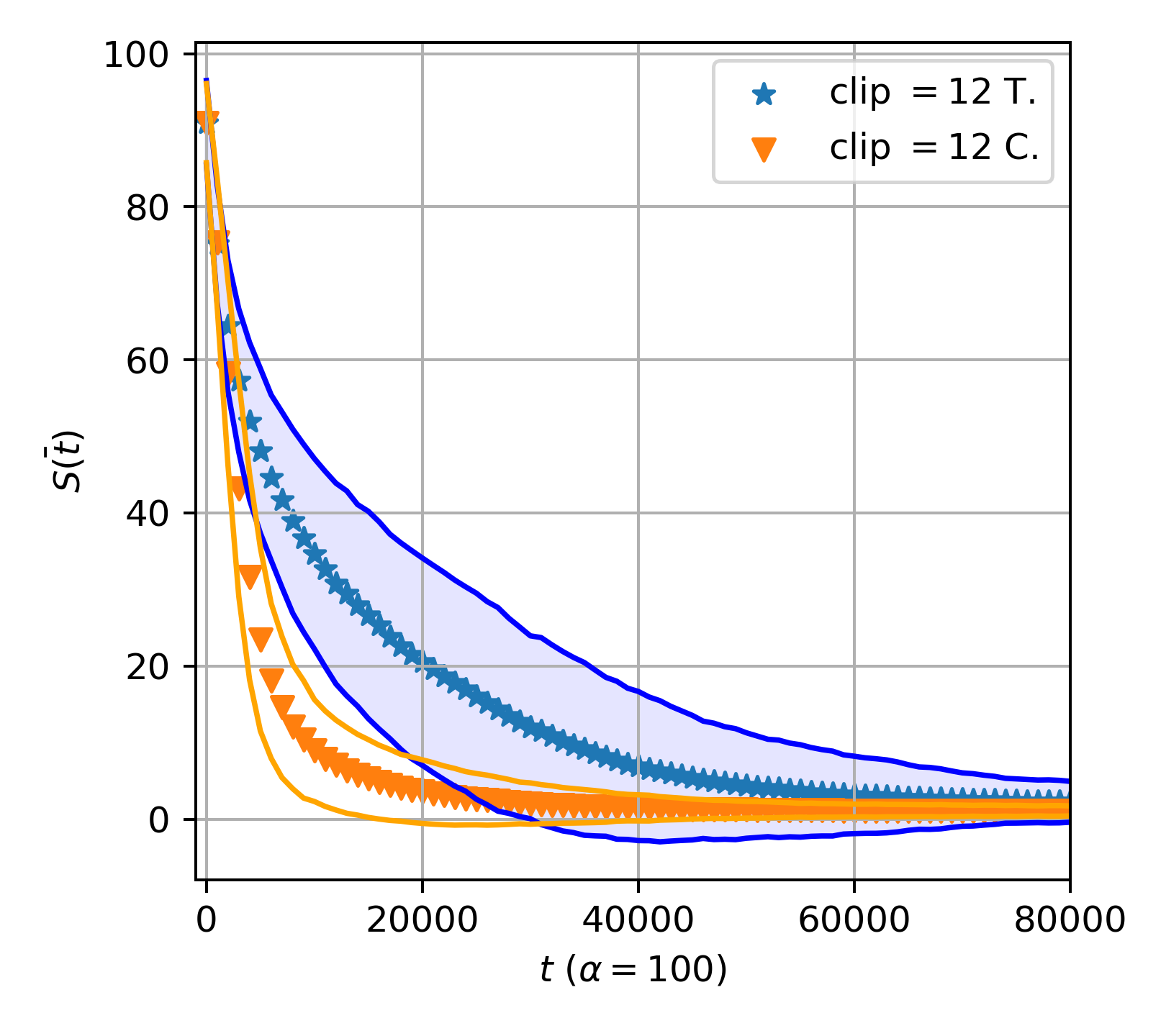}
        \caption{$\alpha=1.0$ (TW)}
    \end{subfigure}%
    \begin{subfigure}{0.33\textwidth}%
        \includegraphics[width=\textwidth]{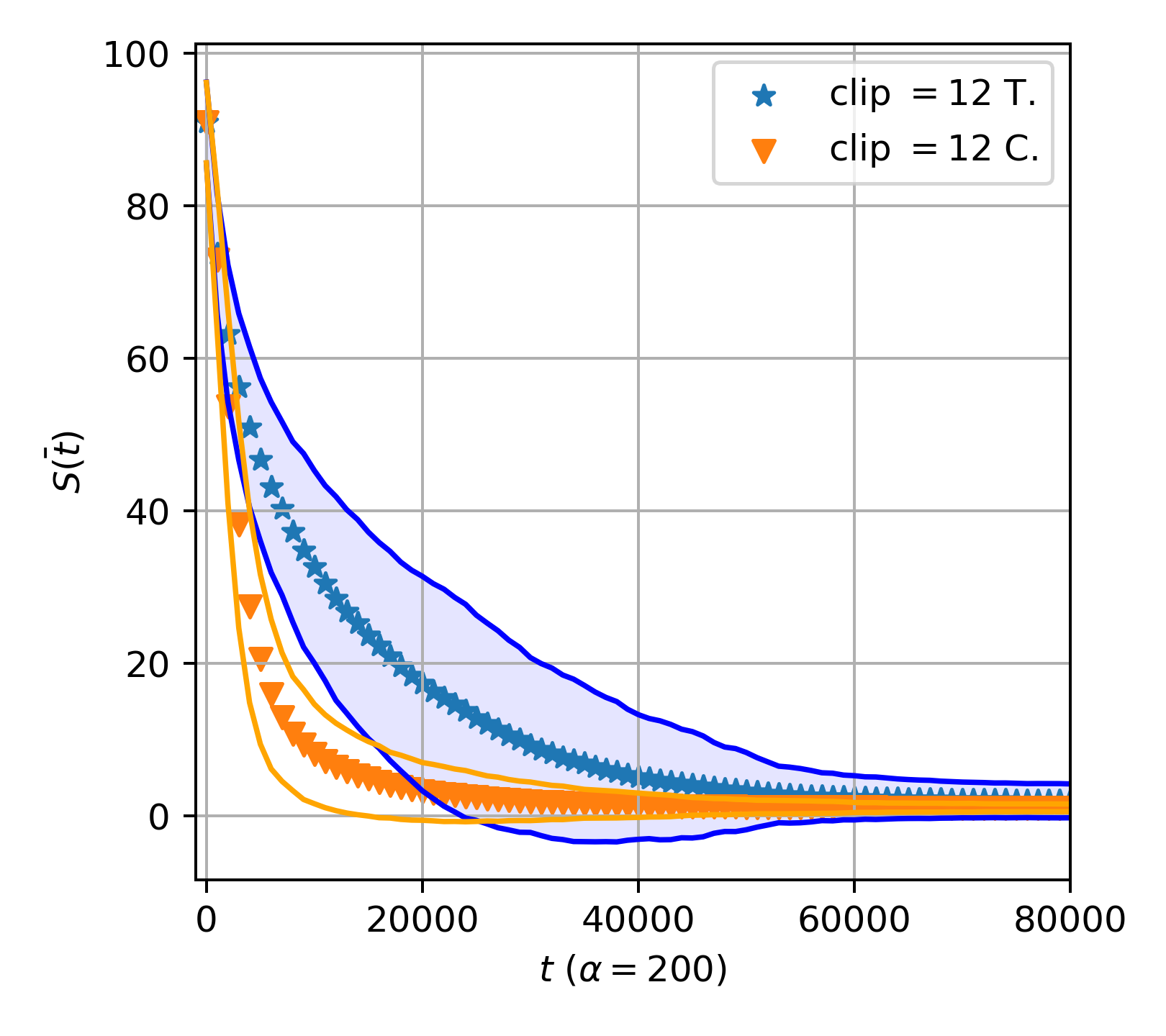}
        \caption{$\alpha=2.0$ (TW)}
    \end{subfigure}%
    \caption{The mean number of groups over time for the simulations with $b=12$ at different values of $\alpha=0.5, 1.0,2.0$ under bounded confidence (BC) and topical weights (TW).}
    \label{fig:SA_gr_12}
\end{figure}

\begin{figure}[h!]
    \centering
    \begin{subfigure}{0.33\textwidth}%
        \includegraphics[width=\textwidth]{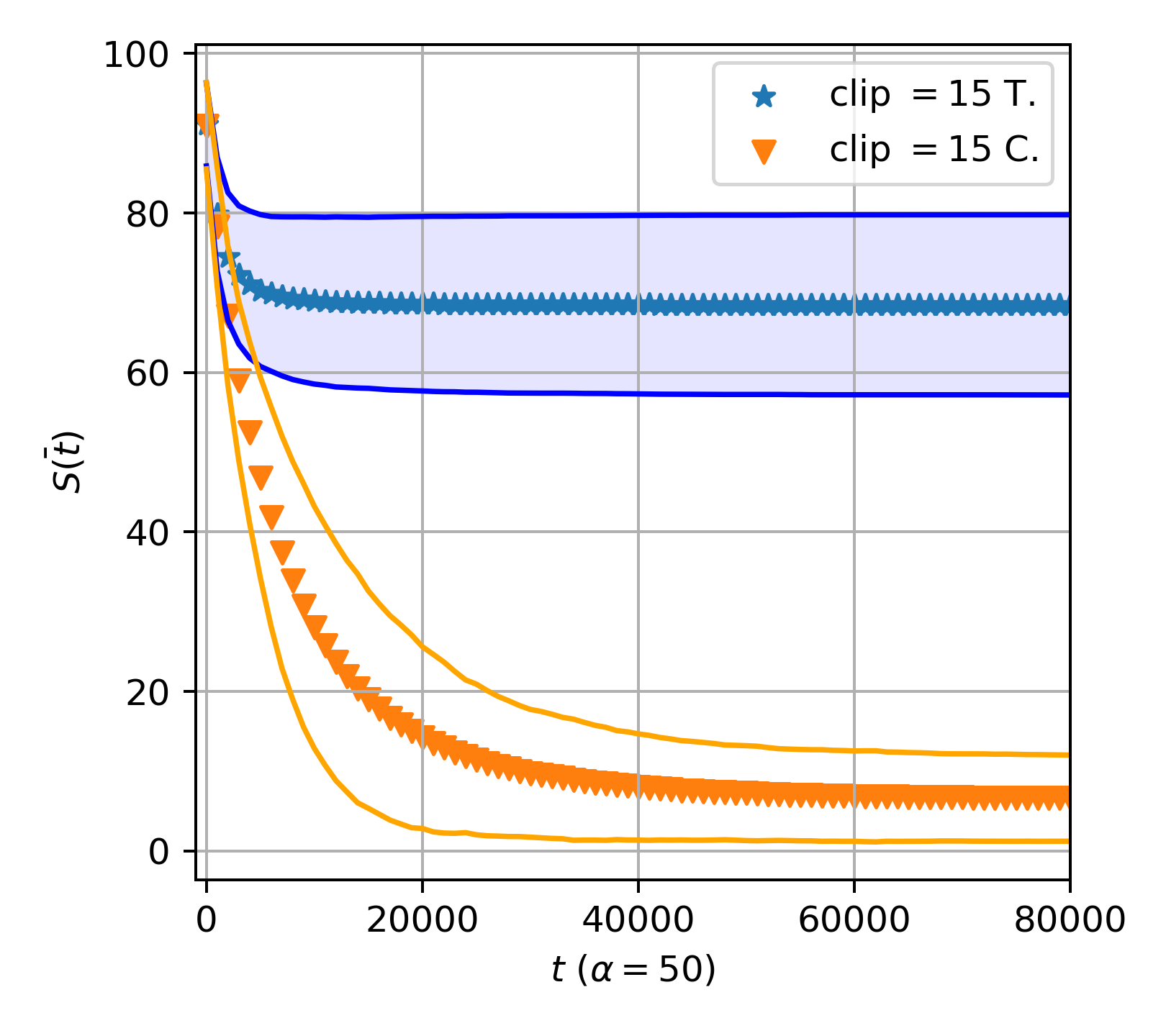}
        \caption{$\alpha=0.5$ (BC)}
    \end{subfigure}%
    \begin{subfigure}{0.33\textwidth}%
        \includegraphics[width=\textwidth]{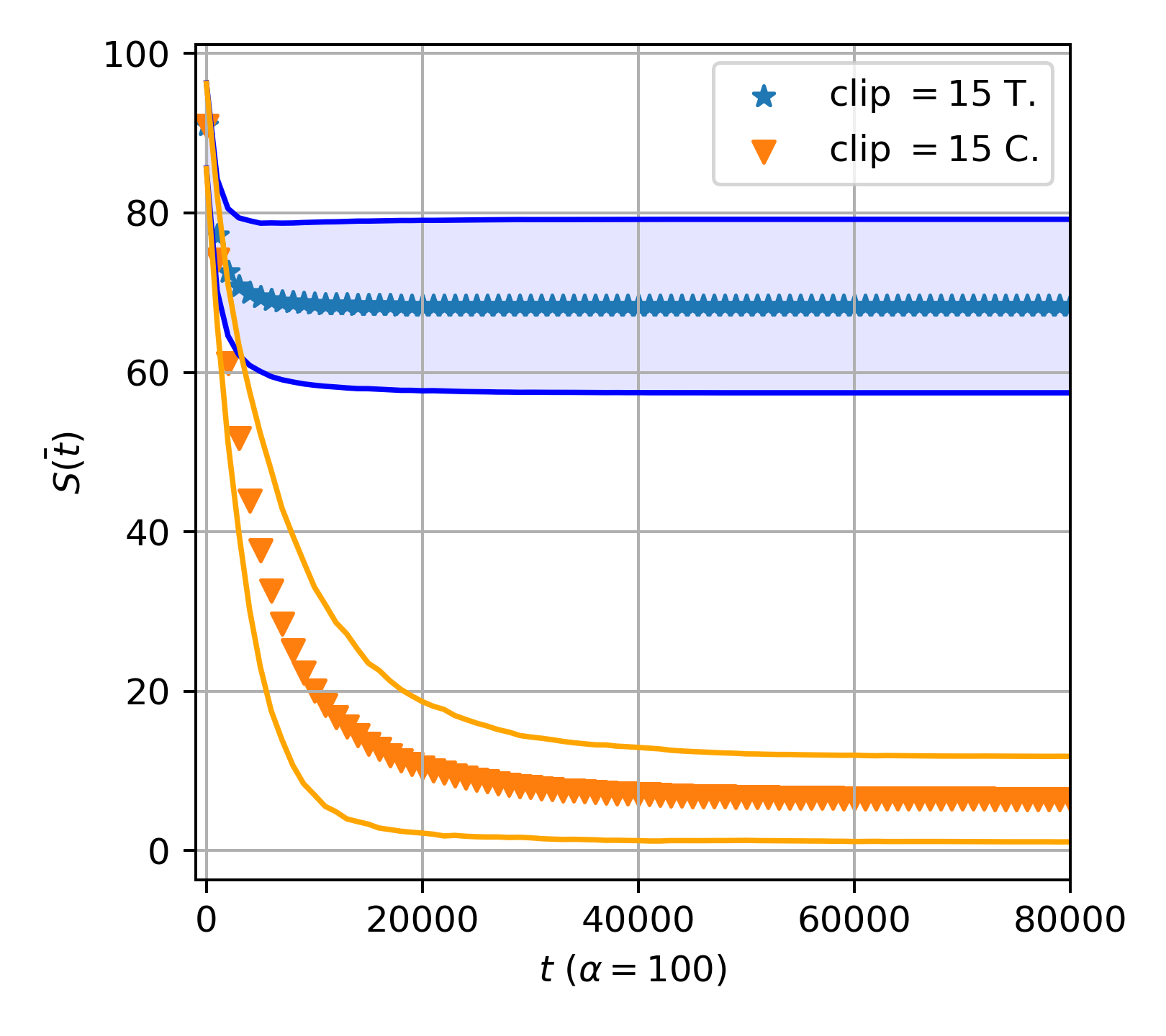}
        \caption{$\alpha=1.0$ (BC)}
    \end{subfigure}%
    \begin{subfigure}{0.33\textwidth}%
        \includegraphics[width=\textwidth]{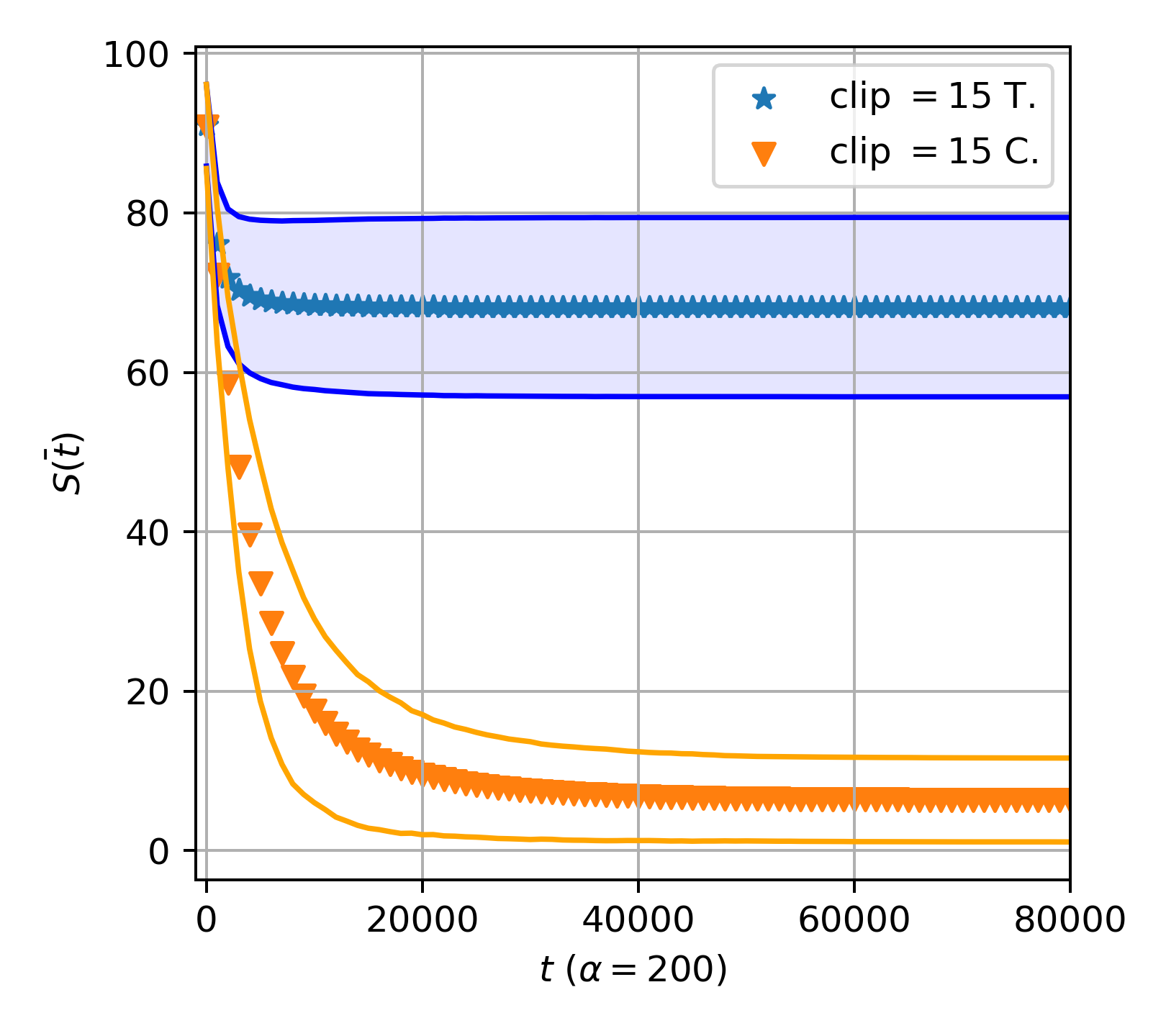}
        \caption{$\alpha=2.0$ (BC)}
    \end{subfigure}\\
    \begin{subfigure}{0.33\textwidth}%
        \includegraphics[width=\textwidth]{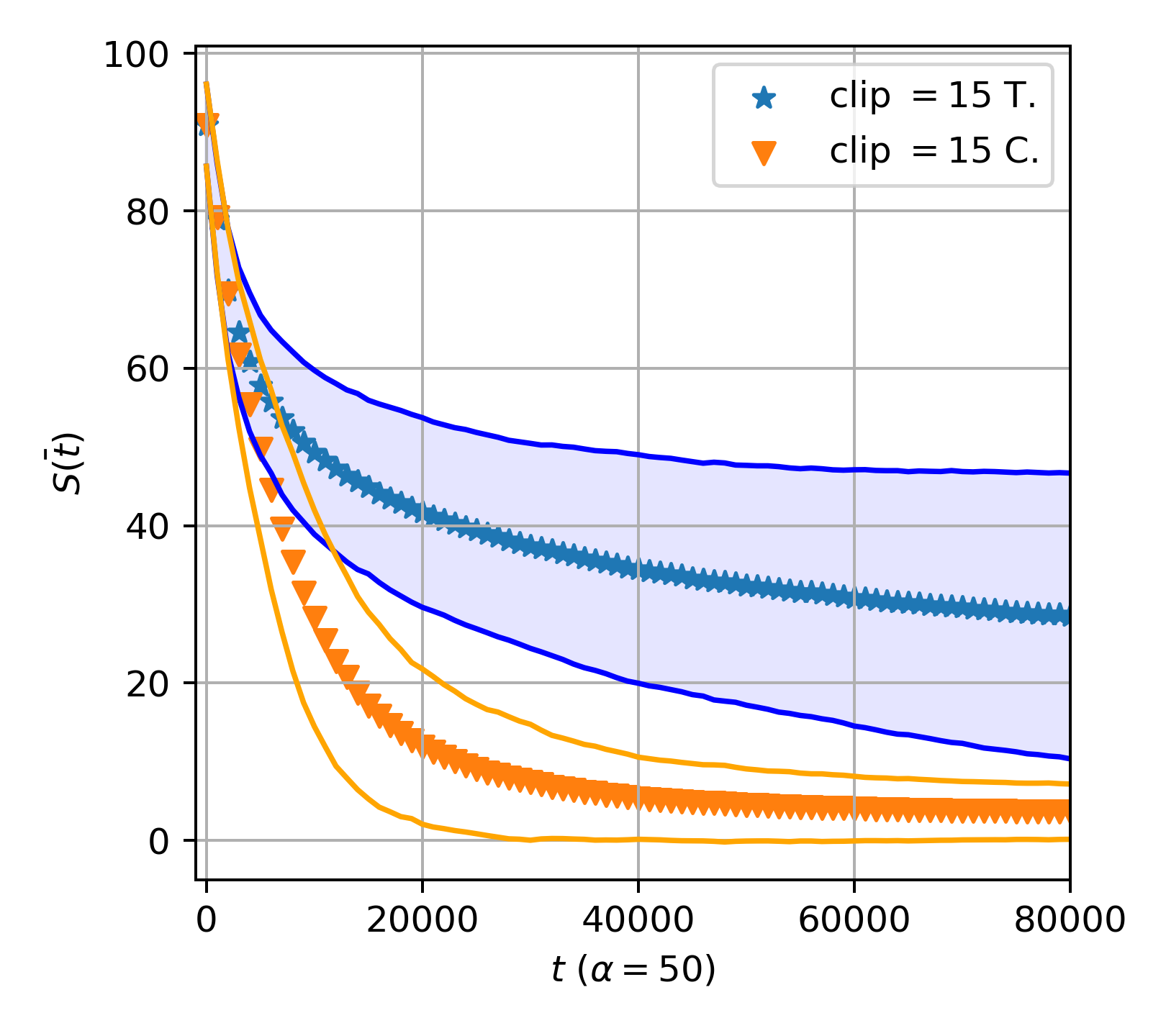}
        \caption{$\alpha=0.5$ (TW)}
    \end{subfigure}%
    \begin{subfigure}{0.33\textwidth}%
        \includegraphics[width=\textwidth]{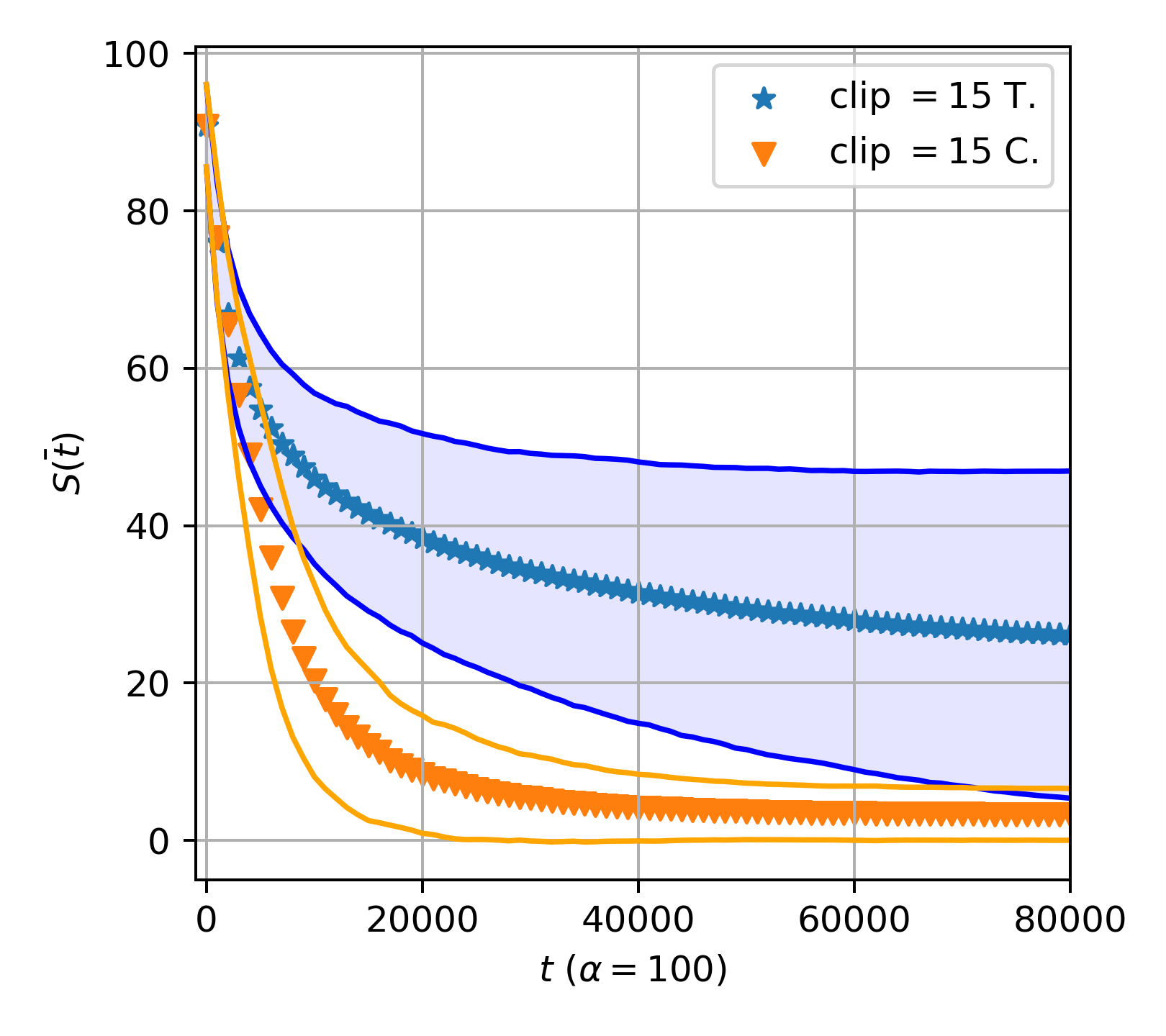}
        \caption{$\alpha=1.0$ (TW)}
    \end{subfigure}%
    \begin{subfigure}{0.33\textwidth}%
        \includegraphics[width=\textwidth]{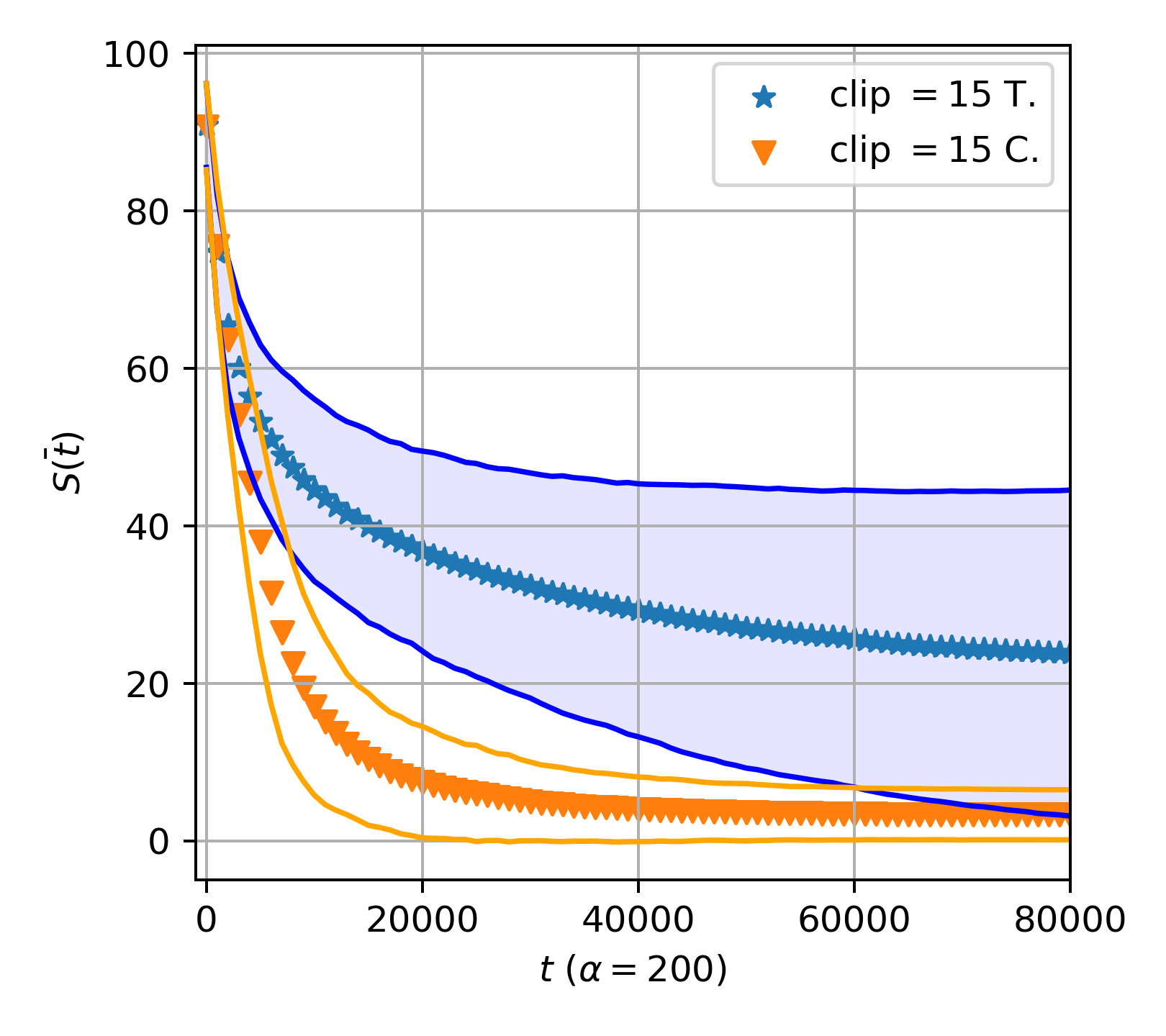}
        \caption{$\alpha=2.0$ (TW)}
    \end{subfigure}%
    \caption{The mean number of groups over time for the simulations with $b=15$ at different values of $\alpha=0.5, 1.0,2.0$ under bounded confidence (BC) and topical weights (TW).}
    \label{fig:SA_gr_15}
\end{figure}

In Figures~\ref{fig:SA_gr_12} and~\ref{fig:SA_gr_15} we plot the number of groups over time for both extensions with varying $\alpha$ at clips $b=12$ and 15 respectively. As mentioned in the discussion of the basic model, the effect of adjusting $\alpha$ is a speeding up or slowing down of dynamics. The effect of the additional extension is uniform across different values of $\alpha$ (lower number of groups later in time). Although this is by no means an exhaustive check on the parameter space, we see that in the region of $b$ which has the most rich dynamics the results are similar for different values of $\alpha$. We conclude that our results are robust to changes in the parameter $\alpha$.

\break
\section{Hamiltonian formulation}
The model and its dynamics are currently formulated in probabilistic terms. This approach aligns with a substantial body of existing literature. However, an alternative framework for studying opinion dynamics draws on analogies with Ising models, which originate from the statistical physics of solid and soft matter as well as spin glasses. In the social science context, such models are commonly referred to as {\it Hierarchical Ising Opinion Models} (HIOM) (cf.\ \cite{maasetal2020}).
In the original Ising model, the Hamiltonian represents the energy associated with a particle flipping its spin from `up' to `down' or vice versa, where these directions are defined relative to an external field or a mean field collectively generated by the ensemble of particles. This mechanism can result in macroscopic polarization of the system, in which all spins align in the same direction.

In such models the approach to establishing interaction probabilities is to first write down the Hamiltonian 
of the interactions, which is equivalent to the total energy for all of the particles. An interaction will change the state of particles and hence the value of that energy after summing over all particles in the entire system.
From statistical physics the concept of the Boltzmann distribution is adopted, which describes the overall 
likelihood of a system to be in a state with a given energy as $\exp(-H/kT)/Z$, where $Z$ is the partition function and $T$ the temperature of the system. 
If an interaction leads to a decrease in the total energy, the resulting state becomes more probable — that is, the likelihood of such a transition is high. Conversely, if the interaction increases the system’s energy, the probability of that transition is correspondingly low.
There exist several ways to implement these transition probabilities in algorithms. In Glauber dynamics 
(\cite{Glau1963}) the probability $p\left(\Delta{\boldsymbol s}\right)$ for a change $\Delta{\boldsymbol s}$ is:
\begin{equation}
\label{eq:Boltzmanfac}
p\left(\Delta{\boldsymbol s}\right) = \frac{1}{1+e^{\Delta H\left({\boldsymbol s}\right)/T}}
\end{equation}
alternatively one might also use the Metropolis-Hastings recipe:
\begin{equation}
p(\Delta {\boldsymbol s}) =\begin{cases} 1\ {\rm if}\ \Delta H\le 0 \\ \exp(-\Delta H/T)\ {\rm if}\ \Delta H > 0 \end{cases}
\end{equation}
Note that in \cite{maasetal2020} instead of $1/T$ a factor $-A$ is used in the Boltzmann factor in the denominator of (\ref{eq:Boltzmanfac}).
The constant $A$ in this expression, in the way that it is described in \cite{maasetal2020} is the `attention' of the
agents. This can be specified individually or as a single constant for all agents in the system. If the attention of 
an agent were to be $0$, then any change in $H$ will have no influence on a transition probability and it remains at
$1/2$. If the attention $A$ is very high then even a small decrease in $H$  will produce a very small value of the 
Boltzmann factor and therefore a high probability of the state to change.
The Hamiltonian in the Ising model case is particularly simple:
\begin{equation}
H({\boldsymbol s}) = -\sum_i \tau_i s_i  - \sum_i\sum_j \omega s_i s_j
\end{equation}
in which the $s_i$ can only have the values $\pm 1$, and the sign and value express the relative strengths of
the coupling of the spins to the external field and to neighbouring spins and whether the material is ferromagnetic or anti-ferromagnetic. 

It is of interest to point out that the likelihood $P_n$ of the
system having made some transition after $n$ independent steps (pairwise interactions) satisfies:
\begin{eqnarray}
\label{eq:ntransits}
P_n &=& 1- \prod\limits_{i=1}^n (1-p_i) \nonumber\\
&=& 1 - \prod\limits_{i=1}^n (1-\frac{1}{1+e^{\Delta H_i/T}}) \nonumber\\
&=& 1 - \prod\limits_{i=1}^n \frac{e^{\Delta H_i/T}}{1+e^{\Delta H_i/T}} \nonumber\\
&\approx& 1 -  \exp\left({\sum\limits_{i=1}^n\Delta H_i/T}\right) \approx \frac{1}{1+\exp\left(\sum\limits_{i=1}^n\Delta H_i/T\right)}
\end{eqnarray}
where the approximate equality uses the notion that the only transitions worth accounting for are the most likely ones: those for which $\Delta H \ll 0$. Expression (\ref{eq:ntransits}) demonstrates that, at least approximately, the Hamiltonian for $n$ independent interactions is the sum of the Hamiltonians for each pairwise interaction.

The model presented in the main text, the probability is
given by Eq. (\ref{eq:fa}), or by Eq. (\ref{eq:fa_bc}) in the bounded confidence variant. 
The distance in opinion space is given by either (\ref{eq:disC}) or (\ref{eq:disT}) depending on whether the opinion space is a cube or a torus respectively.
In the current setting of agents on a grid, where at every timestep pairs of neighbours interact, the change in state $\Delta {\boldsymbol s}$ translates to a change
in distance in opinion space between neighbours i.e. $\Delta {\boldsymbol s} \rightarrow \Delta d$.
With the probabilities given, it is possible to follow the opposite reasoning and establish what the corresponding 
Hamiltonian looks like if for instance the Glauber recipe applies:
\begin{equation}
\Delta H = -T\ln\left(\frac{p(\Delta {\boldsymbol s})}{1-p(\Delta {\boldsymbol  s})}\right) 
\end{equation}
To illustrate the effect of these choices it is of interest to look at the behaviour of $\Delta H$ with $d$. If $d$ is small enough, then the agents remain within the region of bounded confidence. For the purpose of elucidating the behaviour of the Hamiltonian as 
a function of $d$ it is convenient to set the limit of probability for $d\downarrow 0$ to a constant value just below $1$ so that 
$\Delta H$ does not diverge for $d\downarrow 0$ :
\begin{equation}
\label{eq:indivprob}
p_0 = 1-\epsilon
\end{equation}

\begin{figure}[h!]
    \centering
    \begin{subfigure}{0.49\textwidth}
    \centering
        \includegraphics[width=0.8\textwidth]{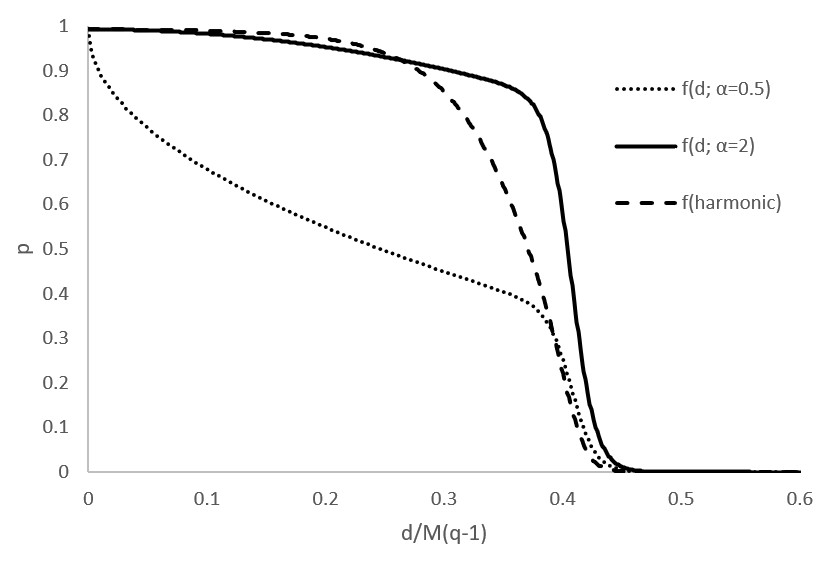}
        \caption{transition probabilities}\label{fig:Probties}
    \end{subfigure}%
    \begin{subfigure}{0.49\textwidth}
    \centering
        \includegraphics[width=0.8\textwidth]{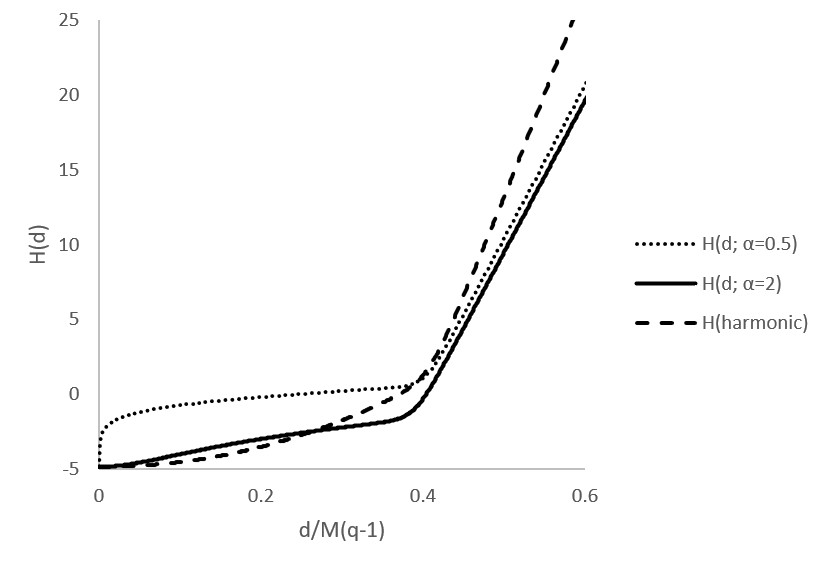}
        \caption{associated Hamiltonians}\label{fig:Hamtnians}
    \end{subfigure}%
    \caption{(a): the transition probabilities as a function of opinionspace distance $d$, scaled by $M(q-1)$, using either the standard shape option (dotted and solid) or the alternative eq. (\ref{eq:monotfb}) (dashed) for the function $f$. Right panel: the equivalent Hamiltonian, which, using Glauber dynamics, would induce those probabilities. The parameters chosen for this example are either $\alpha=1/2$ (dotted) or $\alpha=2$ (solid) for function $f$, and $b = 1/\sqrt{6}\approx 0.408$. For shape (\ref{eq:monotfb}) the parameters $\beta = 7.2$ and $\sigma_b = 0.2$ for the function $f_b$. In setting these particular choices it is deliberate that the values of the functions match for $d\downarrow 0$ and that the probabilities drop away near the same value for $d$.}\label{fig:PandH}
\end{figure}

It is clear that then
\begin{eqnarray}
\label{eq:approxH}
\Delta H &=& -T\ln \left(\frac{f(d)}{1-f(d)}\right) = -T\ln \left[\frac{1-\left(\frac{d}{M(q-1)}\right)^{\alpha}}{\frac{\epsilon}{1-\epsilon} +\left(\frac{d}{M(q-1)}\right)^{\alpha}} \right] 
\ \ {\rm for}\ d< M(q-1) - b
\end{eqnarray} 
Now it becomes clear that the choice for the shape of $f$, for $0<\alpha<1$ leads to a cusp-like behaviour in 
$\Delta H$ for $d\downarrow 0$ and a flat (i.e. $0$ derivative with respect to $d$) behaviour for $\alpha >1$.

For illustrative purposes it is worthwhile to consider an alternative parametrization of a monotonically 
decreasing function $f$:
\begin{equation}
\label{eq:monotfb}
f_b (d) = \frac{e^{(1-\beta d^2)/\sigma_b}}{1+e^{(1-\beta d^2)/\sigma_b}}
\end{equation}
with the parameters $\beta>M^2(q-1)^2$ and $0<\sigma_b<1$ (preferably small). The reason is that for this shape of the probability function, the Hamiltonian satisfies:
\begin{eqnarray}
\label{eq:approxHtwo}
\Delta H &=& -T\ln \left(\frac{f(d)}{1-f(d)}\right) 
=\frac{T}{\sigma_b}\left(\beta d^2-1\right)
\end{eqnarray}
which is immediately recognisable as the Hamiltonian for a harmonic oscillator. The implication is that the behaviour of transition probabilities with distance in opinion space
for a choice of $\alpha\approx 2$ is similar to
a system governed by a harmonic potential which is quite well-studied in statistical physics, like the Ising system.

The harmonic potential, a system of harmonic oscillators, is a useful approximation for ideal gas particles or particles in a solid state lattice. 
For low temperatures the existence of a phase transition of a gas of bosonic particles to a Bose-Einstein condensate is modelled in this way (cf. \cite{Mullin1997}), and for lattices some of the properties of phonons are well represented by such a model. One might expect some similar behaviours to also be exhibited in the current setting.

Note that the clipping (bounded confidence) is incorporated in Eq. \ref{eq:monotfb}: the clipping distance $d_{clip} = M(q-1) -b \approx 1/\sqrt{\beta}$ and the `sharpness' of the transition is governed by the $\sigma_b$ parameter. 
A smaller value of $\sigma_b$ corresponds to a sharper transition, so a hard clipping as in fig \ref{fig:clip} of the main paper would correspond to $\sigma_b\downarrow 0$.
Setting $\beta$ small enough would move any transition outside of the values that $d$ can take, or (alternatively) setting $\sigma_b$ quite large would make any clip so soft as to be almost unnoticeable.
From Eq. (\ref{eq:approxHtwo}) it can be seen that decreasing the temperature of the system, at a fixed value of $\sigma_b$ (and $\beta$) would have the same effect as increasing $\sigma_b$
at a fixed temperature, i.e. (gradually) removing any clipping. We have seen that without the bounded confidence (no clipping) the system always reaches consensus: everyone `condenses' into the same state. This appears to be the equivalent of a Bose-Einstein condensate for (the equivalent of) very low temperatures. For higher temperatures, stronger clipping, in particular a sharp drop at small distances so that the drop in interaction probability is substantial, there is the equivalent of finite/higher temperatures so that the system not always ends as condensation into a single state. 
The harmonic oscillator correspondence of the Hamiltonian requires also considering the boundary conditions of the opinion space. Recall that a harmonic oscillator differential equation, together with Neumann or Dirichlet type conditions on the boundary (of a cube), produces a discrete spectrum of allowed eigenstates/oscillations (standing waves). In a toroidal space with periodicity conditions the same occurs, although the energy levels and eigenstates differ between a toroidal and cubic space. Translated to the present situation one might expect to obtain a larger number of subgroups if the system is in a higher energy eigenstate. The random initial conditions may well put the system in a superposition of multiple eigenstates. It might even be that the likelihood for the number of groups that the system ends up in is skewed towards the lower eigenstates (lower number of subgroups) unless something quite specific is done to the initial conditions. Such exploration is beyond the scope of the present paper.

\section{Rewiring on circulant graphs}\label{app:circ}
In this appendix we briefly demonstrate the effect of network rewiring starting from a circulant graph on the dynamics in our model. 

When running these numerical simulations we keep previously adjusted parameters constant. In particular $\alpha = 2$, each agent has personal topical weights per opinion dimension, and the number of nearest neighbours is 4, while the total number of agents is 100. Circulants with 4 nearest neighbours are chosen because most of the agents in the grid have 4 nearest neighbours. However, in this circulant the shortest paths between agents are considerably longer than in the grid. There is also more clustering as the all nodes are a part of 3 triangles before rewiring, whereas in the grid, the nodes are not part of any triangles before rewiring.

\begin{figure}[htb]
    \centering
    \begin{subfigure}{0.33\textwidth}
        \includegraphics[width=0.85\textwidth]{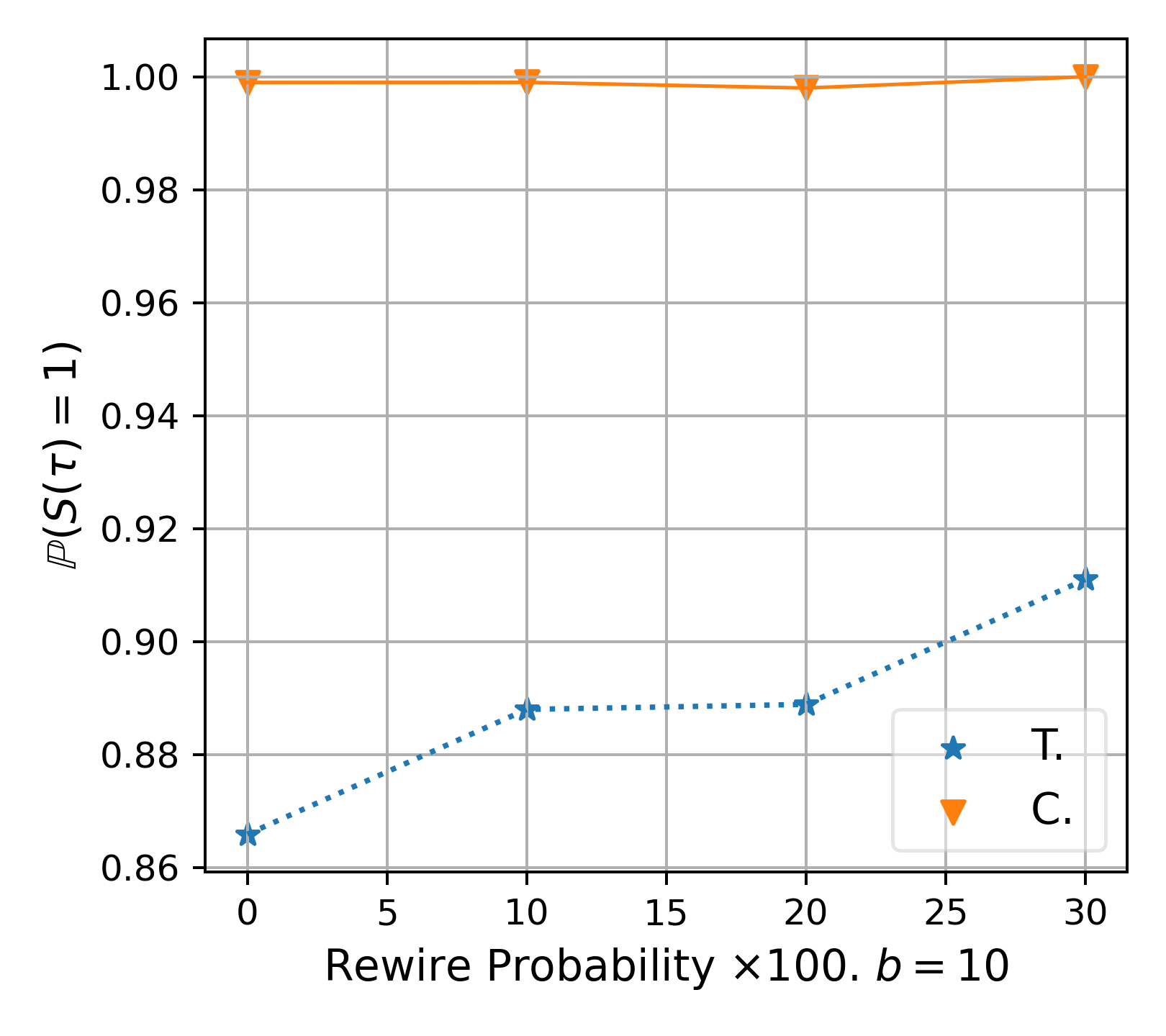}
    \caption{$b=10$}
    \label{fig:WC_con_rw_b10}
    \end{subfigure}%
    \begin{subfigure}{0.33\textwidth}
        \includegraphics[width=0.85\textwidth]{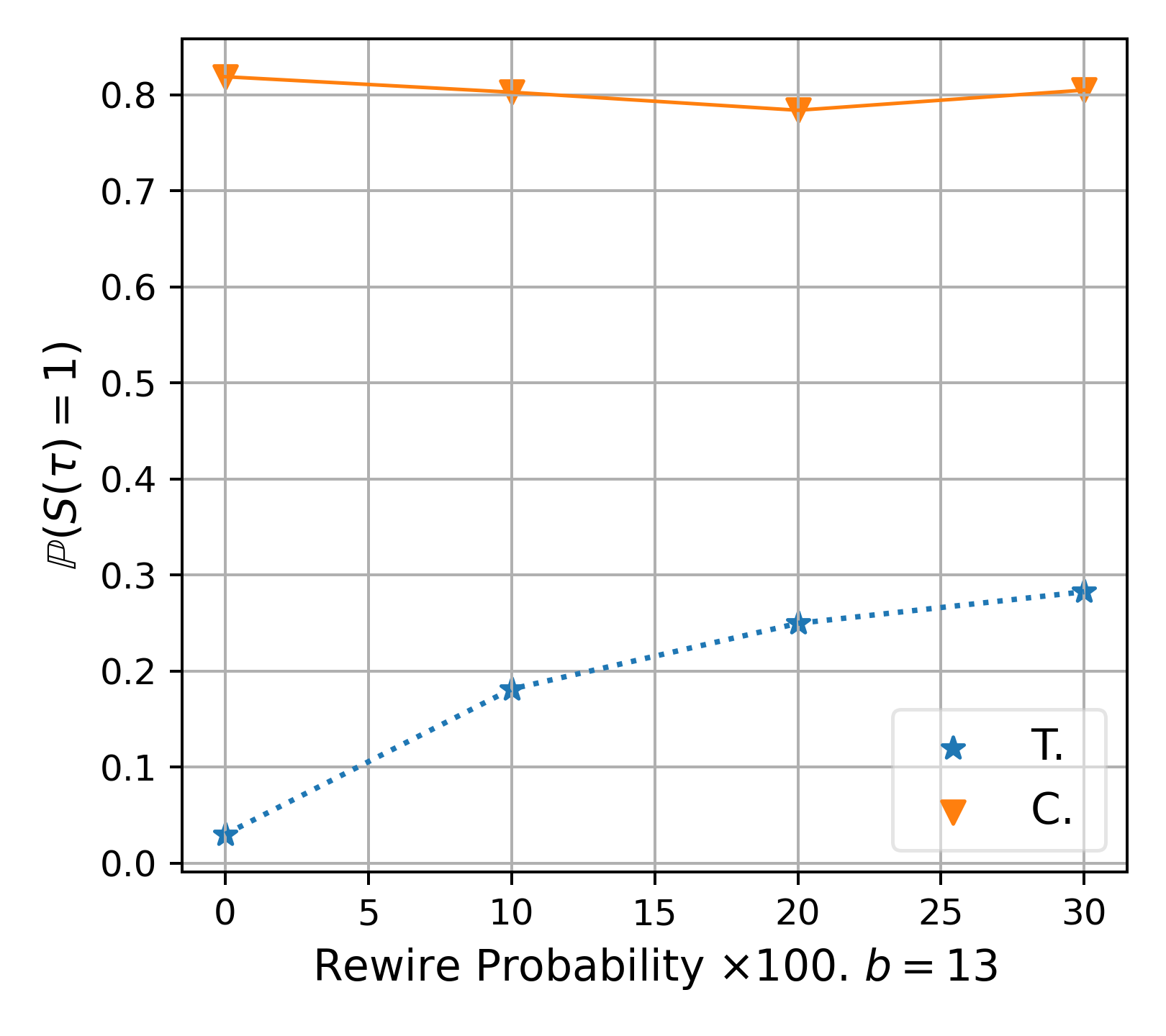}
    \caption{$b=13$}
    \label{fig:WC_con_rw_b13}
    \end{subfigure}%
        \begin{subfigure}{0.33\textwidth}
        \includegraphics[width=0.85\textwidth]{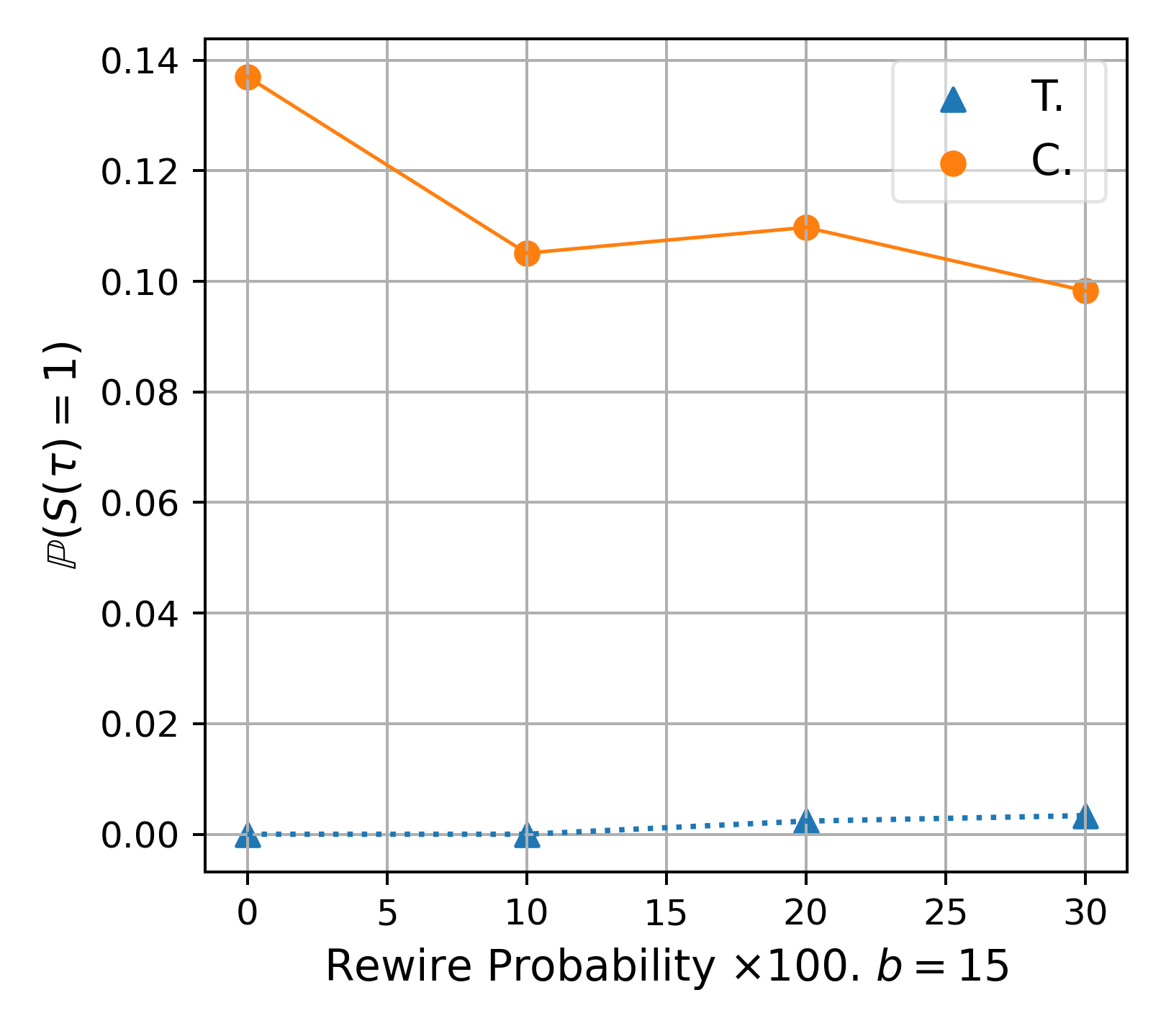}
    \caption{$b=15$}
    \label{fig:WC_con_rw_b15}
    \end{subfigure}%
    \caption{The estimated probability of consensus at differing levels of bounded confidence. }
\end{figure}

\begin{figure}[htb]
    \centering
        \begin{subfigure}{0.33\textwidth}
        \includegraphics[width=0.85\textwidth]{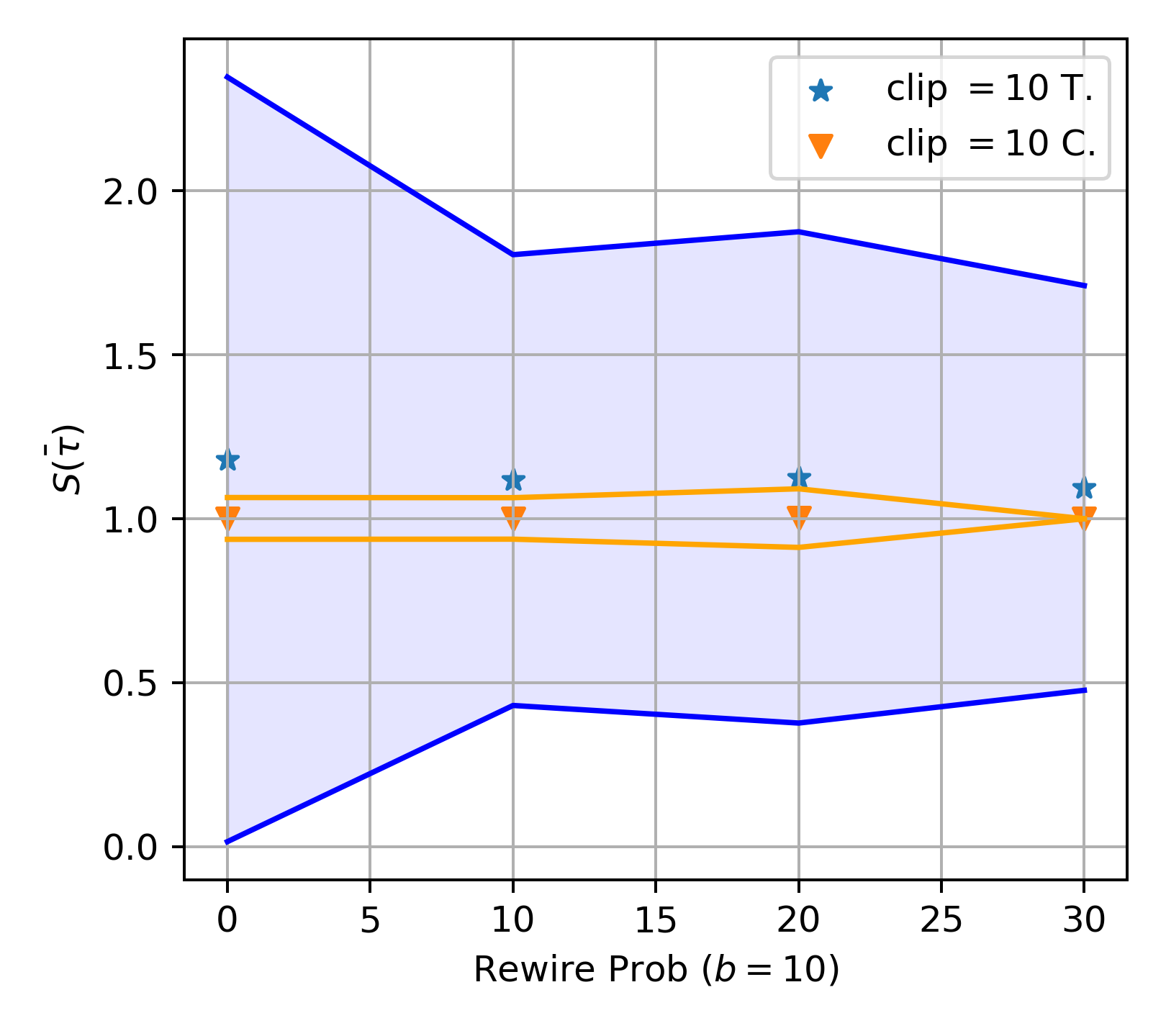}
    \caption{$b=10$}
    \label{fig:WC_grps_rw_b10}
    \end{subfigure}%
    \begin{subfigure}{0.33\textwidth}
        \includegraphics[width=0.85\textwidth]{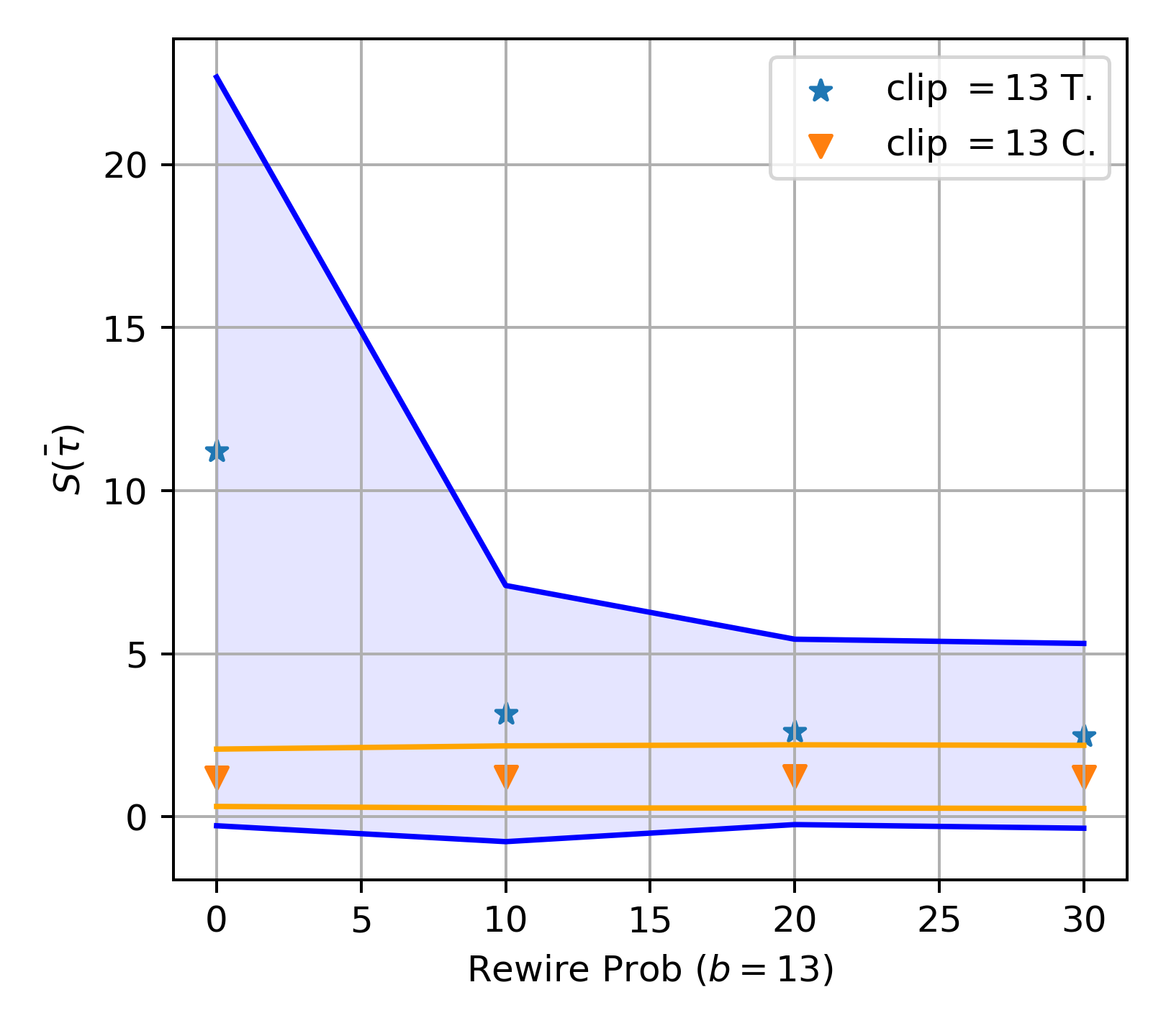}
    \caption{$b=13$}
    \label{fig:WC_grps_rw_b13}
    \end{subfigure}%
        \begin{subfigure}{0.33\textwidth}
        \includegraphics[width=0.85\textwidth]{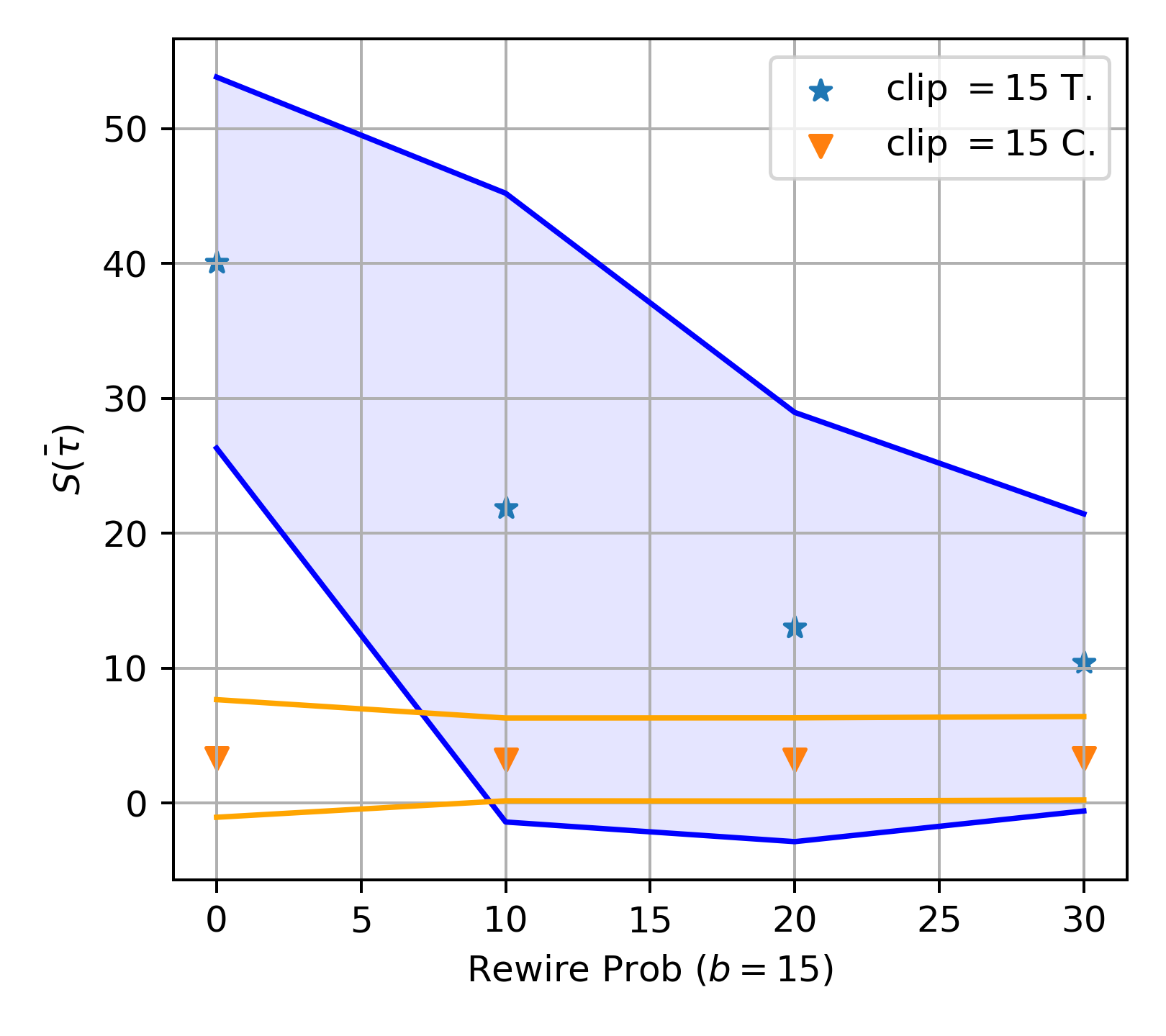}
    \caption{$b=15$}
    \label{fig:WC_grps_rw_b15}
    \end{subfigure}%
    \caption{The estimated mean number of groups at the termination of the simulation at different levels of bounded confidence.}
\end{figure}

In Figures~\ref{fig:WC_con_rw_b10}--\ref{fig:WC_grps_rw_b15} we plot the likelihood of consensus as well as the number of groups at the termination of the simulation (of those runs that terminated before the time limit and thus have converged to steady state). We see that for the toroidal opinion space the effect is much more akin to what we would have expected rewiring to cause. More rewiring leads to more consensus and smaller  number of groups in steady state. In the circulant graphs this works by bringing close-knit communities that are initially far away from one another much closer together.

\begin{figure}
    \centering
    \includegraphics[width=0.6\linewidth]{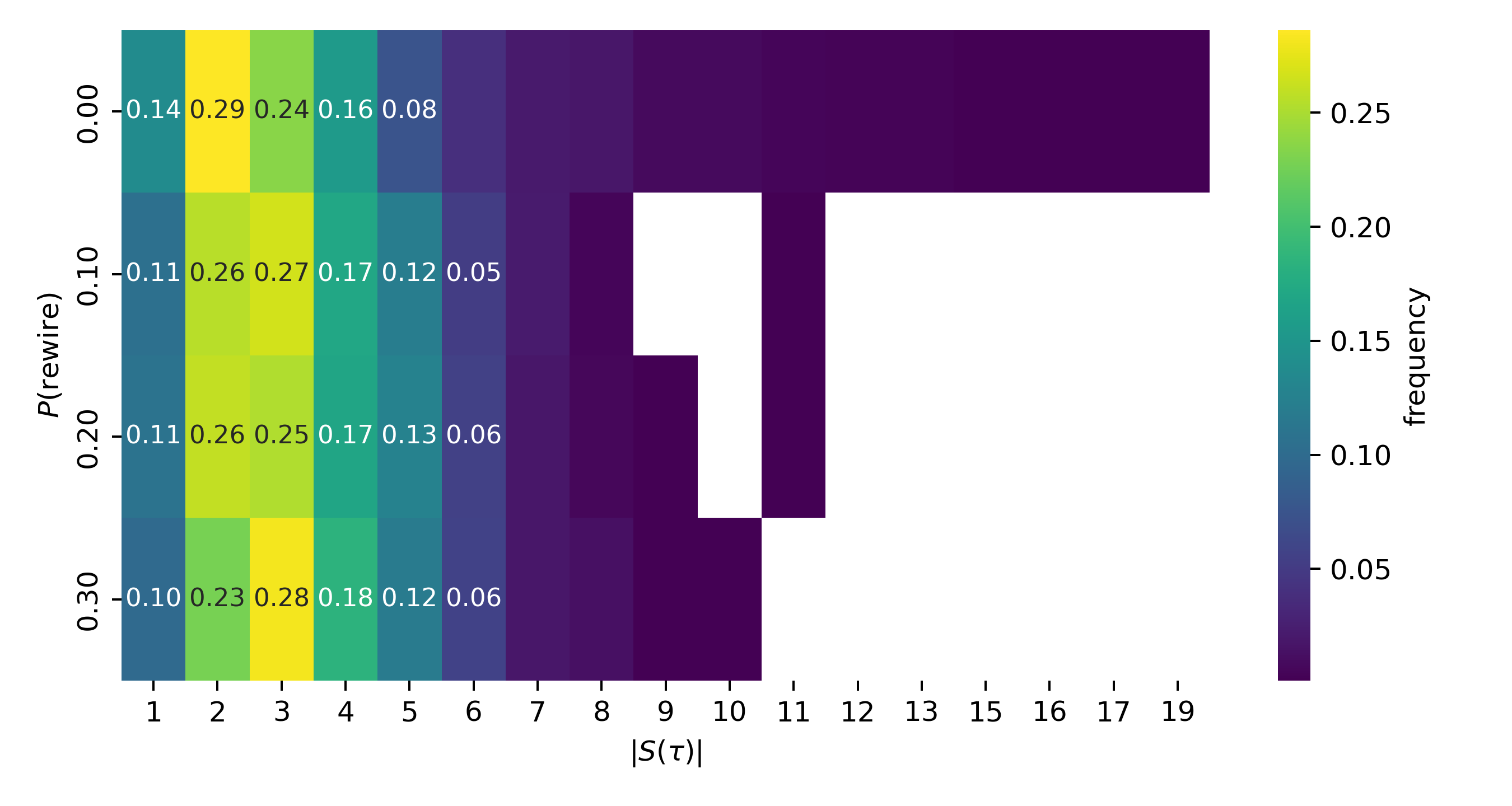}
    \caption{The frequency of $|S(\tau)|$ in the cubic opinion space under bounded confidence $b=15$ and rewiring from circulant networks. Note that empty squares imply no simulation runs with that outcome, and the annotations are suppressed for values $<0.05$ for legibility.}
    \label{fig:HM_grps}
\end{figure}

We see the unexpected outcome that in the cubic opinion space, under a bounded confidence of $b=15$, the rewiring decreases the likelihood of consensus. This is similar to the rewiring studied in \S\ref{sec:rewire_con} starting from the 2-dimensional grid. The expected number of groups however, is steady. We inspect this further by plotting the number of groups at steady state in heat-map format in Figure~\ref{fig:HM_grps}.

We observe that while the likelihood of consensus is going down as the rewiring increases, so does the likelihood of the more extreme number of groups in steady state. Thus rewiring on a circulant under this much bounded confidence breaks up the local community structure at locations where edges are cut, while reinforcing them in places where the edges are rewired.

\begin{figure}[htb]
    \centering
     \begin{subfigure}{0.8\textwidth}
    \includegraphics[width=0.9\linewidth]{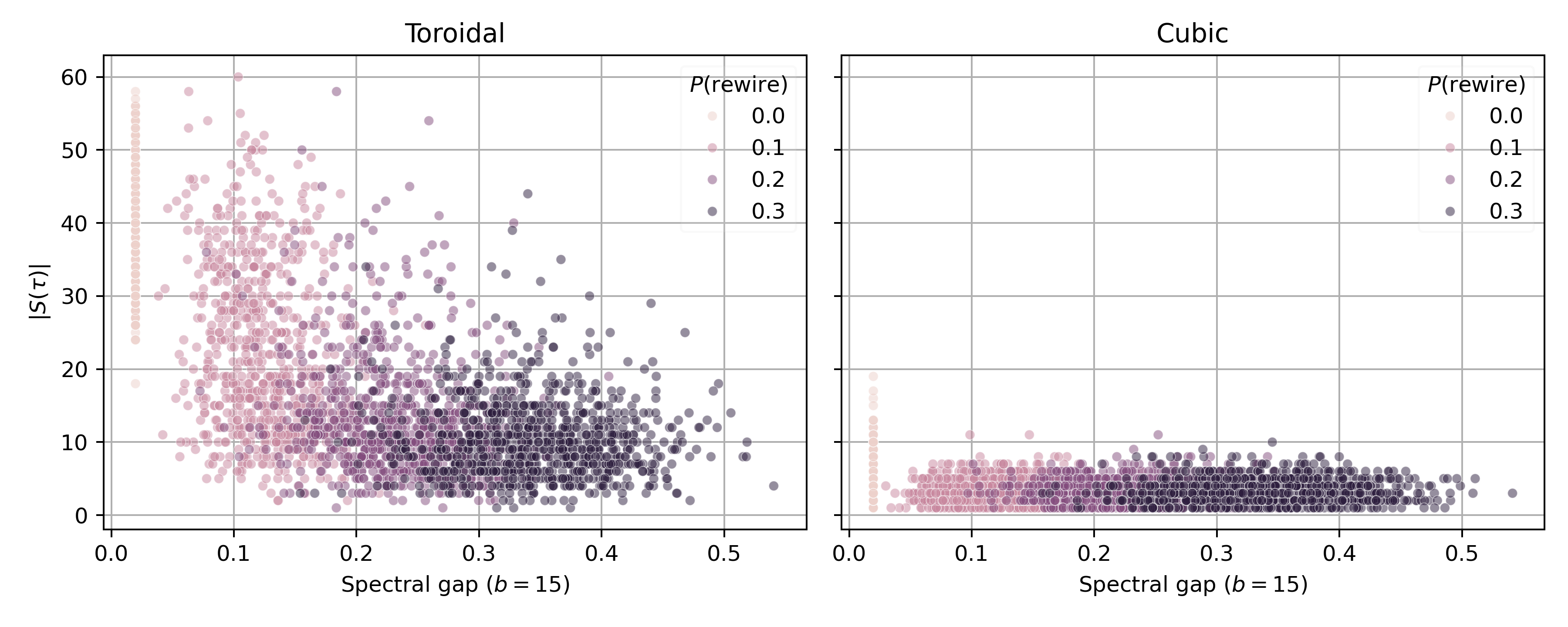}
 \caption{Rewiring from the circulant graph}
    \label{fig:CircRW_SG_Sb15}
    \end{subfigure}\\
         \begin{subfigure}{0.8\textwidth}
    \includegraphics[width=0.9\linewidth]{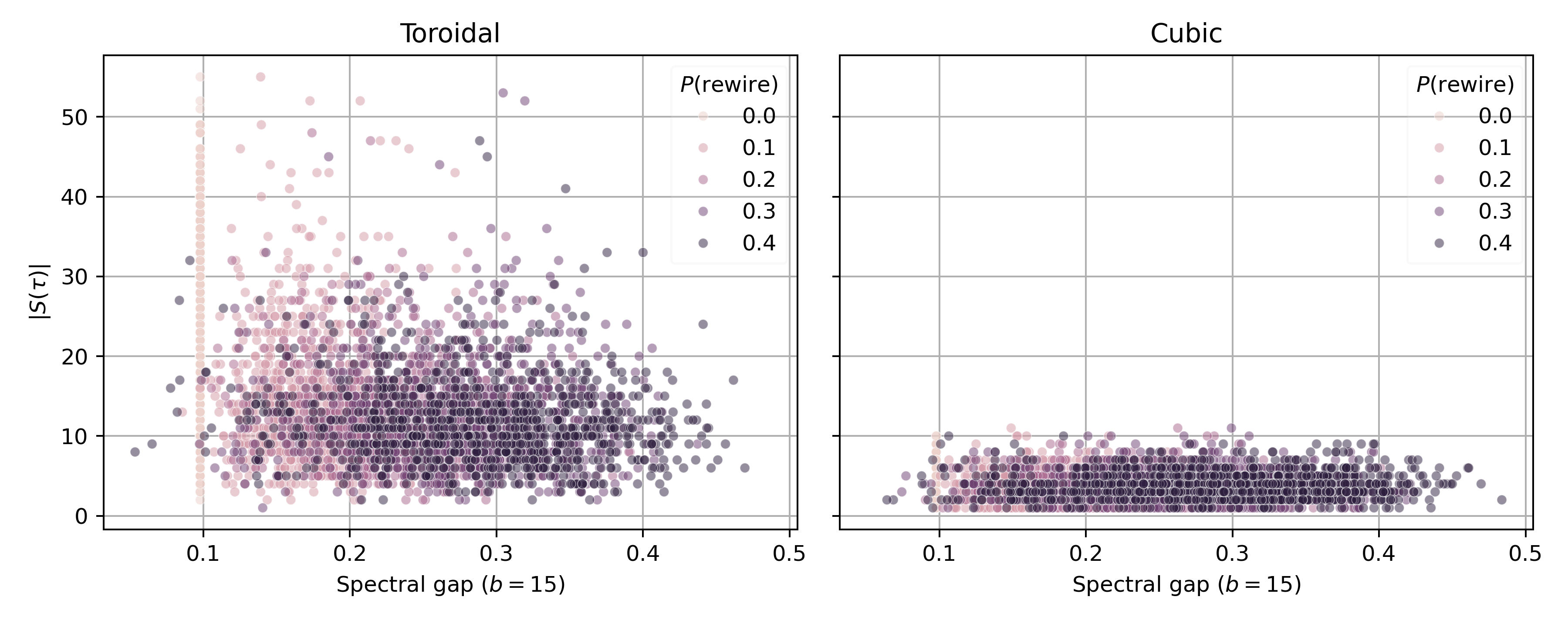}
    \caption{Rewiring from the 2-dimensional grid}
    \label{fig:GridRW_SG_Sb15}
    \end{subfigure}%
       \caption{The number of groups at termination  under rewiring from the (a) circulant network and (b) the grid network in the toroidal and cubic opinion spaces for $b=15$.}
       \label{fig:SG_grps_both}
\end{figure}

Finally, as last illustration of how the rewiring has a different effect when starting from a circulant compared to the 2-dimensional grid we plot the number of groups at termination of the simulation against the spectral gap (defined as the second largest eigenvalue of the Laplacian of the graph) for $b=15$ and the rewiring probabilities $p\in\{0, 0.1, 0.2, 0.3\}$ in Figure~\ref{fig:SG_grps_both}. For the toroidal opinion space we see a clear correlation with a greater spectral gap and a lower number of groups in steady state. This correlation is much weaker for the grid graph.

When starting from the circulant, increasing the probability of rewiring has the effect of increasing the spectral gap. A larger spectral gap typically implies faster mixing times because the graph somehow better connected. This increased connectivity also may explain the decrease in the number of groups in steady state. 

\begin{figure}[!htb]
    \centering
        \begin{subfigure}{0.45\textwidth}
        \includegraphics[width=0.85\textwidth]{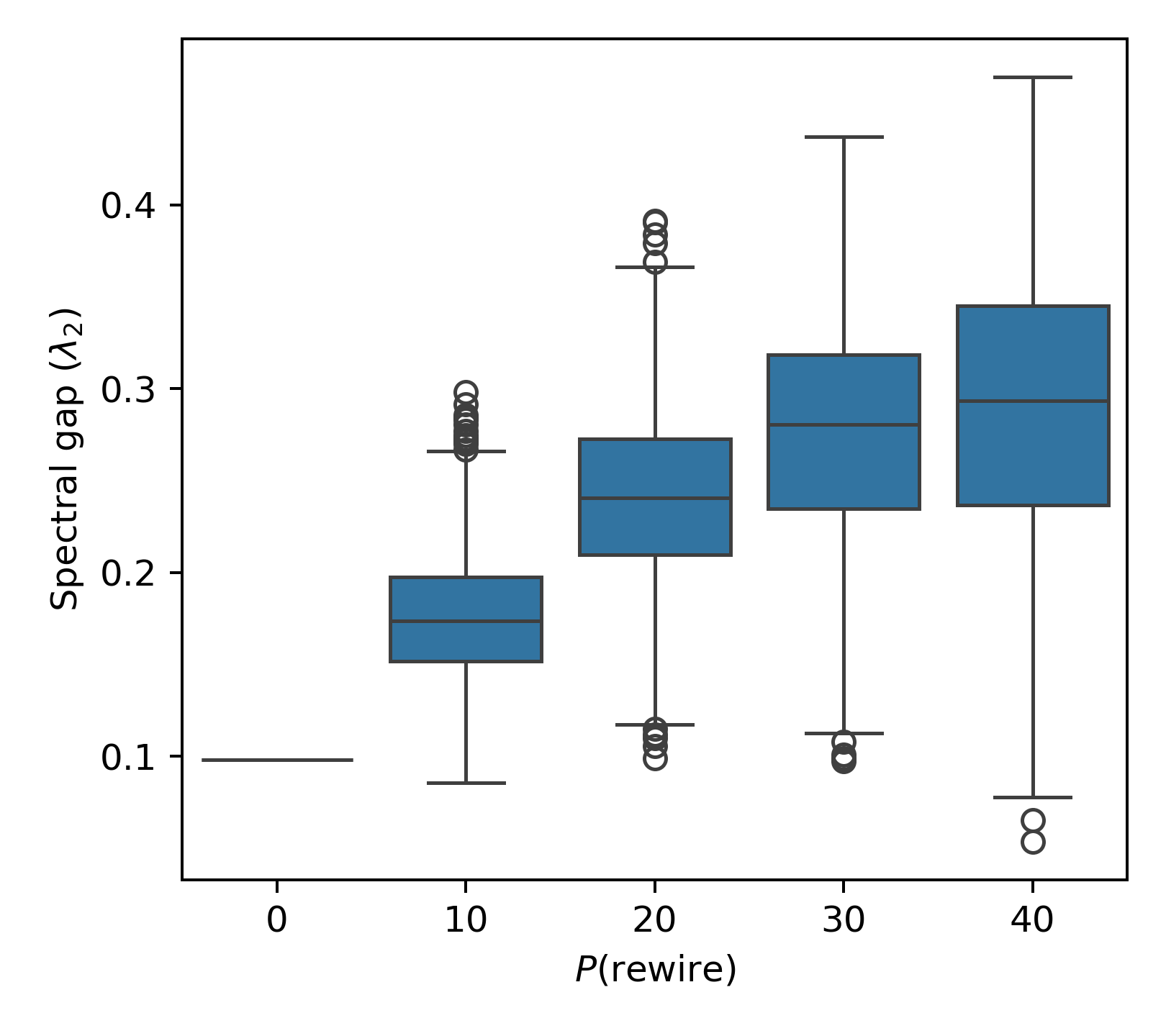}
    \caption{Grid}
    \label{fig:SpecGap_Grid}
    \end{subfigure}%
    \begin{subfigure}{0.45\textwidth}
        \includegraphics[width=0.85\textwidth]{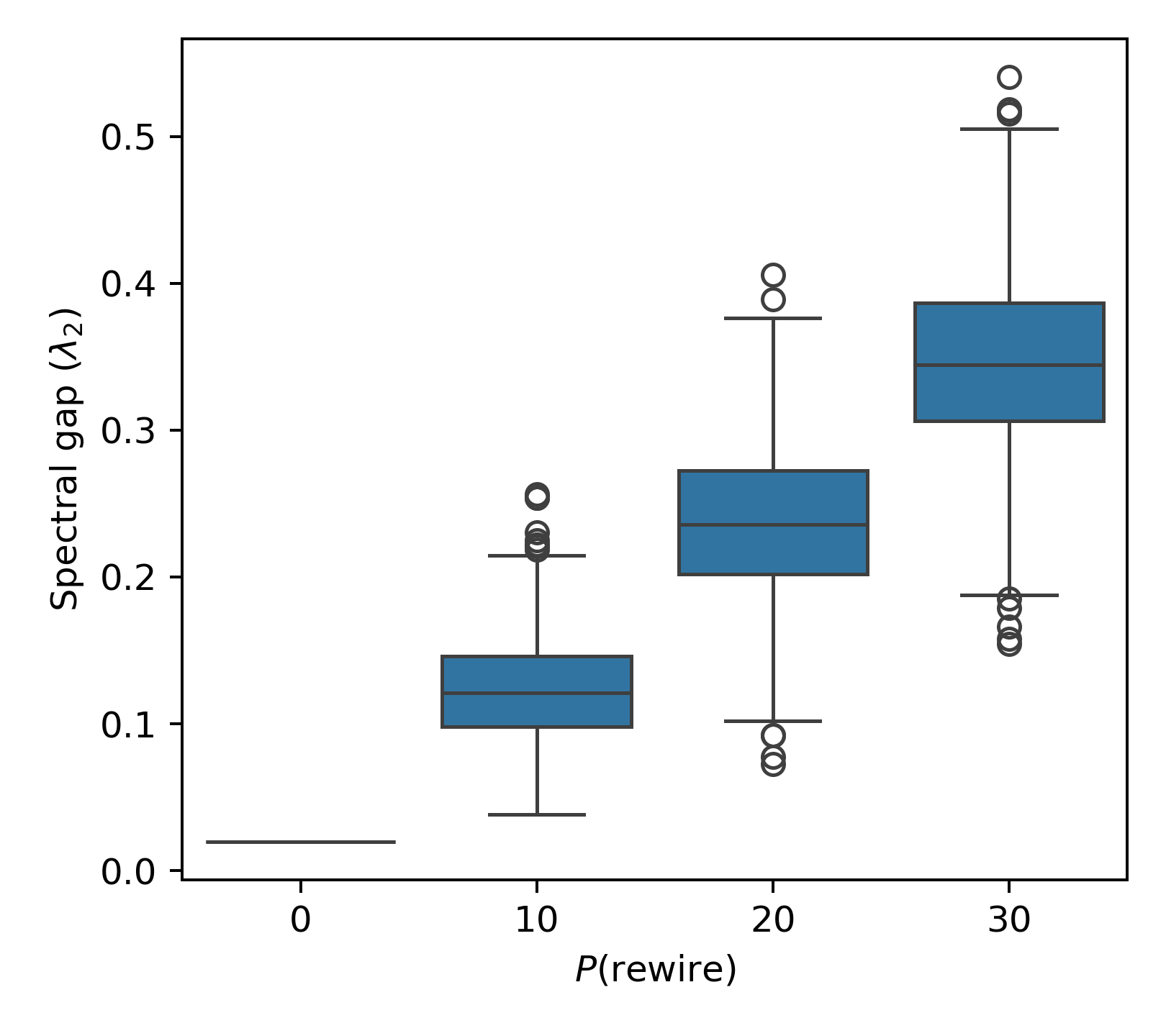}
    \caption{Circulant}
    \label{fig:SpecGap_Circ}
   \end{subfigure}%
    \caption{Boxplots showing how rewiring the grid and the circulant graphs changes the spectral gap which is defined as the second eigenvalue of the Laplacian of the graph.}
    \label{fig:SpectralGapRewiring}
\end{figure}

In Figure~\ref{fig:SpectralGapRewiring} depicting boxplots of the spectral gap for different values of rewiring probability $p$ starting from the 2-dimensional grid and the circulant. We see that rewiring the grid results in a greater spread of spectral gaps with an increasing maximum but also an increasing minimum. Rewiring the circulant, in contrast, leads much more uniformly to an increasing spectral gap. This is a possible explanation for why rewiring increases the likelihood of consensus in the circulant network but not in the grid.

\section{Additional figures illustrating the effect of rewiring the grid on time to steady-state}\label{app:omitFig}
For completeness we include figures omitted from the main text here. In particular, the intermediate value of $b=13$ shows similar time-dynamics as $b=10$ and $b=15$. In Figure~\ref{fig:TW_RW_b13} the absorption time $\tau$ is plotted against the spectral gap of the rewired grid used as the network.

\begin{figure}
    \centering
        \begin{subfigure}{0.33\textwidth}%
        \includegraphics[width=\textwidth]{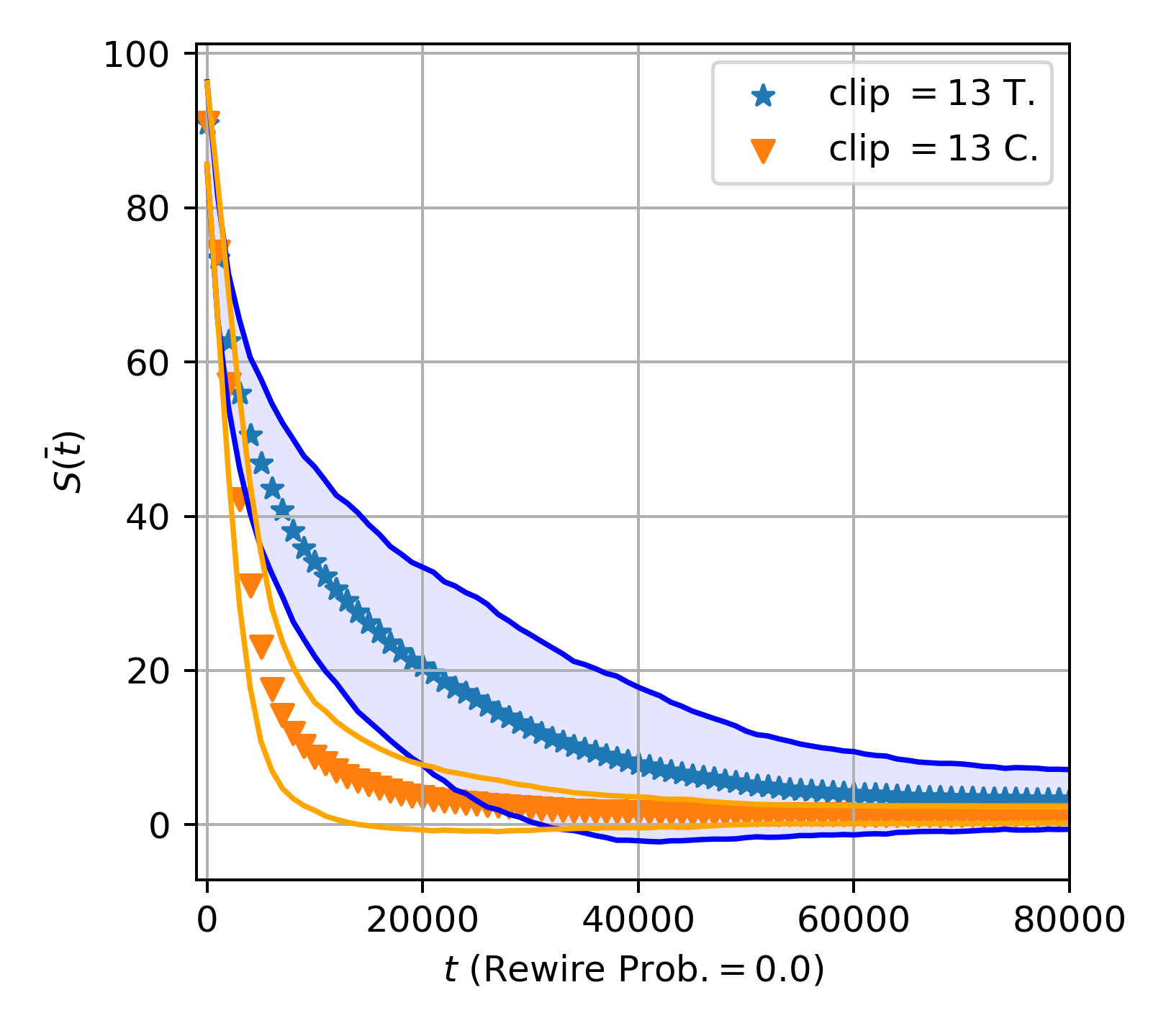}
        \caption{$b=13$, $p=0.0$}
        \label{fig:grps_RW_b13_p0}
    \end{subfigure}%
    \begin{subfigure}{0.33\textwidth}%
        \includegraphics[width=\textwidth]{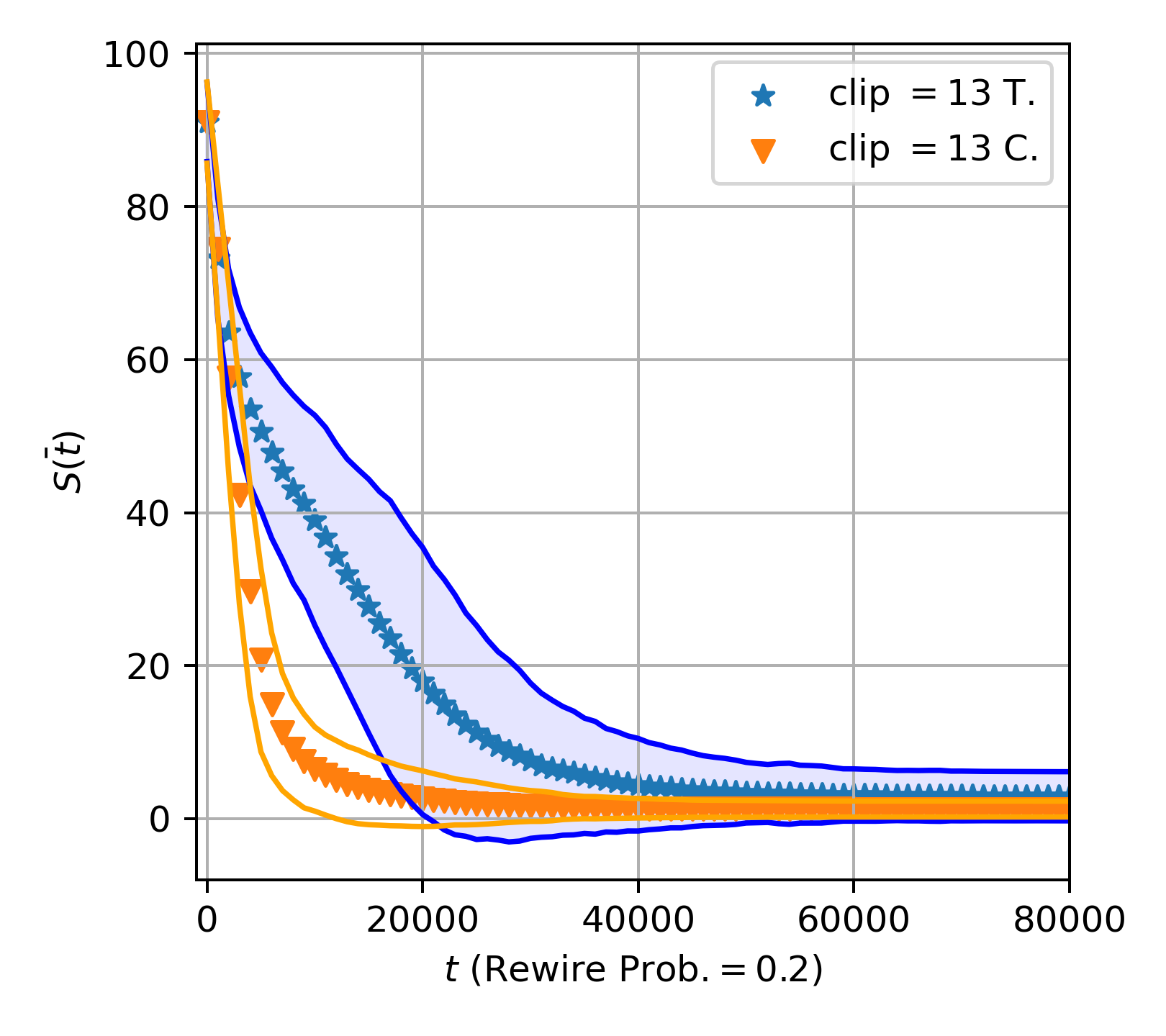}
        \caption{$b=13$, $p=0.2$}
        \label{fig:grps_RW_b13_p2}
    \end{subfigure}%
    \begin{subfigure}{0.33\textwidth}%
        \includegraphics[width=\textwidth]{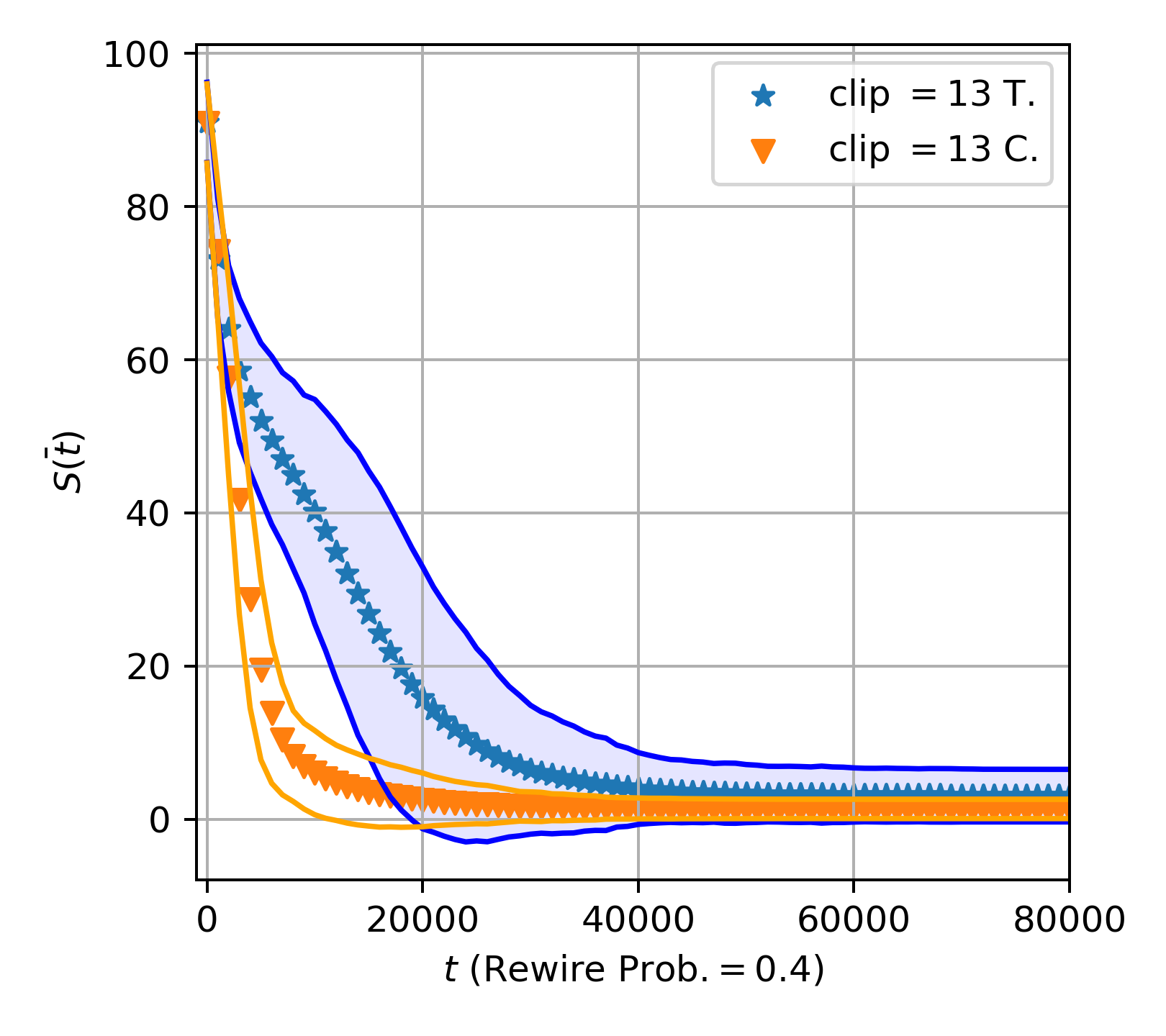}
        \caption{$b=13$, $p=0.4$}
        \label{fig:grps_RW_b13_p4}
    \end{subfigure}\\
    \caption{The mean number of groups over time for the simulations for $b=13$ and rewiring probability $p\in\{0.0, 0.2, 0.4\}$ with $\alpha=2$ on weighted cubic and weighted toroidal opinion spaces (marked C.\ and T.\ respectively).}
    \label{fig:grps_RW_time_b13}
\end{figure}

In Figures~\ref{fig:grps_RW_b13_p0}--\ref{fig:grps_RW_b13_p4} we plot the mean (over simulations) number of groups against simulated time for $b=13$ under rewiring of the grid network.

\begin{figure}[htb!]
    \centering
    \begin{subfigure}{0.85\textwidth}
        \includegraphics[width=0.8\textwidth]{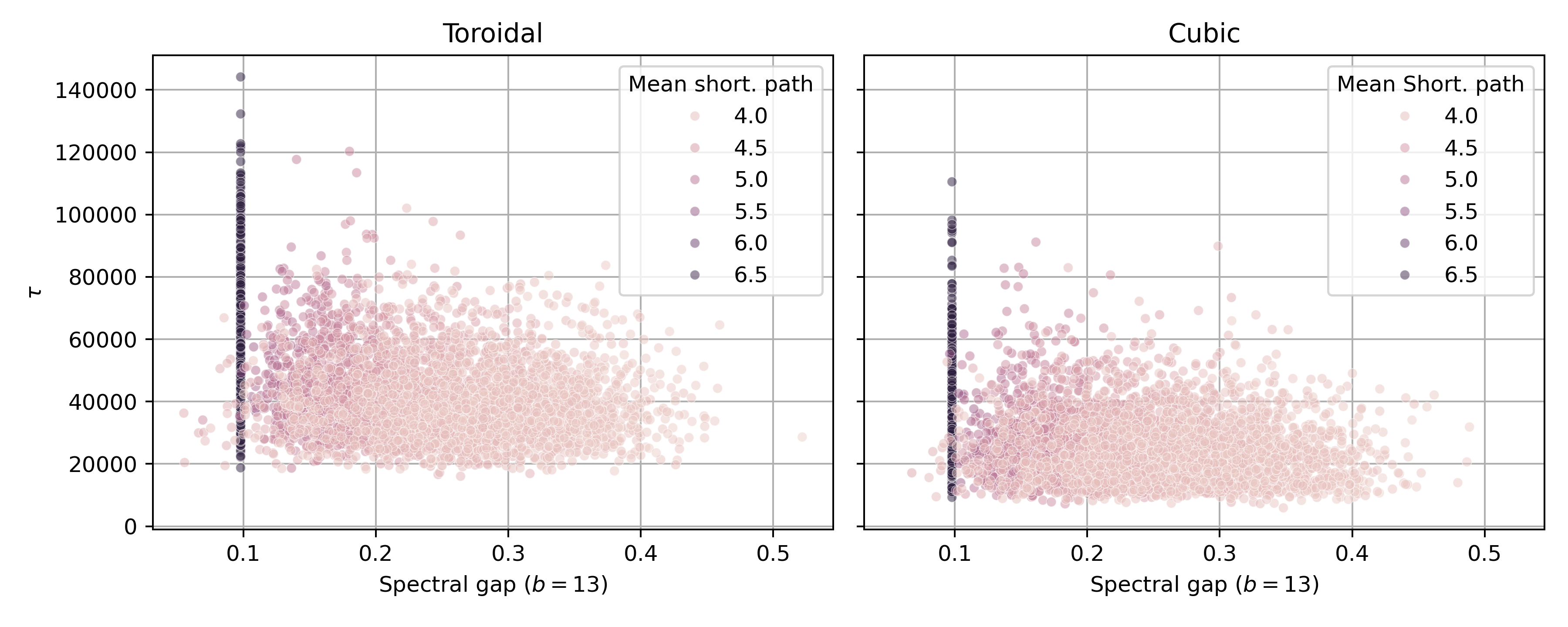}
    \caption{Spectral gap and termination time $\tau$}
    \label{fig:TW_RW_b13}
    \end{subfigure}\\
    \begin{subfigure}{0.85\textwidth}
        \includegraphics[width=0.9\textwidth]{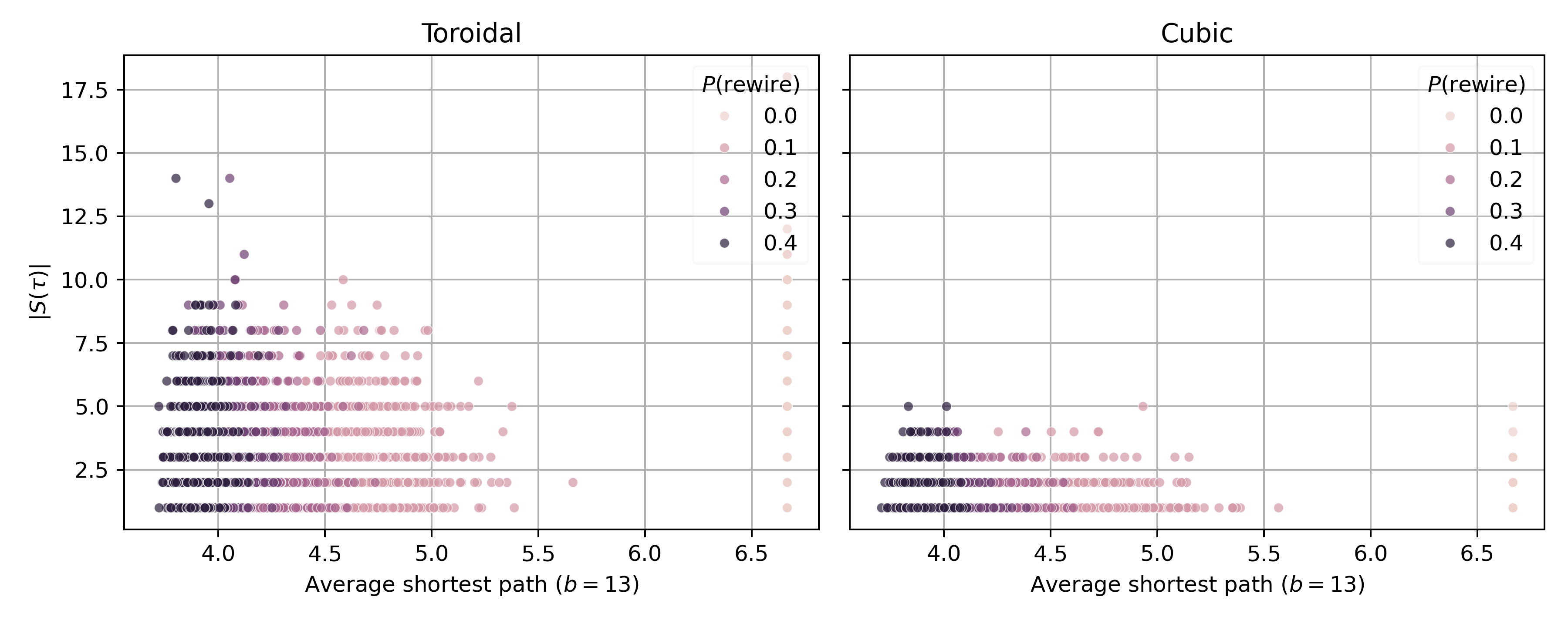}
    \caption{Shortest path and number of groups $|S(\tau)|$}
    \label{fig:S_RW_b13}
    \end{subfigure}\\
    \caption{The effect of the spectral gap  of the network on the termination time $\tau$ of the simulation and the effect of the shortest path on the number of groups at termination for $b=13$.}
    \label{fig:TW_RW_b13}
\end{figure}

\end{document}